\documentclass[none]{imsart}

\RequirePackage{amsthm,amsmath,amsfonts,amssymb} \RequirePackage{natbib} \RequirePackage[colorlinks, citecolor=blue, urlcolor=blue]{hyperref}%% uncomment this for coloring bibliography citations and linked URLs
\RequirePackage{graphicx}%% uncomment this for including figures
\usepackage{amsmath,amsfonts,amssymb}
\usepackage{amsthm,amsmath,amsfonts,amssymb}
\usepackage{mathtools}
\mathtoolsset{showonlyrefs=true}
\usepackage[version=3]{mhchem}
\usepackage{ulem}
\usepackage{nicefrac}
\usepackage{bbm,qtree,forest}
\usepackage[caption = false]{subfig}
\usepackage{multirow} 

\startlocaldefs
\endlocaldefs

\begin{document}

\begin{frontmatter}
\title{Bayesian inference of Latent Spectral Shapes}
%\title{A sample article title with some additional note\thanksref{T1}}
\runtitle{Latent Spectral Shapes}
%\thankstext{T1}{A sample of additional note to the title.}

\begin{aug}
%%%%%%%%%%%%%%%%%%%%%%%%%%%%%%%%%%%%%%%%%%%%%%%
%% Only one address is permitted per author. %% % Only division, organization and e-mail is %% % included in the address.                  %% % Additional information can be included in %% % the Acknowledgments section if necessary. %% % ORCID can be inserted by command:         %% % \orcid{0000-0000-0000-0000}               %%
%%%%%%%%%%%%%%%%%%%%%%%%%%%%%%%%%%%%%%%%%%%%%%%
\author[A,*]{\fnms{Hiu Ching }~\snm{Yip}}, \author[B,*]{\fnms{Daria}~\snm{Valente}}, 
\author[A]{\fnms{Enrico}~\snm{Bibbona}}, \author[B]{\fnms{Olivier}~\snm{Friard}}
\author[A]{\fnms{Gianluca}~\snm{Mastrantonio}}  \and 
\author[B]{\fnms{Marco}~\snm{Gamba}}

%%%%%%%%%%%%%%%%%%%%%%%%%%%%%%%%%%%%%%%%%%%%%%
%% Addresses                                %%
%%%%%%%%%%%%%%%%%%%%%%%%%%%%%%%%%%%%%%%%%%%%%%
\address[A]{Politecnico di Torino, Italy}
\address[B]{Universit\`a di Torino, Italy}
\address[*]{co-first authors}

\end{aug}
 
\begin{abstract} 

This paper proposes a hierarchical spatial-temporal model for modelling the spectrograms of animal calls. The motivation stems from analyzing recordings of the so-called grunt calls emitted by various lemur species.
Our goal is to identify a latent spectral shape that characterizes each species and facilitates measuring dissimilarities between them. The model addresses the synchronization of animal vocalizations, due to varying time-lengths and speeds, with non-stationary temporal patterns and accounts for periodic sampling artifacts produced by the time discretization of analog signals. The former is achieved through a synchronization function, and the latter is modeled using a circular representation of time. To overcome the curse of dimensionality inherent in the model's implementation, we employ the Nearest Neighbor Gaussian Process, and posterior samples are obtained using the Markov Chain Monte Carlo method.
We apply the model to a real dataset comprising sounds from 8 different species. We define a representative sound for each species and compare them using a simple distance measure. Cross-validation is used to evaluate the predictive capability of our proposal and explore special cases. Additionally, a simulation example is provided to demonstrate that the algorithm is capable of retrieving the true parameters.
\end{abstract}

\end{frontmatter}

%%%%%%%%%%%%%%%
% 	Section 1.	Introduction   %
%%%%%%%%%%%%%%%
\section{Introduction}\label{sec:intro}

The study of vocal repertoires is an essential step in understanding animal behavior. Communication plays a crucial role in individual interactions and can provide crucial information about social relations and hierarchical organization within a group, such as  mating behavior, territoriality, group dynamics, and even predator-prey interactions. Understanding these behaviors is fundamental for conservation efforts and ecological research.
Vocalizations are often collected as sound recordings and their analysis is most typically performed using spectrograms, making bioacoustic analysis a form of time-frequency analysis. Processing and learning from such high-dimensional data require statistical methods or automated algorithms.
An overview of the contemporary computational methods in bio-acoustic analysis is summarized in the recent survey by \cite{sainburg}. The most common practice is the application of feature engineering methods, which generally involve manual or automatic selection of a set of basis-features for quantitative comparison. There are numerous types of such basis-features in bio-acoustic such as pitch, amplitude envelope, Wiener entropy, and others.  \cite{kershenbaum2016} provided additional background on several paradigms of features and suggested a protocol to analyze them with the appropriate methodologies. Studies conducted by \cite{gamba2007} and \cite{gamba2016} are examples of bioacoustic analysis that manually select features such as pitches, behaviors and sex of the vocal animals to be independent predictors or responses variables for linear model fitting. 
Analyzing the full acoustic dataset without feature selection often involves dimensionality reduction and clustering and for instance, \cite{valente2019} implemented a dimensionality reduction method to construct a probabilistic weighted graph using the distances between the sound intensity points of the spectrograms.
%For instance, a study implemented dimensionality reduction to construct a probabilistic weighted graph using distances between sound intensity points of spectrograms, which was then projected to a lower-dimensional space for clustering. Unsupervised methods avoid the pitfalls of arbitrarily selecting features but may not offer detailed insights into the spectral shape of acoustic structures.
%This paper proposes a spatial temporal non-parametric model for acoustic data in spectrogram representation that can be used by bio-acoustic researchers.
There are a few papers where GP have been proposed for analyzing bioacoustic signals and modeling spectrograms. For instance, in \cite{tavakoli}, GPs are used in linguistics to model the variation of the British accent across Great Britain. In \cite{WilkinsonW}, a spectral mixture of GPs is utilized for denoising and source separation tasks of audio signals.

The motivating data of this work is a set of lemur vocal signals that were recorded in Madagascar over 12 years. The lemurs of Madagascar are particularly fascinating because they evolved in a diverse and isolated environment from the extensive continental shelves. Moreover, among lemurs there is one of the few ``singing primates'', Indri indri, a species possessing rhythmic capabilities similar to that of human musicality \citep{DEGREGORIO2021R1379} and a rich vocal repertoire \citep{valente2019}. Lemurs can emit different signals with different aims (e.g., alarm, territorial defense, group cohesion and coordination, anti-predator mobbing) and their classification and
interpretation have been subject to several investigations \citep{pozzi2010,valente22}. Among the various vocalizations observed in lemurs, we focus on the``grunt'', a nasal vocalization mostly ubiquitous among lemurs  \citep{Macedonia94} and emitted in various contexts, including group movements in forests \citep{sperber2017}. Interestingly, while the grunt is shared across several lemur species, each species produces its own distinct version due to differences in vocal tract morphology \citep{gamba2012}. Such acoustic divergence has been hypothesized to allow discrimination across species,  and to have played a role in speciation. Therefore, investigating 
grunting behavior could provide insights into the unique evolutionary history of these primates \citep{Zimmermann2017}. Moreover, grunts belong to the so-called close calls, i.e. low-intensity, low-frequency calls emitted when individuals are in close spatial proximity with one another, common in many primate species. Being lemurs among the most basal primates, investigating their close-calling behavior could shed light on the evolutionary roots of primate gregarious behavior, as well as social and cognitive abilities \citep{Pfluger}.

%The motivating data of this work is a set of vocal signals of lemurs that were recorded in Madagascar over $12$ years. The lemurs of Madagascar are particularly fascinating because they evolved in a diverse and isolated environment from the extensive
%continental shelves. Moreover, lemurs are known to be the only "singing primates" that possess rhythmic capabilities similar to that of human musicality, and a huge vocal repertoire \cite{gamba2016}. They can emit different types of calls with different aims: alarm, location markers, group cohesion, predator mobbing. The classification and interpretation of these vocalizations have been subject to several investigations \citep{pozzi2010}.
%Among the various vocalizations observed in lemurs, we focus on the "grunt" sound \textcolor{red}{To Marco and Daria: A couple of lines to explian why the grunt is interesting}, a nasal vocalization common among true lemurs and emitted in various situations, including group movements in forests \cite{sperber2017}. While the grunt sound is shared across several lemur species, each species produces its own distinct version due to differences in vocal tract morphology \cite{gamba2012}.
 
The primary aim of this paper is to infer a representation of the grunt sound that allows for quantitative comparison across different species. To achieve this goal, two important factors must be considered. Firstly, each grunt call has its own specific duration and different samples of the same call may be emitted at different speeds, with the speed possibly changing during the call itself, resulting in accelerated vocal signals. Secondly, periodic artifacts may arise when processing the signal with the Short Time Fourier Transform (STFT) using a fixed time window \citep{Graps1995,kumar1997}. However, the periodicity of these artifacts only appears in the original real-time frame and not in the stretched common frame.
We address these two effects by taking a weighted average of two independent stationary Gaussian processes (GPs), each defined to take into account one of the two aforementioned factors. We introduce a \textit{synchronization function} in the first GP to align and stretch the signal, while a correlation function based on a cyclic representation of time is used to model the cyclic artifact. It should be noted that, even if correlation function based on a cyclic component have been already used, to the best of our knowledge, this is the first time that the cyclic period is considered a random variable, see for example  \cite{mastrantonio} or \cite{Shinichiro2017}.
By formalizing the model under a Bayesian framework, we are then able to obtain a realization of a spectrogram, that we call \textit{latent spectral shape} of the sound or \textit{representative} sound, which  can be considered representative of the grunt calls emitted by a species.

The large size of the data, makes the inference challenging from a computational point of view: this problem is often referred to as the "Big $n$ problem" \citep{Jona2013b}. The first step we employed to solve this issue is to use the Nearest Neighbor Gaussian Process (NNGP) method introduced by \cite{datta}, which is well-suited for Bayesian analysis of large spatial data. The NNGP requires setting the maximum number of neighbors to be considered in the approximation, and although it is generally assumed that 10 to 15 give reasonable results \citep{datta}, our model is more complex than the usual GP. Therefore, we prefer to have a larger set of neighbors, but, on the other hand, computational costs increase. For this reason, instead of using the entire dataset, we decided to work with a subset of the most informative observations, which are the longer ones.
The model is estimated using real data, and for each species, we present the estimate of the latent sound. Additionally, we provide a method to compare and measure distances between these sounds using a quadratic distance. We discuss the results and use this distance to construct a phylogenetic tree, which is then compared with established trees from the literature. We employed a cross-validation setting to test the utility of all components of our model in describing the data. While the results vary across species, in most cases, introducing our features improves predictive performance. Using a simulated example, we demonstrate that all parameters of our model can be learned from the data.

The paper is organized as follows. Section \ref{sec:data} contains a description of the available dataset. In Section \ref{sec:model}, the proposed model is explained. Section \ref{sec:compd} discusses the issues involved in the implementation of the model, including identifiability, priors selection, and the sampling of the  latent spectral shape. Section \ref{sec:apply} presents the analysis of the results obtained from the application of the model on the dataset. Section \ref{sec:conclude} concludes the paper and discusses some future work. A simulation example is shown in the Appendix.

%%%%%%%%%%%%%%%
% 	Section 2.	The data set   %
%%%%%%%%%%%%%%%	

% coordinates := time coordinate (dimension) + frequency coordinate (dimension) 
% each time point in time coordinate 

\section{The data set}\label{sec:data}
\begin{table}[t]
	\begin{tabular}{|c|cccccccc|}
	\hline 
	 & EC & ER & FL & FU & II & MA & MO & PD \\
	 \hline 
	 \# recordings		& 1966  & 6594  & 1174  &1041 &1144 & 3615 & 1492 & 145 \\
	 min length			& 0.025 & 0.020 & 0.020  &0.021 &0.011 & 0.020 &0.031 & 0.027 \\ 
	 max length			& 4.016 & 0.374 & 0.474 &0.513&0.238&2.057&1.135& 0.105 \\  \hline
	 \# length > 0.1 		& 1311  & 4531  & 542   & 464  & 25  & 1599 & 1277  & 2 \\
	 \# length > 0.2 		& 412    & 180   &15   & 33    & 4    & 204   & 268    & 0 \\
	 \# length > 0.3 		& 87      & 12     & 2    & 13   & 0    & 47     & 73      &  0 \\
	\hline 
	\end{tabular}\caption{The ``grunt'' call-type for each species: (from top to bottom) the number of recordings, the minimum and maximum recorded time-length, and the number of recordings that last longer than $0.1$, $0.2$, and $0.3$ seconds, respectively.}
	\label{tab1}
\end{table}
\begin{figure}[t]
	{\subfloat[ER - sound 2]{\includegraphics[scale=0.25]{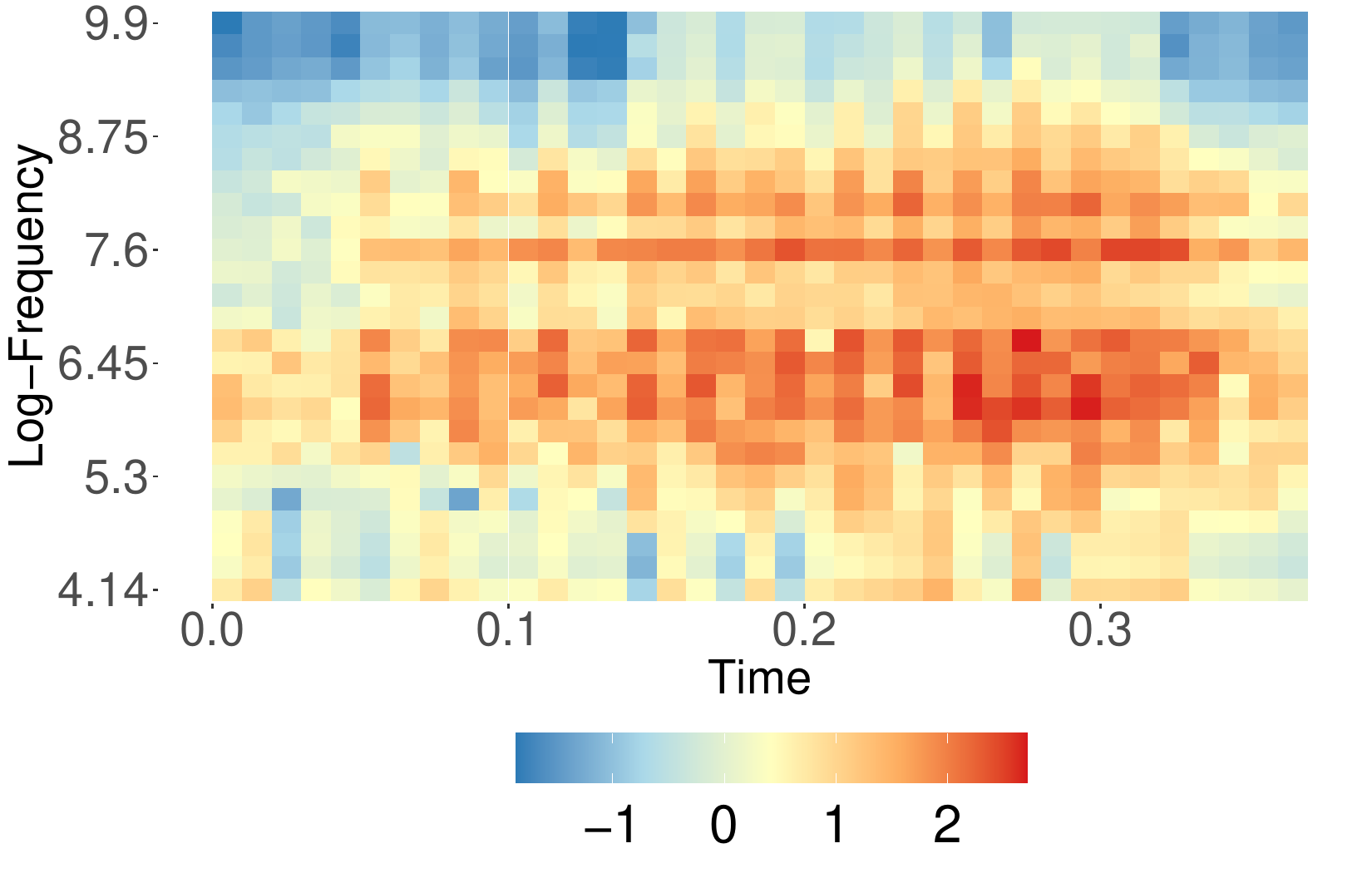}}} 
	{\subfloat[ER - sound 26]{\includegraphics[scale=0.25]{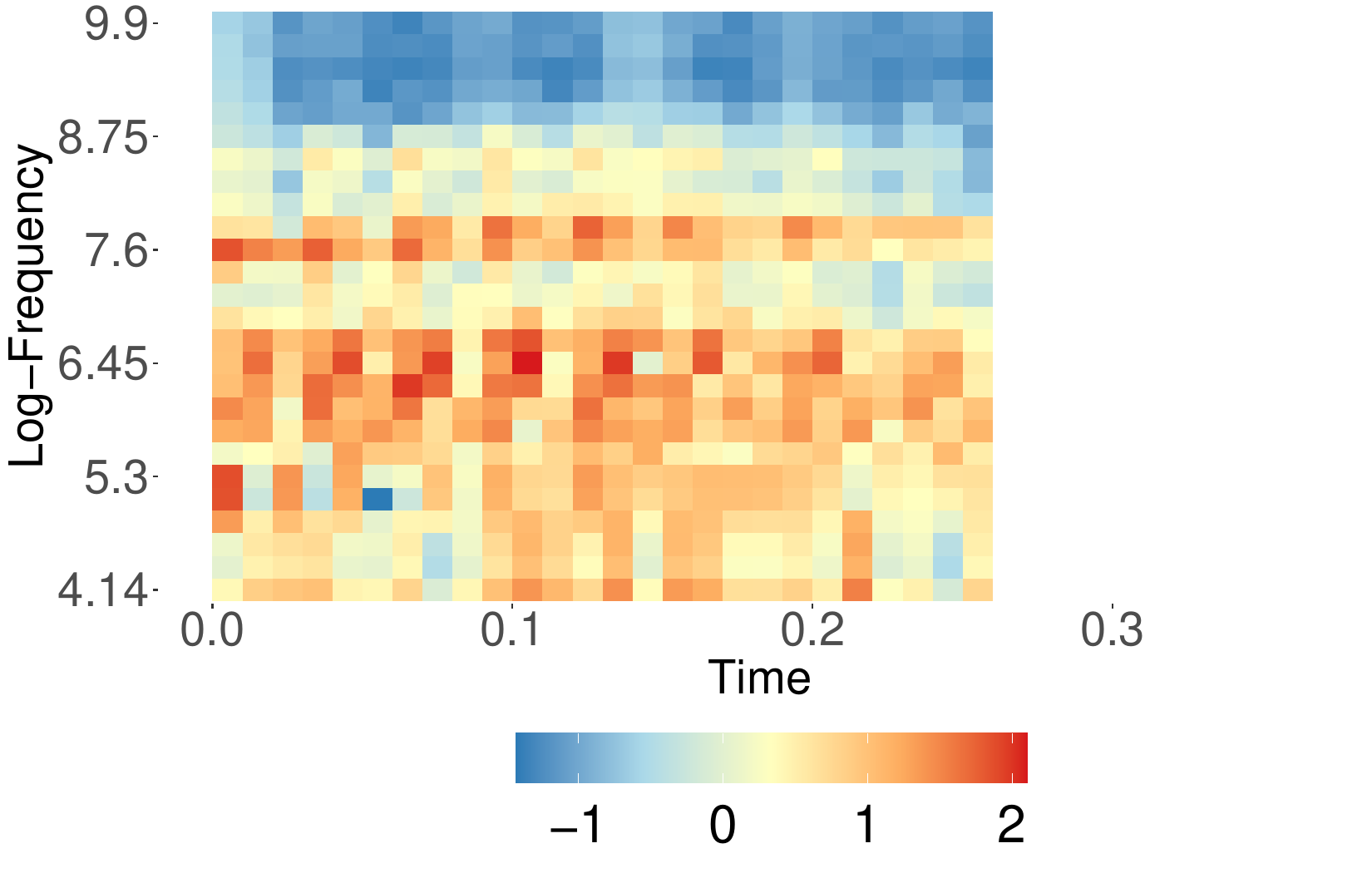}}}
	\caption{Spectrograms of  two recorded signals of species ER. The $x$-axis and $y$-axis correspond to the discretized time and frequency domains, respectively. The two figures have the same  $x$-axis.}
	\label{fig:data1}
 \end{figure}
 \begin{figure}[t]
	{\subfloat[MO - sound 2]{\includegraphics[scale=0.25]{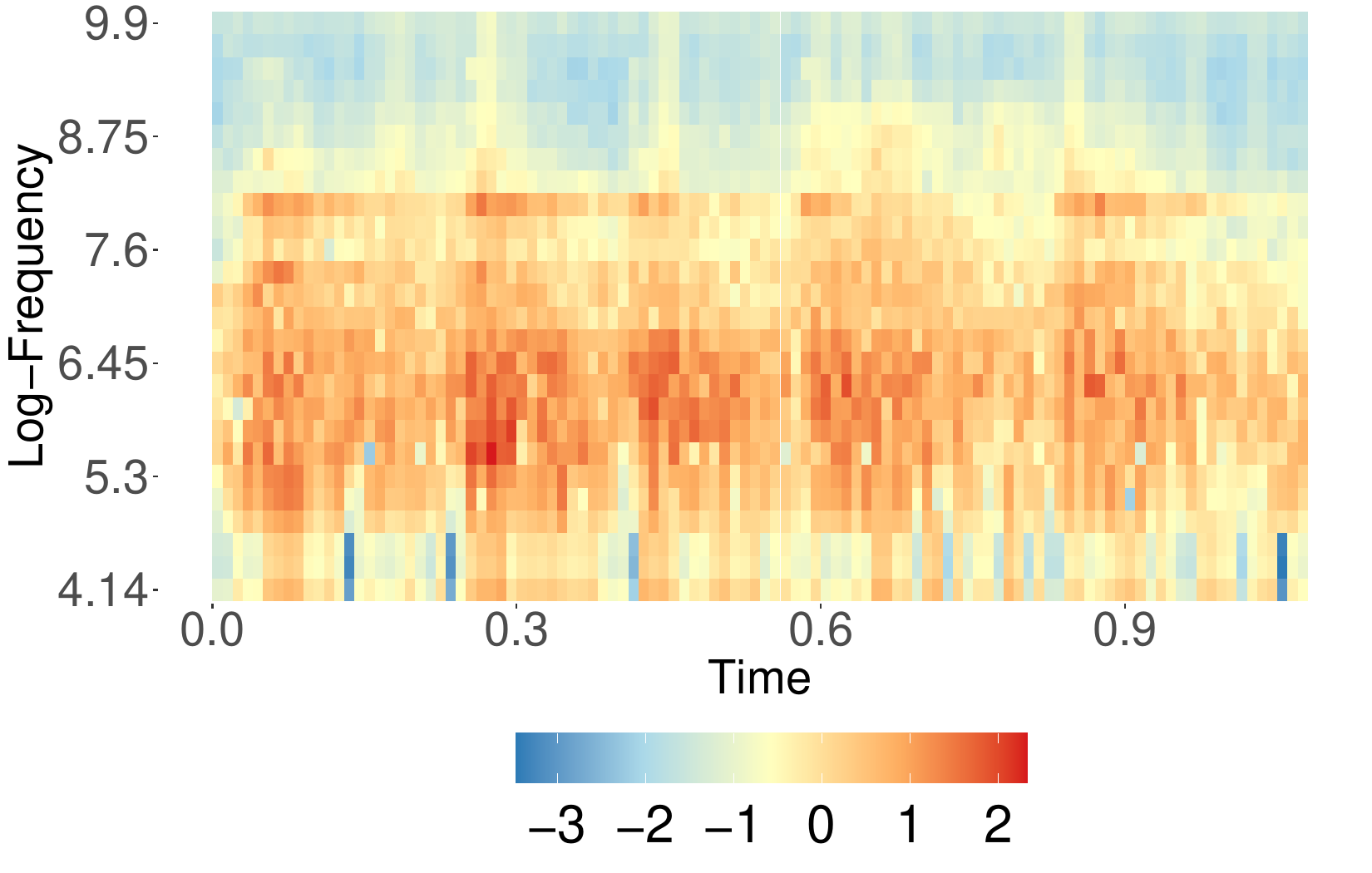}}} 
	{\subfloat[MO - sound 100]{\includegraphics[scale=0.25]{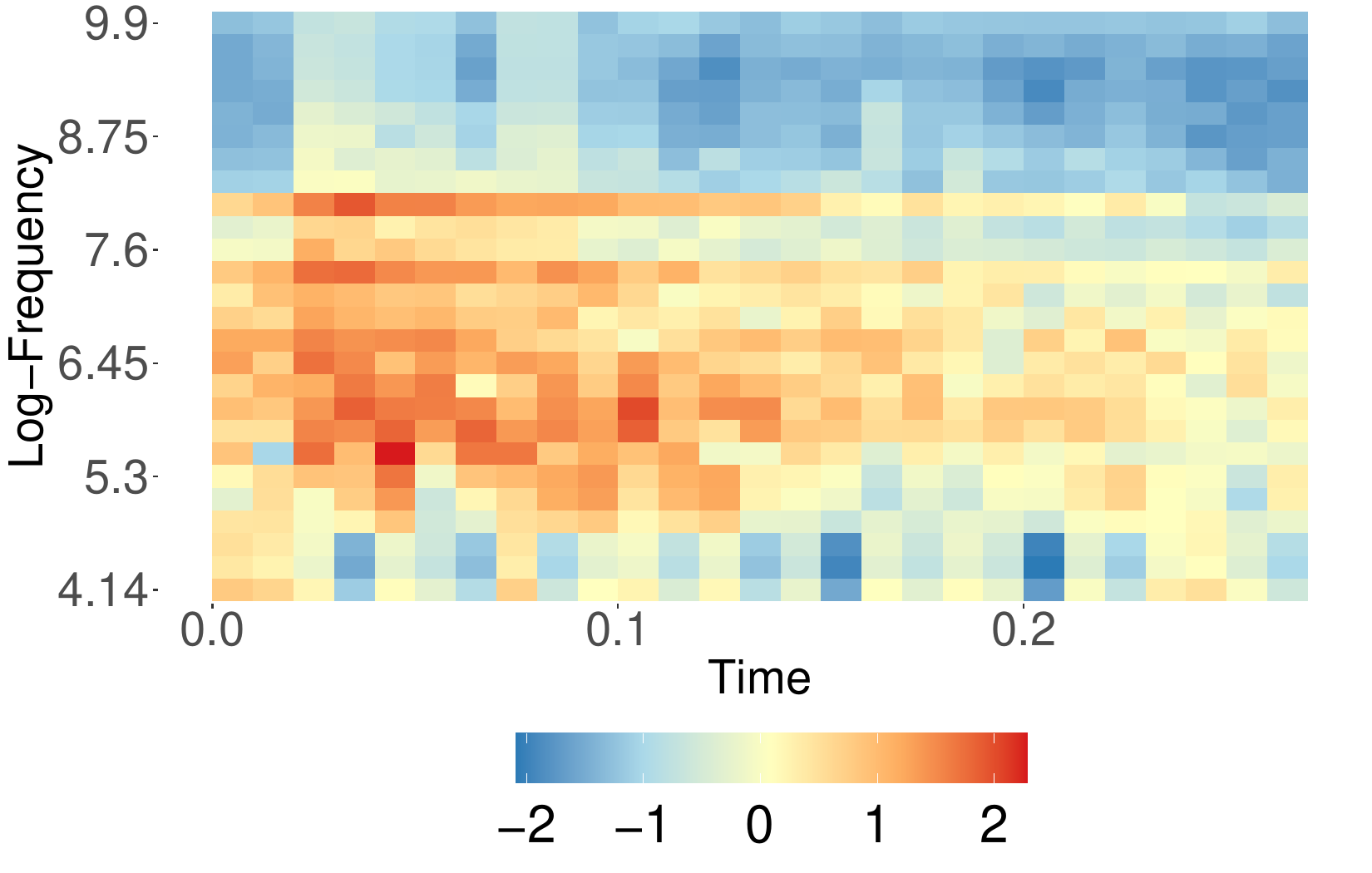}}}
	\caption{pectrograms of  two recorded signals of species MO. The $x$-axis and $y$-axis correspond to the discretized time and frequency domains, respectively. The figures have different   $x$-axis.}
	\label{fig:data2}
 \end{figure} 
 
%\textcolor{red}{
%The motivating dataset comprises vocal signals from lemurs, recorded over 12 years in Madagascar at four different forest sites: Analamazaotra Special Reserve, Mantadia National Park, Mitsinjo Forest Station, and Maromizaha Forest New Protected Area. Recordings were made using Sennheiser shotgun ME 66 and ME 67, and AKG CK 98 microphones. The signals were captured at a sampling frequency of 44.1 kHz with solid-state digital audio recorders including the Marantz PMD671, SoundDevices 702, Olympus S100, and Tascam DR-100MKII, with a 16-bit amplitude resolution. Vocalizations were recorded from a distance of 2-10 meters, and efforts were made to ensure the microphone was oriented towards the vocalizing animal.
%Each sound was categorized  into one of the  vocal types identified in  \cite{Maretti2010}, e.g., clacsons, hums, roars, based on its  acoustic  characteristic and evolution. 
%Among these types there is the ``grunts'', which is the main focus of this work. The grunts were emitted by numerous individuals across eight different species from two taxonomic families, Lemuridae and Indriidae. All Lemuridae species were congeneric, including Eulemur coronatus (EC), Eulemur rubriventer (ER), Eulemur flavifrons (FL), Eulemur fulvus (FU), Eulemur macaco (MA), and Eulemur mongoz (MO). The two Indriidae species were Indri indri (II) and Propithecus diadema (PD).}

The motivating dataset comprises vocal signals from eight different lemur species belonging to two taxonomic families, Lemuridae and Indriidae. All Lemuridae species were congeneric, including Eulemur coronatus (EC), Eulemur rubriventer (ER), Eulemur flavifrons (FL), Eulemur fulvus (FU), Eulemur macaco (MA), and Eulemur mongoz (MO). Indriidae species were Indri indri (II) and Propithecus diadema (PD). We sampled numerous individuals over 19 years in Madagascar and various institutions maintaining lemurs ex-situ. We recorded Indri indri at four forest sites from 2005 to 2018: Analamazaotra Special Reserve, Mantadia National Park, Mitsinjo Forest Station, and Maromizaha Forest New Protected
Area. We recorded Propithecus diadema in the Maromizaha Forest New Protected Area from 2014 to 2016. We collected vocalizations of E. fulvus from 28 individuals (17 males and 11 females). We recorded the vocal samples in the western dry forest of Ankatsabe-Analabe; Mariarano, Mahajanga II District, Boeny region, (Madagascar), in the humid forest of Maromizaha; Anevoka, Alaotra-Mangoro region, (Madagascar), and in the following zoos: Banham Zoo (United Kingdom), Colchester Zoo (United Kingdom), Linton Zoo (United Kingdom), Mulhouse Zoo (France), Parco Natura Viva (Italy), and Parc Botanique et Zoologique de Tsimbazaza (Madagascar). We collected the audio recordings of E. rubriventer from 16
individuals (7 males, nine females) at the Mulhouse Zoological and Botanical Garden (France), from 24 individuals (15 males, nine females) recorded in five European zoos in 1999: Apenheul Primate Park (Netherlands), Kölner Zoo (Germany), Linton Zoo (United Kingdom), Banham Zoo (United Kingdom), Twycross Zoo (United Kingdom). A third corpus comprised vocalizations recorded from seven captive individuals (five males, two females) at the Parc Botanique et Zoologique de Tsimbazaza (Madagascar) in 2005. The dataset of E. mongoz calls comprised vocalizations of 54 individuals (32 males, 22 females), collected from 2008 and 2102 in Madagascar and Comoros. We recorded the mongoose lemurs'
vocalizations in Antsilahiza (Mitsinjo District, 80 km from Katsepy; Madagascar), in the New Protected Area of Bombetoka-Belemboka (Madagascar), and Anktsabe-Analabe forest (Mariarano, Madagascar). All these sites are located in the Boeny region. In Comoros, we recorded vocalizations in Dziani (Anjouan Island), Dzitso (Anjouan Island), and Ouallah (Mohéli Island). In addition, we also collected vocalizations from 19 captive individuals (twelve males, seven females) in 1999 in the following zoos: Mulhouse Zoo (France), Parc Botanique et Zoologique de Tsimbazaza (Madagascar), Banham Zoo (United Kingdom), Linton Zoo (United Kingdom), Colchester Zoo (United Kingdom), and Parco Natura Viva
(Italy). We recorded  Eulemur coronatus calls from 37 captive lemurs (27 males, 10 females) hosted in the following zoos: Mulhouse Zoo (France), Parc Botanique et Zoologique de Tsimbazaza (Madagascar), Banham Zoo (United Kingdom), Linton Zoo (United Kingdom), and Twycross Zoo (United Kingdom). Details can be found  in \cite{gamba23008}. We chose vocalizations of E. flavifrons from a dataset of 37 individuals (22 males, 15 females) recorded in the following zoos: Apenheul Primate Park (Netherlands), Banham Zoo (United Kingdom), Linton Zoo (United Kingdom), Mulhouse Zoo (France), Parco Natura Viva (Italy), and Parc Botanique et Zoologique de Tsimbazaza (Madagascar). We recorded E. macaco
from 21 individuals (13 males, 8 females) in the following zoos: Lemur Reserve (Madagascar), Parc Botanique et Zoologique de Tsimbazaza (Madagascar), Parco Natura Viva (Italy), and St Louis Zoo (U.S.A). We made recordings using Sennheiser shotgun ME 66 ME 67 and AKG CK 98 microphones. The signals were captured at a sampling frequency of 44.1 kHz with solid-state digital audio recorders, including the Marantz PMD671, SoundDevices 702, Olympus S100, and Tascam DR-100MKII, with a 16-bit amplitude resolution. We recorded vocalizations from a distance of 2-10 meters and ensured the microphone was oriented towards the vocalizing animal. Each sound was categorized  into one of the  vocal types
identified in  \cite{Maretti2010}, e.g., clacsons, hums, roars, based on its  acoustic  characteristic and evolution. 
Among these types there is the "grunt", which is the main
focus of this work. 
Each grunt emitted by an individual lemur has been  converted into a spectrogram, using the short-time Fourier transform implemented in the Praat 6.0.28 software \citep{praat}. The frequency  is divided into one-third octave bands ranging from 63 to 20,000 Hertz, evenly segmented into 26 bins in log-scale with a step size of 0.23 in log-frequency. Meanwhile, the temporal axis is discretized with a constant time-step of 0.01 seconds.
While all spectrograms have the same number of log-frequency bins, the number of time bins may vary due to differences in recording lengths, as summarized in Table \ref{tab1}. 
Given that shorter sounds are more challenging to analyze and convey less information, the analysis in the paper focuses solely on the 100 longest sounds available for each species, striking a balance between precision and efficiency.
Figures \ref{fig:data1} and \ref{fig:data2} depict examples of these spectrograms.

The sounds of different species are assumed independent, and hence we can define the model for a single species and, for simplicity in notation, we omit indices indicating species, and all introduced notation pertains to the recorded signals of this single species. Let $N$ denote the total number of recorded signals, $H$ denote the number of log-frequencies, which remains constant across all signals, and $T_i$ denote the number of observed time-points in the $i$-th recorded signal, where $i = 1,\dots,N$. Define $\mathcal{T}_i$ and $\mathcal{H}$ as the sets of time log-frequency points for the $i$-th spectrogram, which, in our case, are defined as
\begin{equation}
\mathcal{T}_i = \{ 0.01(k-1) \ | \ k = 1 , \dots , T_i \}, \qquad \mathcal{H} = \{0.23k+\log 63 \ | \ k = 1, \dots, H\},
\end{equation}
and let $l_i = \max{\mathcal{T}i}$ represent the length of signal $i$. Each point in the $i$-th spectrogram, denoted as $y_{i,j,k}$ is measured in dB SPL, and can be seen as  the realization of a two-dimensional process $\mathcal{Y}_i(\cdot, \cdot) \in \mathbb{R}$, existing in continuous time and continuous space, evaluated at $(t_j,h_k)$, where $t_j$ denotes the $j$-th element of $\mathcal{T}_i$ and $h_k$ denotes the $k$-th element of $\mathcal{H}$. Additionally, let $\mathbf{y} = (\mathbf{y}_1^\top, \mathbf{y}_2^\top, \dots, \mathbf{y}_N^\top)^\top$ denote the collection of all recorded signals, where the elements of $\boldsymbol{y}_i$ are sorted in ascending order of time and increasing values of log-frequencies within each time coordinate.

%%%%%%%%%%%%%%
% 	Section 3.  The Model   %
%%%%%%%%%%%%%%
\section{The Model}\label{sec:model}

The two-dimensional process $\mathcal{Y}_i(t,h)$ is assumed to be a noisy version of a process $\mathcal{A}_i(t,h)$, modelled as follows:
\begin{align}\label{eq:model}
\mathcal{Y}_i (t,h) & = \mu_i + \mathcal{A}_i \left(t, h \right) + \epsilon_i(t,h), \\
\mathcal{A}_i (t,h) & = \sigma \left( \sqrt{\lambda} \mathcal{W}_1 \big( \psi(t|\boldsymbol{\chi}_i), h \big)+\sqrt{1-\lambda} \mathcal{W}_2(t,h) \right),
%\epsilon_i(t,h) & \stackrel{ind}{\sim} \text{GP} (0, \tau_i)
\end{align}
Here, $\mathcal{W}_1(\cdot,\cdot)$ and $\mathcal{W}_2(\cdot,\cdot)$ are two unknown functions $ \mathbb{R}^+\times \mathbb{R} \rightarrow \mathbb{R}$, which we estimate assuming they are realizations of GPs, $\psi(t|\boldsymbol{\chi}_i)$ is a function what we will introduce in Section \ref{subsec:W1},  and $\epsilon_i(t,h)$ is an error term assumed to follow a Gaussian distribution. Hence, we have
\begin{align}\label{eq:model_v2}
%\mathcal{Y}_i (t,h) & = \mu_i + \mathcal{A}_i \left(t, h \right) + \epsilon_i(t,h) \
%\mathcal{A}_i (t,h) & = \sigma \left( \sqrt{\lambda} \mathcal{W}_1 \big( \psi(t|\boldsymbol{\chi}_i), h \big)+\sqrt{1-\lambda} \mathcal{W}_2(t,h) \right) \
\mathcal{W}_1(d,h) & \sim \text{GP} \left( 0,\mathcal{C}^g( \cdot, \cdot ; \boldsymbol{\theta}) \right), \\
\mathcal{W}_2(t,h) & \sim \text{GP} \left( 0,\mathcal{C}^c( \cdot, \cdot ; \boldsymbol{\theta}) \right),\\
\epsilon_i(t,h) & \stackrel{ind}{\sim} \text{GP} (0, \tau_i),
\end{align} 
where $\mathcal{C}^g( \cdot, \cdot ; \boldsymbol{\theta})$ and $\mathcal{C}^c( \cdot, \cdot ; \boldsymbol{\theta}) $ are correlation functions.
Here, $\mu_i \in \mathbb{R}$ represents the mean sound intensity level and $\tau_i \in \mathbb{R}_{> 0}$ represents the variance. For convenience in notation, we group them together into $\boldsymbol{\eta}_i = (\mu_i, \tau_i^2)$, and define $\boldsymbol{\eta} = (\boldsymbol{\eta}_1, \boldsymbol{\eta}_2, \dots, \boldsymbol{\eta}_N)^\top$.
The terms $\mu_i$ and $\epsilon_i(t,h)$ can be regarded as nuisance parameters in this context as they account for factors such as environmental noise, individual variability, and variations in the global intensity of the sound due to unknown distances from the receiver.

From \eqref{eq:model} and \eqref{eq:model_v2}, we can see that the process $\mathcal{A}_i(t,h)$ is defined as the weighted average of two GPs, which are introduced to serve distinct purposes. The component $\mathcal{W}_1(d,h)$ operates as a stationary process defined over a latent time dimension $d$ and log-frequency $h$. Here, $d$ represents a dimension to which the actual time of each recording is non-linearly mapped by a signal-specific transformation encoded in the \textit{synchronization function}  $\psi(t|\boldsymbol{\chi}_i)$ (see Section \ref{subsec:W1}).  Time  $d$ is termed the \textit{shared-time} and $\mathcal{W}_1(d,h)$ is referred to as the \textit{shared-time process} or \textit{shared-time component}. The parametric construction of the synchronization function and its correlation function $\mathcal{C}^g(\cdot, \cdot ; \boldsymbol{\theta})$ will be discussed in Section \ref{subsec:W1}.
On the other hand, the component $\mathcal{W}_2(t,h)$ is an additional GP utilized exclusively for modeling periodic sampling artifacts, as detailed in Section \ref{subsec:W2} along with the correlation function $\mathcal{C}^c(\cdot, \cdot ; \boldsymbol{\theta})$. We refer to $\mathcal{W}_2(t,h)$ as the \textit{artifact-process} or \textit{artifact-component}.

It is noteworthy that although $\mathcal{A}_i(t,h)$ is call-specific, $\mathcal{W}_1(d,h)$ and $\mathcal{W}_2(t,h)$ are shared across all spectrograms. Consequently, the distinction between two latent sounds, $\mathcal{A}_i(t,h)$ and $\mathcal{A}_{i'}(t,h)$, relies solely on different synchronization functions.
Hence, a representative sound for the entire set of spectrograms emitted by a species could be encapsulated into a new $\mathcal{A}_{\ell}(t,h)$, with $\ell = N+1$, where its length $l_{\ell}$ and synchronization function $\psi(t|\boldsymbol{\chi}_{\ell})$ must be selected carefully to be able to interpret $\mathcal{A}_{\ell}(t,h)$, which it will be discussed in Section \ref{sec:abst}. We will refer to $\mathcal{A}_{\ell}(t,h)$
 as the  \textit{latent spectral shape} of the sound.

%%%%%%%%%%%%%%  
% 	Sub-section 3.1		W1  %
%%%%%%%%%%%%%%
\subsection{The shared-time component $ \mathcal{W}_1(d,h) $ and the synchronization function $ \psi(t;\boldsymbol{\chi}_i) $}\label{subsec:W1}
\begin{figure}[t]
	{\subfloat[]{\includegraphics[scale=0.25]{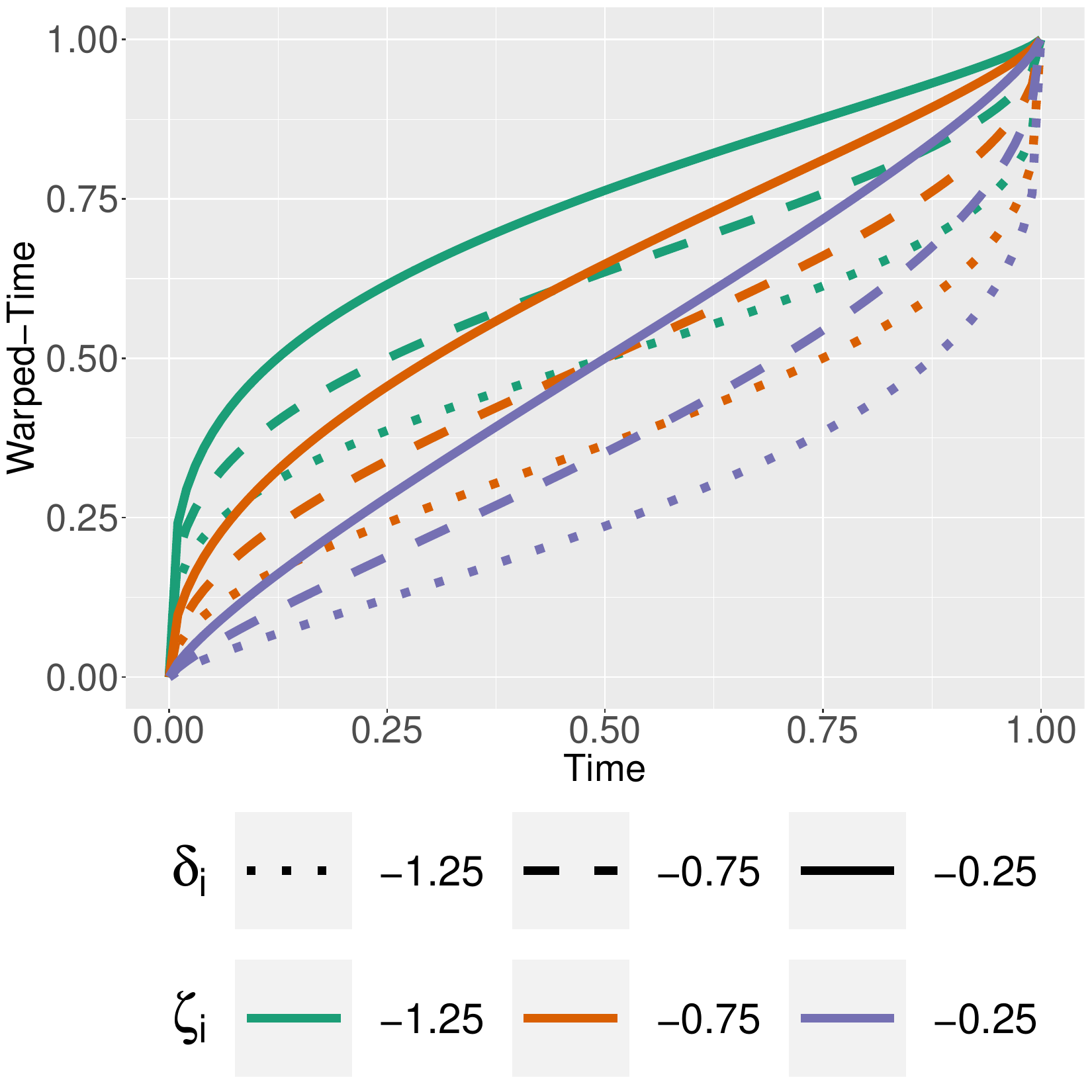}}} 
	{\subfloat[]{\includegraphics[scale=0.25]{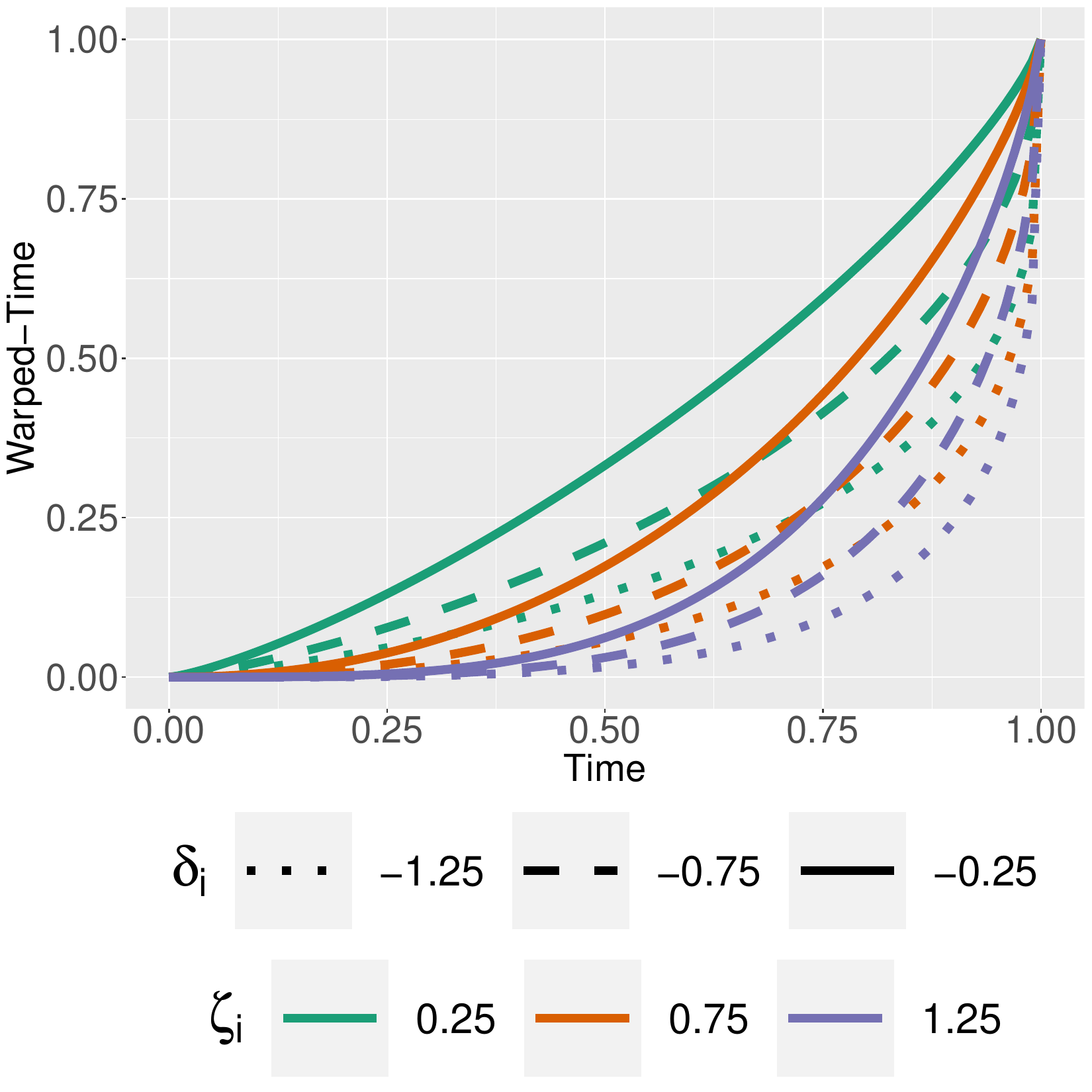}}} \\
	{\subfloat[]{\includegraphics[scale=0.25]{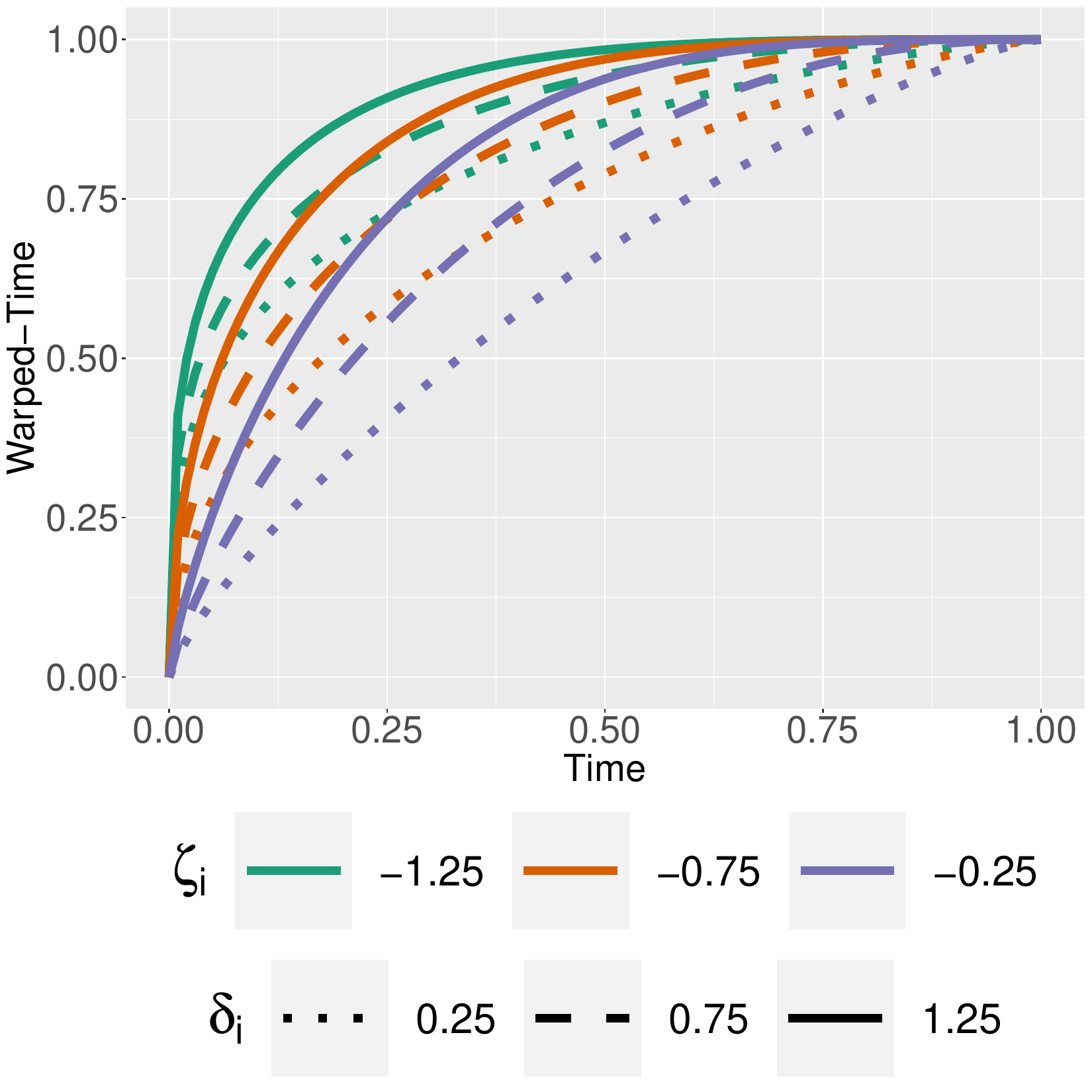}}} 
	{\subfloat[]{\includegraphics[scale=0.25]{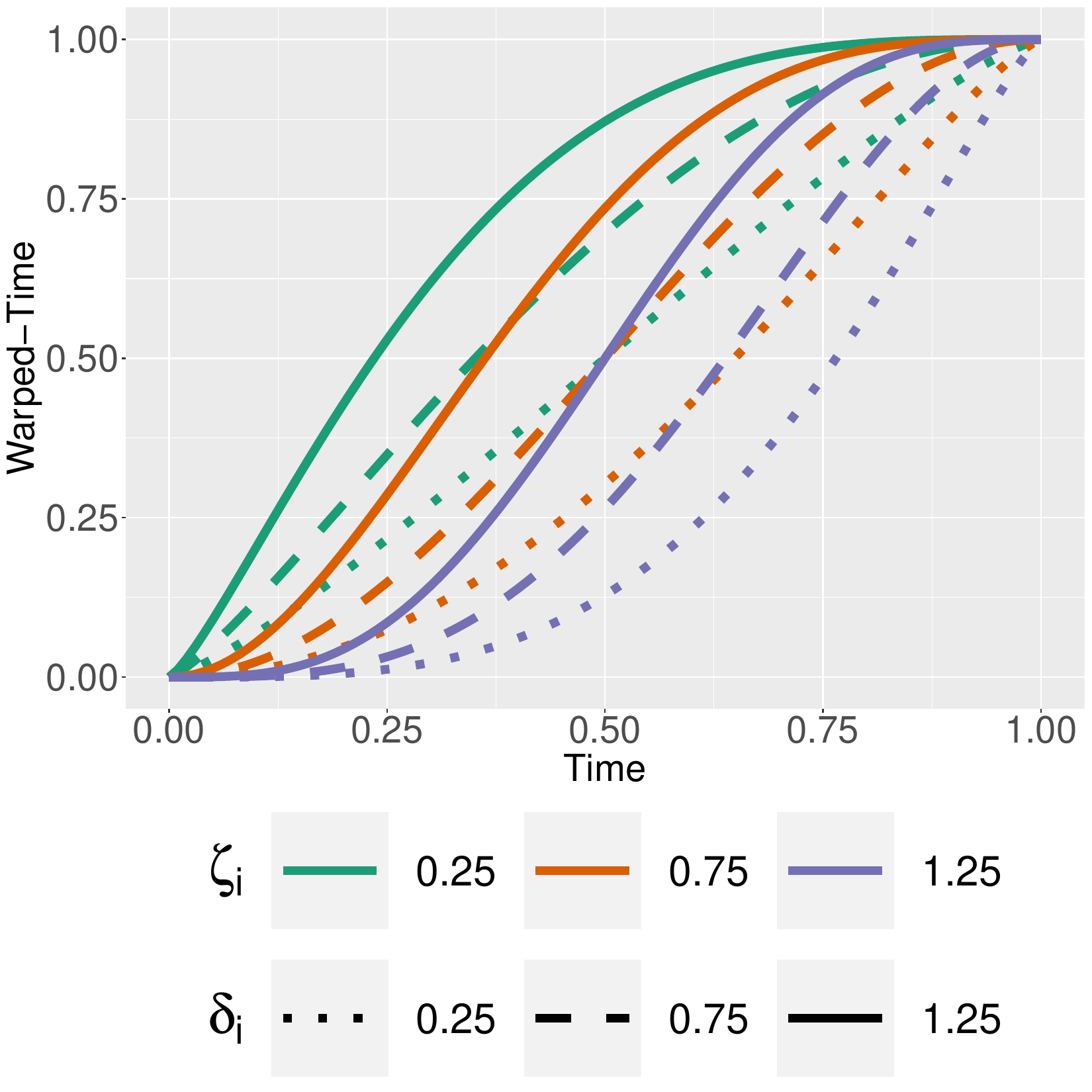}}} 
	\caption{Time-warping function $\Omega(q|\boldsymbol{\xi}_i)$ under different parametric values.}\label{fig:warpf}
 \end{figure}

From Figure \ref{fig:data1}, several notable features of the data become apparent. Regarding the time dimension, it is evident that the recordings exhibit varying durations, and they may not be perfectly aligned. For instance, Figure \ref{fig:data1} (b) could correspond to a shifted portion of Figure \ref{fig:data1} (a), or it might represent the same sound emitted at a higher speed. Alternatively, the actual sound in Figure \ref{fig:data1} (a) might commence around 0.05 seconds instead of 0. A similar analysis can be applied to Figure \ref{fig:data2}.
Hence, it becomes necessary to introduce a family of functions capable of synchronizing the sounds along the time dimension, which is defined as follows:
\begin{equation} \label{eq:psi} 
	\psi(t ; \boldsymbol{\chi}_i) = \alpha_i + \beta_i  \Omega \left(\frac{t}{l_i} ; \boldsymbol{\xi}_i \right) l_i,
\end{equation}
where $\boldsymbol{\chi}_i = ( \alpha_i , \beta_i, \boldsymbol{\xi}_i) $, and $\Omega(q ; \boldsymbol{\xi}_i)$ represents the non-linear \textit{time-warping function}, defined as the Beta cumulative distribution function (CDF):
\begin{equation} \label{eq:beta}
	\Omega\left(q ; \boldsymbol{\xi}_i\right) = \frac{\Gamma(\exp \zeta_i+\exp \delta_i)}{\Gamma(\exp \zeta_i)\Gamma(\exp \delta_i)}	\int_{0}^{q} x^{\exp \zeta_i-1}(1-x)^{\exp \delta_i-1} d x, \,\qquad q \in [0,1],
\end{equation}
with warping parameters $\boldsymbol{\xi}_i = ( \zeta_i, \delta_i )$.
This synchronization function maps the real-time interval $[0,l_i]$ of the $i$-th spectrogram to the interval $[\alpha_i, \alpha_i+\beta_il_i]$ of the shared-time, introducing a non-linear warping encoded by the function $\Omega(q ; \boldsymbol{\xi}_i)$. 
In principle, various choices for the warping function can be made, as long as it remains continuous and strictly increasing in $[0,1]$, satisfying the boundary conditions $\Omega(0 ; \boldsymbol{\xi}_i) = 0$ and $\Omega(1 ; \boldsymbol{\xi}_i) = 1$. We selected the Beta CDF due to its ease of interpretation and flexibility, as it has few parameters to avoid over-parametrization.

Figure \ref{fig:warpf} presents several examples of the time-warping function under different parameterizations. Notably, a special case of equation \eqref{eq:beta} arises when $\delta_i = \zeta_i = 0$, resulting in $\Omega\left(q ; \boldsymbol{\xi}_i\right) = q$, a linear function where the Beta CDF reduces to the Uniform CDF.
It is evident from Figure \ref{fig:warpf} that if $|\zeta_i| \gtrapprox 0.75$ or $|\delta_i| \gtrapprox 0.75$, the time-warping becomes so severe that the derivative of $\Omega\left(q ; \boldsymbol{\xi}_i\right)$ evaluated at $q \approx 0$ or $q \approx 1$ approaches either $0$ or $\infty$. This would imply that the sound can be emitted very slowly and/or extremely fast, which is unjustifiable from an application perspective. Therefore, constraints on the values of the warping parameters are necessary, and this will be discussed in Section \ref{subsec:priors} where the prior distributions over the warping parameters are introduced.

%The correlation function $\mathcal{C}^g( \cdot, \cdot ; \boldsymbol{\theta})$ for the shared-time component is defined as 
%\begin{equation} \label{eq:gnei}
%	\mathcal{C}^g( |h-h'|, |d-d'| ; \boldsymbol{\theta}) = \frac{1}{\phi_d |d - d'| + 1} \exp \left ( - \frac{\phi_h |h-h'|}{(\phi_d |d - d'| + 1)^{\rho/2} } \right )
%\end{equation}
%where $\phi_h>0$, $\phi_d>0$ and $\rho \in [0,1]$ are some of the entries  of the vector $\boldsymbol{\theta}$. This is a parametrization of the Gneiting correlation function formulated in the work of \cite{gneiting} and is chosen for its non-separability and flexibility. % depends only on the shared-time lag and log-frequency lag as a consequence of stationarity within the shared-time dimension. 
%The non-separable parameter $\rho$ is the time-frequency interaction parameter, while $\phi_d$ and $\phi_h$ are the time and frequency decays respectively. 
%It should be noted that there are  identifiability issues in the model above, that will be discussed in details in Section \ref{subsec:identifiable} and Section \ref{subsec:priors}

The correlation function $\mathcal{C}^g( |h-h'|, |d-d'| ; \boldsymbol{\theta})$ for the shared-time component is defined as follows:
\begin{equation} \label{eq:gnei11}
	\mathcal{C}^g( |h-h'|, |d-d'| ; \boldsymbol{\theta}) = \frac{1}{\phi_d |d - d'| + 1} \exp \left ( - \frac{\phi_h |h-h'|}{(\phi_d |d - d'| + 1)^{\rho/2} } \right ).
\end{equation}
Here, $\phi_h>0$, $\phi_d>0$, and $\rho \in [0,1]$ represent some entries of the vector $\boldsymbol{\theta}$. This is a parametrization of  the Gneiting correlation function  introduced by \cite{gneiting}, selected here for its non-separability and flexibility. The non-separable parameter $\rho$ represents  the time-frequency interaction, while $\phi_d$ and $\phi_h$ denote the time and frequency decays, respectively.
It should be noted that the model above presents identifiability issues, which will be   addressed in Section \ref{subsec:identifiable} and Section \ref{subsec:priors}.

%%%%%%%%%%%%%% 
% 	Sub-section 3.2		W2  %
%%%%%%%%%%%%%% 
\subsection{The artifact-component $\mathcal{W}_2(t,h)$}\label{subsec:W2}
\begin{figure}[t]
{\subfloat[]{\includegraphics[scale=0.25]{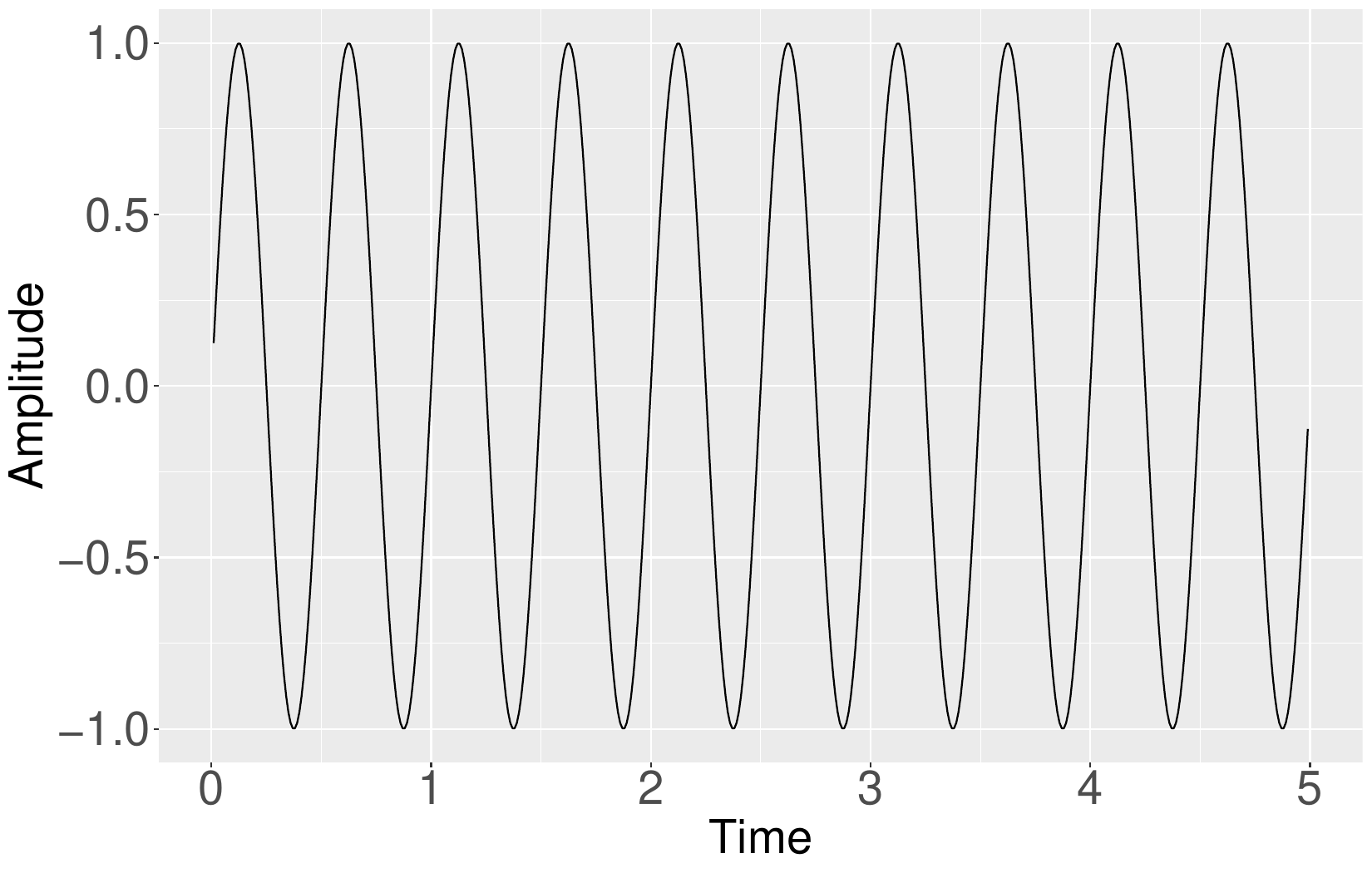}}} 
{\subfloat[]{\includegraphics[scale=0.25]{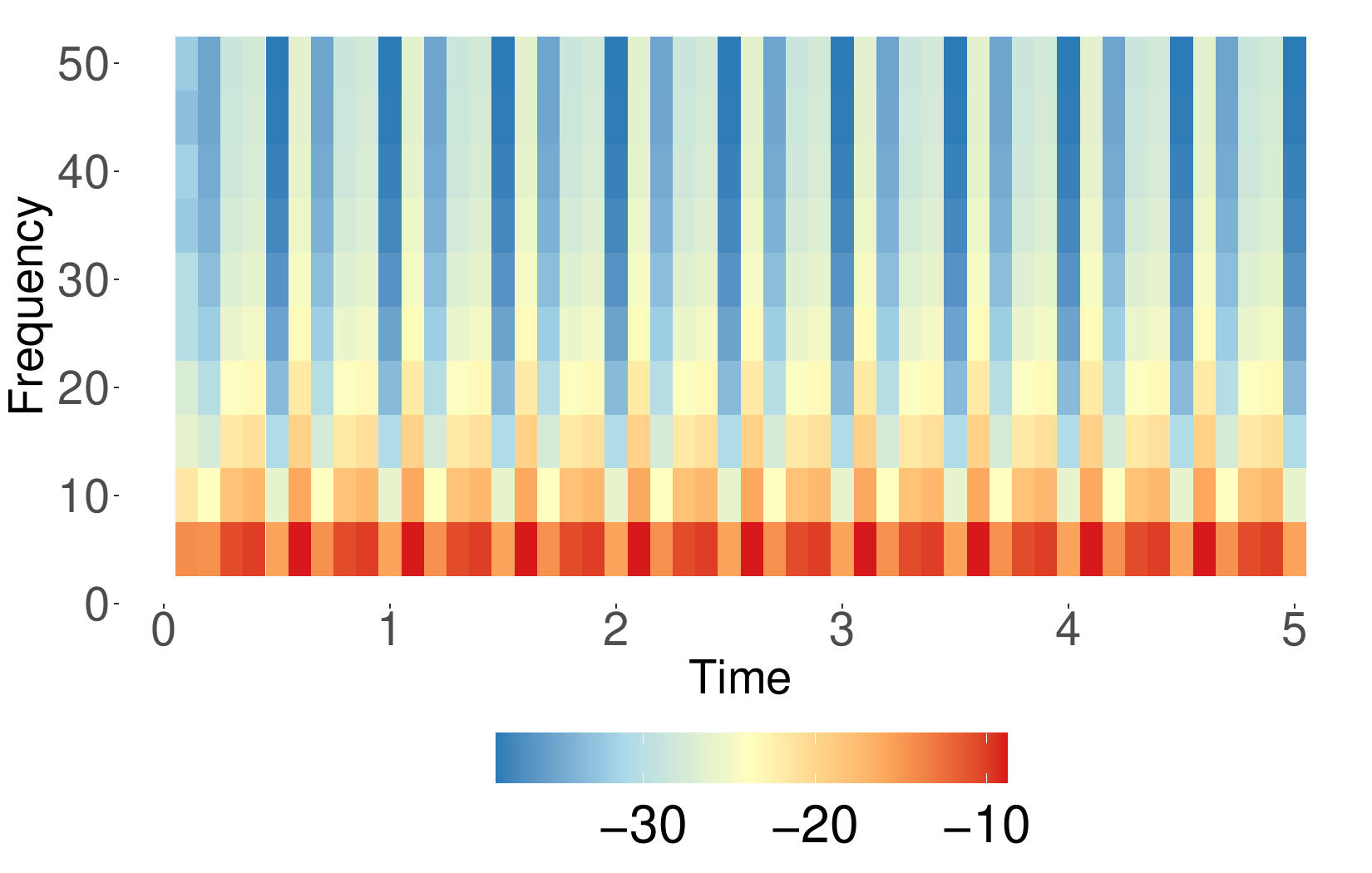}}} 
% {\subfloat[]{\includegraphics[scale=0.25]{sp2}}} 
%  {\subfloat[]{\includegraphics[scale=0.25]{sp3}}} 
\caption{Example of the periodic sampling artifacts caused by the windowing effect. Panel (a) is a periodic signal generated by $\displaystyle y(t) = sin(2\pi f_o t)$ with frequency $f_o = 2$.  Panel (b) is the spectrogram obtained by applying the short-time Fourier Transform with a time-step of $0.1$ seconds on the periodic signal.
% The periodicity in the sound intensities of (b) occurs because the $0.1$ time-step do not lead to the true waveform cycles in (a) being captured by the sampled frequencies.
}\label{fig:simfreq}
\end{figure}

As shown in Figure \ref{fig:simfreq}, the selection of the time window for the short-time Fourier Transform can induce the appearance of artifacts, particularly when the time window is relatively small compared to the waveform cycles \citep{Graps1995,kumar1997}. These artifacts manifest as approximately cyclical variations in intensity along the real-time axis.
Upon examining Figures \ref{fig:data1} and \ref{fig:data2}, it appears that both spectrograms exhibit temporal patterns, with a period of 0.02-0.04, seconds displaying oscillations along the time domains, potentially attributable to these artifacts. The artifact component $\mathcal{W}_2(t,h)$, defined on the real time $t$ rather than the shared-time, is designed to model this phenomenon. Consequently, it is specified by a correlation function $\mathcal{C}^c( \cdot, \cdot ; \boldsymbol{\theta})$ based on circular time-distance:
\begin{equation} 
	\Delta_c(|t-t'| ; \gamma) = \text{min}\{ |t-t'| \ \text{mod} \ \gamma, \ \gamma - |t-t'| \ \text{mod} \ \gamma \}. 
\end{equation}
This choice of periodic distance function restricts the circular distance between any two time coordinates to a circular scale with period $\gamma$, ensuring $ \Delta_c(|t-t'|  ; \gamma) \in [0,\gamma/2] \ \forall \ t,t' $. Subsequently, this circular distance is employed to define the Gneiting circular correlation function, akin to the approaches of \cite{Shinichiro2017} and \cite{mastrantonio}, resulting in
\begin{equation} \label{eq:gnei22}
	\mathcal{C}^c(|t-t'|, |h-h'| ; \boldsymbol{\theta}) = \frac{1}{\phi_c \Delta_c(|t-t'| ; \gamma) + 1} \exp \left ( - \frac{\phi_h |h - h'|}{(\phi_c \Delta_c(|t-t'| ; \gamma) + 1)^{\rho/2}} \right ), % \exp\left( -\phi_c \Delta_c(t,t' ; \gamma)  \right)
\end{equation}
where $\phi_c>0$, and $\gamma>0$ represent additional entries of the vector $\boldsymbol{\theta}$.
To prevent over-parametrization, the frequency decay parameter $\phi_h$ and the non-separability parameter $\rho$ in equation \eqref{eq:gnei22} are shared with those in \eqref{eq:gnei11}. It should be noted that $\gamma$ is a parameter that will be inferred from the data. To the best of our knowledge, this is the first time that estimation of this parameter has been proposed in the literature.

\subsection{The latent spectral shape of the sound $\mathcal{A}_{\ell} (t,h)$} \label{sec:abst}

As delineated in Section \ref{sec:model}, we aim to derive a representative sound that summarizes the information contained in all spectrograms of the same species. This representative sound is   defined as the predicted latent sound of a prospective observation $\mathcal{A}_{\ell} (t,h) $, where $\ell = N+1$. The time duration of this representative sound is denoted by $l_{\ell}$, and its synchronization parameters are represented by $\boldsymbol{\chi}_{\ell} =  ( \alpha_{\ell} , \beta_{\ell}, \boldsymbol{\xi}_{\ell}) $.
For $\mathcal{A}_{\ell}(\cdot, \cdot)$ to serve as a comprehensive representation, it must encompass all the information conveyed by the sample. Therefore, we impose constraints on the synchronization function to satisfy the following conditions:
\begin{equation} \label{eq:cc2}
	\begin{cases}
		\psi(0| \boldsymbol{\chi}_{\ell}) &= \min_{i \in \{1,\dots, N\}} \psi(0| \boldsymbol{\chi}_i), \\
		\psi(l_{\ell}| \boldsymbol{\chi}_{\ell}) &= \max_{i \in \{1,\dots, N\}} \psi(l_i| \boldsymbol{\chi}_i), \\
		l_{\ell}> \gamma.&
	\end{cases}
\end{equation}
These conditions ensure that all points in the set of observed data are represented (the first two conditions) and that at least one cycle of the cyclic component is discernible (the third requirement).

If we define the set $ D $ as follows:
\begin{equation}  \label{eq:constr22}
	D = 
	\begin{cases}
	 	\boldsymbol{\chi}_{\ell} \in \mathbb{R}_+^2 \times \mathbb{R}^2  
	 	\left|
	 	\begin{array}{ll}
	 		\alpha_{\ell}  & = \min\nolimits_{i \in \{1,\dots, N\}}  \alpha_i, \\
			\alpha_{\ell}+\beta_{\ell} l_{\ell} & = \max\nolimits_{i \in \{1,\dots, N\}}  \left(\alpha_i+\beta_i l_i\right),  \\
			l_{\ell} > \gamma &
		\end{array}
		\right\},
	\end{cases} 
\end{equation}
it is easy to see that if $ \boldsymbol{\chi}_{\ell} \in D $ the constraints in \eqref{eq:cc2} are satisfied. Notably, these  determine the parameters $ \alpha_{\ell} $ and $ \beta_{\ell} $, while $ \boldsymbol{\xi}_{\ell} $ remains unconstrained.
Hence, our primary interest lies in the posterior distribution of
\begin{equation} \label{eq:ddss}
	\mathcal{A}_{\ell} (t,h) | \mathbf{y}, \boldsymbol{\chi}_{\ell}  \in D,
\end{equation}
which represents a distribution over the latent sound, conditioned on the data and subject to the constraint \eqref{eq:constr22} being satisfied.
The methodology for obtaining posterior samples from \eqref{eq:ddss}  is detailed in Section \ref{subsec:sampling}.

\section{Computational details} \label{sec:compd}

\subsection{Marginal model}\label{subsec:marginal}

%Although the model in equation \eqref{eq:model} is defined by the introduction of the latent process $ \mathcal{A}_i (t,h) $, and getting posterior sample from it is the main goal of this research, its direct implementation turns out to be complicated. Indeed, each individual $ \mathcal{A}_i (t,h) $ is weighted average of the same two processes $\mathcal{W}_1(\cdot,\cdot)$ and $\mathcal{W}_2(\cdot,\cdot)$, and $\mathcal{W}_1(\cdot,\cdot)$ needs to be evaluated at different time points determined by the individual synchronization function $\psi(t ; \boldsymbol{\chi}_i)$. Due to the dimension of the dataset, this would imply  to sample an impractically large number of random variables. Hence, we decide to marginalize the model with respect to  $\mathcal{W}_1(\cdot,\cdot)$ and $\mathcal{W}_2(\cdot,\cdot)$, which is a common practice in  geo-statistical models. Despite the marginalization, posterior samples of the latent
%process can be obtained after model fitting by standard procedure that will be discussed in Section \ref{subsec:sampling}.
Although the model in equation \eqref{eq:model} is defined by introducing the latent processes $\mathcal{W}_1(d,h)$ and $\mathcal{W}_2(t,h)$, direct implementation poses significant complexity. Evaluating $\mathcal{W}_1(\cdot,\cdot)$ at different time points, determined by the individual synchronization function $\psi(t ; \boldsymbol{\chi}_i)$, would necessitate sampling an impractically large number of random variables due to the dataset's size. 
Consequently, we opt to marginalize the model with respect to $\mathcal{W}_1(\cdot,\cdot)$ and $\mathcal{W}_2(\cdot,\cdot)$, a common practice in geo-statistical models \citep[see, for example, ][]{gelfand2010,mastrantonio2015b}. 
%Despite the marginalization, posterior samples of the latent process can be obtained following model fitting through standard procedures, as discussed in Section \ref{subsec:sampling}.

While the observed processes $ \mathcal{Y}_i (t,h) $ and $ \mathcal{Y}_{i'} (t',h') $, where $ i \neq i' $, are conditionally independent given the latent processes $\mathcal{W}_1(\cdot,\cdot)$ and $\mathcal{W}_2(\cdot,\cdot)$, the marginalization introduces dependence between the signals. Consequently, they form a multivariate GP with a covariance function given by
\begin{align} \label{eq:covy} 
	&\mathcal{C}_{i,i'}^y \big( (t,h), (t',h') ; \boldsymbol{\chi}_i, \boldsymbol{\chi}_{i'}, \boldsymbol{\theta} \big)  	=\\ 
	&\mathcal{C}_{i,i'}^{a} \big( (t,h), (t',h') ; \boldsymbol{\chi}_i, \boldsymbol{\chi}_{i'}, \boldsymbol{\theta} \big)  + \tau_i^2 \mathbb{I}\left( (i,t,g) = (i',t',g') \right) ,
\end{align}
where 
\begin{align} \label{eq:covydddd} 
	\mathcal{C}_{i,i'}^{a} \big( (t,h),& (t',h') ; \boldsymbol{\chi}_i, \boldsymbol{\chi}_{i'}, \boldsymbol{\theta} \big)  = \\
	& \frac{\sigma^2 \lambda}{\phi_d \Delta_d(t,t' ; \boldsymbol{\chi}_i, \boldsymbol{\chi}_{i'}) + 1} \exp \left( - \frac{\phi_h |h-h'| }{(\phi_d \Delta_d(t,t' ; \boldsymbol{\chi}_i, \boldsymbol{\chi}_{i'})  + 1 )^{\rho/2}} \right) +	 \\ 
	& \frac{\sigma^2 (1-\lambda)}{\phi_c \Delta_c(|t-t'| ; \gamma) + 1} \exp \left( - \frac{\phi_h |h-h'| }{(\phi_c \Delta_c(|t-t'| ; \gamma)  + 1 )^{\rho/2}} \right)
\end{align}
with 
\begin{align} \label{eq:distd}
	\Delta_d(t,t' ; \boldsymbol{\chi}_i, \boldsymbol{\chi}_{i'}) 
	& = | \psi(t ; \boldsymbol{\chi}_i) - \psi(t' ; \boldsymbol{\chi}_{i'}) | \\ 
	& =  | \alpha_i + \beta_i \Omega (t/l_i ; \boldsymbol{\xi}_i) l_i - \alpha_{i'} - \beta_{i'} \Omega (t'/l_{i'} ; \boldsymbol{\xi}_{i'}) l_{i'}  |.
\end{align}
%Hence, $\mathcal{C}_{i,i'}^{a} ( (t,h), (t',h') ; \boldsymbol{\chi}_i, \boldsymbol{\chi}_{i'}, \boldsymbol{\theta} )$ 
%can be seen as the  covariance function between the latent sounds $Cov(\mathcal{A}_{i} (t,h), \mathcal{A}_{i'} (t',h'); \boldsymbol{\chi}_i, \boldsymbol{\chi}_{i'}, \boldsymbol{\theta})$. 

This formulation reveals that while stationary processes $\mathcal{W}_1(\cdot,\cdot)$ and $\mathcal{W}_2(\cdot,\cdot)$ are instrumentally employed to formulate the model in equation \eqref{eq:ident}, the spectrograms $ \mathcal{Y}_i (t,h) $ are not assumed to be stationary. This fact is reflected in the covariance \eqref{eq:covy}, which explicitly depends on $t$ and $t'$, not solely on $|t-t'|$. 
Furthermore, an important observation is that the warping parameters \( \boldsymbol{\xi}_i \) and \( \boldsymbol{\xi}_{i'} \) exclusively contribute to the non-stationary part of the covariance function \eqref{eq:covy}. Thus, they can be interpreted as parameters of non-stationarity.

\subsection{Identifiability issues}\label{subsec:identifiable}

The model suffers from identifiability issues. For instance, adding a constant $c \in \mathbb{R}^+$ to all ${\alpha_i}$,  or multiplying all ${\alpha_i,\beta_i}$ by a constant $c \in \mathbb{R}^+$ while dividing the decay rate $\phi_d$ by the same constant, leaves the value of the covariance $\mathcal{C}_{i,i'}^{a}$ in equation \eqref{eq:covydddd} unchanged. To resolve this issue, a constraint can be set as follows:
\begin{equation} \label{eq:constr}
\begin{cases}
\min\{ \alpha_1,\dots ,\alpha_N \} = 0, \\
\max\{ \alpha_1 + \beta_1 l_1, \dots , \alpha_N+ \beta_N l_N \} = 1.
\end{cases}
\end{equation}
It should be noted that the choice of the interval $[0,1]$ is arbitrary.

This condition can be imposed a priori using an appropriate prior distribution, or it can be enforced, after model fitting, by remapping posterior samples to an identifiable version,   by exploiting the covariance's invariance property. We decided to implement the latter.
Let $\alpha^{(b)}, \beta^{(b)},$ and $\phi_d^{(b)}$ represent the $b$-th posterior sample of the parameters. If we compute $\alpha^{(b)}_{\text{min}} = \min\{ \alpha_1^{(b)},\dots,\alpha_N^{(b)} \}$ and $l^{(b)}_{\text{max}} = \max\{ \alpha_1^{(b)} + \beta_1 l_1, \dots, \alpha_N^{(b)} + \beta_N l_N \} - \alpha^{(b)}_{\text{min}}$, 
then we can remap $\alpha^{(b)}, \beta^{(b)},$ and $\phi_d^{(b)}$ to
\begin{equation}\label{eq:ident}
\begin{cases}
\begin{aligned}[b]
\alpha_i^{*(b)} & = \frac{ \alpha^{(b)}_{i} - \alpha^{(b)}_{\text{min}} }{ l^{(b)}_{\text{max}} }  , \quad i=1,\dots,N \\\
\beta_i^{*(b)} & = \frac{ \beta_i^{(b)} }{ l^{(b)}_{\text{max}} } , \quad i=1,\dots,N \\
\phi_d^{*(b)} &= l^{(b)}_{\text{max}} \phi_d^{(b)}.
\end{aligned}
\end{cases}
\end{equation}
Consequently, the remapped posterior samples obtained using equation \eqref{eq:ident} adhere to the identification constraints given by equation \eqref{eq:constr}. Additionally, the covariance function in equation \eqref{eq:covy}, evaluated with $(\alpha_i^{*(b)}, \beta_i^{*(b)}, \phi_d^{*(b)})$, equals the one computed with $(\alpha_i^{(b)}, \beta_i^{(b)}, \phi_d^{(b)})$.

%%%%%%%%%%%%%%%%%%%%
% Subsection 4.2  	   Priors selection     %
%%%%%%%%%%%%%%%%%%%%
\subsection{Priors and hyper-priors specification}\label{subsec:priors}

There are two sets of parameters: the individual ones $\boldsymbol{\eta}_i=(\mu_i, \tau_i,\alpha_i, \beta_i, \boldsymbol{\xi}_i)$ for $i=1,\dots,N$, specific to each spectrogram, and the general ones $\boldsymbol{\theta}=(\sigma^2, \lambda, \phi_d, \phi_c, \phi_h, \rho, \gamma)$. Under the Bayesian framework, a prior distribution must be chosen for each of them.
We assume that the time-warping parameters $\boldsymbol{\xi}_i=(\zeta_i, \delta_i)$ are independent samples from a common parametric distribution that we want to learn from data, acting as a random effect. Although the parameters are theoretically free to vary over $\mathbb{R}$, as mentioned in Section \ref{subsec:W1}, it's hardly justifiable for them to induce severe time-warping, as depicted in Figure \ref{fig:warpf}. Hence, the parameters are constrained to a finite domain such that $\zeta_i \in (-b_{\zeta}, b_{\zeta})$ and $\delta_i \in (-b_{\delta}, b_{\delta})$ for all $i=1,\dots,N$. The prior distributions are given by:
\begin{align}\label{eq:rand1}
	\log\left(\frac{\zeta_i+b_{\zeta}}{b_{\zeta}-\zeta_i}\right) \bigg| m_{\zeta}, v_{\zeta} & \sim \mathcal{N}(m_{\zeta}, v_{\zeta})  	\\
	\log\left(\frac{\delta_i+b_{\delta}}{b_{\delta}-\delta_i}\right) \bigg|  m_{\delta}, v_{\delta} & \sim \mathcal{N}(m_{\delta}, v_{\delta})
\end{align}
for all $i=1,\dots,N$. 
The selection of the prior for the means and variances of the normal distributions must be careful due to the domain restrictions of the warping parameters. For instance, a large $v_{\zeta}$ (commonly considered uninformative) with $m_{\zeta}=0$ leads to a bi-modal distribution over $\zeta_i$, with modes close to the domain boundaries, making it rather informative. To mitigate this, we choose the following distributions for the hyperparameters: 
\begin{align}\label{eq:rand2}
	m_{\zeta} &\sim \mathcal{N}_{(-b_m,b_m)}(m_{0,\zeta},v_{0,\zeta}),	\\
	m_{\delta} &\sim \mathcal{N}_{(-b_m,b_m)}(m_{0,\delta},v_{0,\delta}),	\\
	v_{\zeta} &\sim \text{IG}_{< b_v}(a_{0,\zeta},b_{0,\zeta})	,		\\
	v_{\delta} &\sim \text{IG}_{< b_v}(a_{0,\delta},b_{0,\delta}).
\end{align}
The advantage of the random effects on the warping-function parameters $\boldsymbol{\xi}_i$ is twofold. Firstly, since they can be interpreted as parameters of non-stationarity (see Section \ref{subsec:marginal}), they are expected to be difficult to estimate \citep{gelfand2010}, and a random effect facilitates the estimation. Secondly, as will be shown in Section \ref{subsec:sampling}, the random effect facilitates the sampling of $\mathcal{A}_{\ell} (t,h)$.
For the remaining parameters of the synchronization function we use the following:
\begin{align}\label{eq:priorab}
	\alpha_i & \sim U(a_{\alpha}, b_{\alpha}),    \\
	\tilde{\beta}_i & = \frac{ \beta_i l_i}{ 1-\alpha_i} \sim U(a_{\beta}, b_{\beta}),
\end{align}
where the prior on $\beta_i $ is induced by the one on $\tilde{\beta}_i $.
% It is clear that any $i-$th recorded signal in the shared-time dimension starts at $\alpha_i$ and ends at $\alpha_i+\beta_i l_i$ as per equation \eqref{eq:psi}. 
With the assumption $0 \leq a_{\beta} < b_{\beta} \leq 1$ in equation \eqref{eq:priorab}, each observation in the shared time are constrained in the domain $[0,1]$

The priors for the observation-specific scalar mean $\mu_i$ and variability $\tau_i$ are $\mathcal{N}(m_{\mu}, v_{\mu})$ and $\text{IG}(a_{\tau^2}, b_{\tau^2})$, respectively. The priors of the weight $\lambda$ and of the non-separable parameter $\rho$ are both $\text{U}(0,1)$, while the prior for the variance $\sigma^2$ is $\text{IG}(a_{\sigma}, b_{\sigma})$. For the decay parameters and $\gamma$, the uniform distribution is employed, but some care must be taken when selecting the hyperparameters. In the case of a too small $\gamma$, the periodic distance may be smaller than the minimum distance between time coordinates, resulting in no cyclic dependence on the artifacts at all. On the other hand, if $\gamma$ is too large, e.g., $\gamma > 2\text{max}(l_1,\dots , l_1)$, then the periodic distance is equivalent to the distance between any two time coordinates such that $\Delta_c(|t-t'| ; \gamma)  = |t-t'|$, thereby losing its interpretation as a circular distance and $\gamma$ becomes non-identifiable. Hence, the prior for $\gamma$ is set as $\text{U}(a_{\gamma}, b_{\gamma})$ with $a_{\gamma} = 0.02$, which is twice the minimum temporal distance of $0.01$ (the constant time-step for discretization), and with $b_{\gamma}$ set to twice the median of the observed time-lengths of the recordings.

For  identifiability, the correlation is generally required to be greater than $0.05$ at the minimum observed distance and less than $0.05$ at the largest observed distance \cite[see, for example, ][]{gelfand2010, Banerjee2014,mastrantonio2015c}. To this aim, the so-called practical range is defined to be the observed time-frequency distances at which the correlations in equations \eqref{eq:gnei11} and \eqref{eq:gnei22} equals $0.05$ in the separable case of $\rho=0$, which leads to the practical range for $\phi_h$ being
\begin{equation}\label{eq:pr_h}
\text{pr}_h = -\frac{\log(0.05)}{\phi_h}
\end{equation}
and the other two practical ranges being
\begin{equation} \label{eq:pf_cd}
\text{pr}_c =\frac{1.0-0.05}{0.05\phi_c }, \quad \text{pr}_d =\frac{1.0-0.05}{0.05 \phi_d }.
\end{equation}
Since the minimum and maximum distances in the frequency domain given by data are $|h-h'| = 0.23$ and $(H-1)0.23$, respectively, the prior for the frequency decay is simply defined as $\phi_h \sim \text{U}(0.521, 13.025)$ using equation \eqref{eq:pr_h}. On the other hand, the minimum and maximum distances in the dimensions for the circular-time and the shared-time are themselves random variables. The prior for the circular-time decay $\phi_c$ is thus defined to be conditional on $\gamma$ such that
\begin{equation}
\phi_c|\gamma \sim \text{U}\left( \frac{1.0-0.05}{0.05 (0.5\gamma) }, \frac{1.0-0.05}{0.05 \times 0.01 }\right)
\end{equation}
where $0.01$ and $0.5\gamma$ are the minimum and maximum circular distance, respectively, for any value of $\gamma$. As for the shared-time decay $\phi_d$, the prior is defined to be conditional on $(\beta_1, \beta_2, \dots , \beta_N)$:
\begin{equation}
\phi_d | (\beta_1, \beta_2, \dots , \beta_N) \sim \text{U}\left( \frac{1.0-0.05}{0.05 \max(\{\beta_i l_i\}_{i=1}^N)}, \frac{1.0-0.05}{0.05 \min (\{\beta_i l_i/(T_i-1)\}_{i=1}^N)}\right)
\end{equation}
where $\beta_i l_i/(T_i-1)$ and $\beta_i l_i$ are the minimum and maximum shared-time distances of the $i-$th recorded signal, respectively. Note that the minimum distance here is computed in the absence of time-warping. Even though it is possible to compute the minimum distance under the consideration of time-warping, the rationale for the absence of time-warping is similar to the rationale for the distances used to define the range of the circular-time decay. For instance, if only one recorded signal is severely warped in time, then only very few observed time coordinates will give a distance that is close to the minimum warped distance, which makes the shared-time decay weakly identifiable.

%%%%%%%%%%%%%%%%%%%%%%%%%%
% Subsection 3.3 		Sampling from the abstrac call   %
%%%%%%%%%%%%%%%%%%%%%%%%%%
\subsection{Sampling from the latent spectral shape} \label{subsec:sampling}

As mentioned in Section \ref{sec:abst}, we need to obtain posterior samples from the distribution of the representative sound 
\begin{equation}\label{bho}
	\mathcal{A}_{\ell} (t,h) | \mathbf{y}, \boldsymbol{\chi}_{\ell}  \in D.
\end{equation}
We can easily see that, if we indicate as $\mathbf{a}_{\ell}$ a realization of $\mathcal{A}_{\ell} (t,h)$ corresponding to a given set of time and frequencies, then we have the following joint distribution:
\begin{equation}\label{eq:joints}
	\left(\begin{array}{c}
		\mathbf{a}_{\ell}\\
		\mathbf{y}\\
	\end{array}\right)\bigg|  \boldsymbol{\eta},\boldsymbol{\theta}, \boldsymbol{\chi}_\ell \sim N \left( 
	\left(\begin{array}{c}
		\mathbf{0}\\
		\boldsymbol{\nu}\\
	\end{array}\right) , 
	\left(\begin{array}{cc}
		\boldsymbol{\Sigma}_{a}& \boldsymbol{\Sigma}_{a,y}\\
		\boldsymbol{\Sigma}_{a,y}^T& \boldsymbol{\Sigma}_{y}\\
	\end{array}\right) 
	\right),
\end{equation}
where $\boldsymbol{\nu} = (\boldsymbol{\nu}_1^\top, \boldsymbol{\nu}_2^\top, \dots , \boldsymbol{\nu}_N^\top)^\top$ with $\boldsymbol{\nu}_i = \mu_i\mathbf{1}_{T_i H}$, $\boldsymbol{\Sigma}_{y}$ is the covariance between the observed data that has elements computed using \eqref{eq:covy}, while $\boldsymbol{\Sigma}_{a}$ and $\boldsymbol{\Sigma}_{a,y}$ are computed using \eqref{eq:covydddd}. The conditional distribution of $\mathbf{a}_{\ell}$ can be derived by \eqref{eq:joints} with the standard results from the multivariate normal:
\begin{equation}\label{eq:joints2}
	\mathbf{a}_{\ell}|\mathbf{y}, \boldsymbol{\eta},\boldsymbol{\theta}, \boldsymbol{\chi}_\ell  \sim N\left(\boldsymbol{\Sigma}_{a,y}\boldsymbol{\Sigma}_{y}^{-1}(\mathbf{y} - \boldsymbol{\nu}), \boldsymbol{\Sigma}_{a}- \boldsymbol{\Sigma}_{a,y}\boldsymbol{\Sigma}_{y}^{-1}\boldsymbol{\Sigma}_{a,y}^T\right).
\end{equation}

We can use equation \eqref{eq:joints2} to sample from \eqref{bho}, since the latter can be seen as conditioning of the former over $\boldsymbol{\chi}_{\ell}  \in D$. The conditioning requires that $\alpha_\ell=0$, and $\beta_\ell l_\ell=1$, while, on the other hand, we have to marginalized with respect to the warping parameters. Moreover, the third point of \eqref{eq:cc2} can be satisfied by simply selecting $l_{\ell} >b_{\gamma}$, which is the right end limit of the uniform prior distribution for $\gamma$. This means that we are interested in the distribution
\begin{equation}
	\mathcal{A}_{\ell} (t,h) | \mathbf{y}, \alpha_\ell=0, \beta_\ell l_\ell=1.
\end{equation}
By exploiting the random effect distribution over the warping parameter, see equation \eqref{eq:rand1}, samples from the posterior of interest can be drawn by a standard Monte Carlo procedure, since
\begin{align} \label{eq:predic2}
	&f ( \mathbf{a}_\ell |\mathbf{y}, \alpha_\ell=0, \beta_\ell l_\ell=1) = \\&\iiint  f (\mathbf{a}_\ell | \boldsymbol{\xi}_{\ell},\boldsymbol{\theta}, \boldsymbol{\eta},\boldsymbol{y}, \alpha_\ell=0, \beta_\ell l_\ell=1 ) f ( \boldsymbol{\xi}_{\ell}|\boldsymbol{\theta}, \boldsymbol{\eta} ) f( \boldsymbol{\theta}, \boldsymbol{\eta}|\boldsymbol{y} ) d \boldsymbol{\theta} d \boldsymbol{\eta} d \boldsymbol{\xi}_{\ell},
\end{align}
where $f( \boldsymbol{\theta}, \boldsymbol{\chi}|\boldsymbol{y} )$ is the posterior distribution of all model parameters, $ f ( \boldsymbol{\xi}_{\ell}|\boldsymbol{\theta}, \boldsymbol{\chi} )$ is the distribution of the random effects \eqref{eq:rand1}, and $f (\mathbf{a}_\ell | \boldsymbol{\xi}_{\ell},\boldsymbol{\theta}, \boldsymbol{\chi},\boldsymbol{y}, \alpha_\ell=0, \beta_\ell l_\ell=1 )$ is the density of \eqref{eq:joints2}.

\subsection{Nearest neighbors Gaussian process}\label{subsec:NNGP}
The most obvious problem in the implementation of the marginalized model is the so-called Big $n$ Problem caused by the computation of the multivariate Gaussian likelihood, which requires the inversion of the covariance matrix of dimension $\prod_{i=1}^N T_iH$. To be able to estimate the model, here one of the methods for efficiently and accurately approximating the density of a realization of a GP, namely the NNGP from the work of \cite{datta}, is adopted.

Let $\boldsymbol{y}_{i}^{1:j}$ be the vector composed of the first $j$ elements of $\boldsymbol{y}_i$ (by convention $\boldsymbol{y}_{i}^{1:0}$  is an empty vector), and, for any k and for any set of indices $\{i_1, \ldots,i_k\}$, the vector $\boldsymbol{y}_{ \{ i_1, \ldots,i_k \} }$ is defined as $\boldsymbol{y}_{\{i_1, \ldots,i_k\}}=\{\boldsymbol{y}_{i_1},\ldots,\boldsymbol{y}_{i_k}\}$. Define $\pi$ an arbitrary permutation of the integer vector $\{1,2,\dots, N\}$ and let $\pi_i$ be the $i-$th element of $\pi$. Then, the joint density of the observed data can be decomposed as: 
\begin{align}\label{f(y)}
	f(\boldsymbol{y} | \boldsymbol{\theta}, & \boldsymbol{\chi} )  =  \prod_{j=1}^{n_{\pi_1}}f(y_{\pi_1,j}| \boldsymbol{y}_{\pi_1}^{1:j-1}, \boldsymbol{\theta}, \boldsymbol{\chi})\prod_{i=2}^N  \prod_{j=1}^{n_{\pi_i}}f(y_{\pi_i,j}| \boldsymbol{y}_{\pi_i}^{1:j-1}, \boldsymbol{y}_{\{\pi_1, \ldots,\pi_{i-1}\}}, \boldsymbol{\theta}, \boldsymbol{\chi})
\end{align}
which is valid for any permutation. The idea of the NNGP is that, if the covariance function is monotonic with respect to the distances used to compute it, then only the closest neighborhoods are strongly correlated with each observation, therefore we can approximate the likelihood by removing from the conditioning set all observations whose correlation is small. 
Then equation \eqref{f(y)} can be approximated by:
\begin{equation}\label{f(y)2}
	f(\boldsymbol{y} | \boldsymbol{\theta}, \boldsymbol{\chi} ) 
	\approx  \prod_{i=1}^N \prod_{j=1}^{n_{\pi_{i}}} f(y_{\pi_{i},j}| \mathcal{N}_{\pi_{i},j},\boldsymbol{\theta}, \boldsymbol{\chi})
\end{equation}
where $\mathcal{N}_{\pi_{1},1} = \emptyset$, and $ \mathcal{N}_{\pi_{i},j} $ is the subset of variables that are kept in the conditional set of  $ y_{\pi_{i},j} $, named \textit{neighbors-set}. %To have a good approximation, the elements in the neighbors-set must have a high correlation with $y_{\pi_{i},j}$.

We opt to include a maximum of $4k$ neighbors, divided into four distinct groups. These groups consider both correlation functions and dependencies between points within the same spectrogram, as well as dependencies between a spectrogram and the previous one in the given permutation.
In more details,  
to select the elements of the neighbors-set, we first define $\mathcal{M}_{\pi_{i},j}^{g}$ as the set containing the $\min(k,j-1)$ entries of $\boldsymbol{y}_{\pi_i}^{1:j-1}$ that are most correlated with $y_{\pi_{i},j}$ in terms of the covariance \eqref{eq:gnei11}. 
Then we define $\mathcal{R}_{\pi_{1},j}^{g}=\emptyset$ and for all $i\geq 1$, $\mathcal{R}_{\pi_{i},j}^{g}$ as the set of the $k$ entries of $\boldsymbol{y}_{\pi_{i-1}}$ that are most correlated with $y_{\pi_{i},j}$ in terms of the covariance \eqref{eq:gnei11}. 
%Sets $\mathcal{M}_{\pi_{i},j}^{c}$ and $\mathcal{R}_{\pi_{i},j}^{c}$ are defined in the same way, but using the covariance \eqref{eq:gnei22} instead, but elements that are already in $\mathcal{M}_{\pi_{i},j}^{g}$  and $\mathcal{R}_{\pi_{i},j}^{g}$  are not condifered as potential elements of these two set.
Sets $\mathcal{M}_{\pi_{i},j}^{c}$ and $\mathcal{R}_{\pi_{i},j}^{c}$ are defined similarly, but using the covariance \eqref{eq:gnei22}. However, elements that already belong to $\mathcal{M}_{\pi_{i},j}^{g}$ and $\mathcal{R}_{\pi_{i},j}^{g}$ are not considered as potential elements of these two sets, i.e., $\mathcal{M}_{\pi_{i},j}^{c} \bigcap \mathcal{M}_{\pi_{i},j}^{g} = \mathcal{R}_{\pi_{i},j}^{c} \bigcap \mathcal{R}_{\pi_{i},j}^{g} = \emptyset$.
The full neighbor set of $y_{\pi_{i},j}$ is then defined as:
\begin{equation}\label{N_ij}
	\mathcal{N}_{\pi_i,j} = \mathcal{M}_{\pi_{i},j}^{g} \cup\mathcal{R}_{\pi_{i},j}^{g} \cup  \mathcal{M}_{\pi_{i},j}^{c} \cup\mathcal{R}_{\pi_{i},j}^{c} .
\end{equation}
It is important to note that the neighbor sets change with each iteration of the MCMC algorithm as the parameters in the covariances are updated, similarly to the approach proposed by \cite{datta2}. Additionally, to account for the variability introduced by the choice of the arbitrary permutation $\pi$, we treat it as an additional parameter with a uniform prior distribution over the space of permutations. This parameter is then updated using a Metropolis step to ensure thorough exploration of the parameter space and convergence of the algorithm.

\section{Application on real bio-acoustics}\label{sec:apply}

\begin{figure}[t]
{\subfloat[$\sigma^2$]{\includegraphics[scale=0.16]{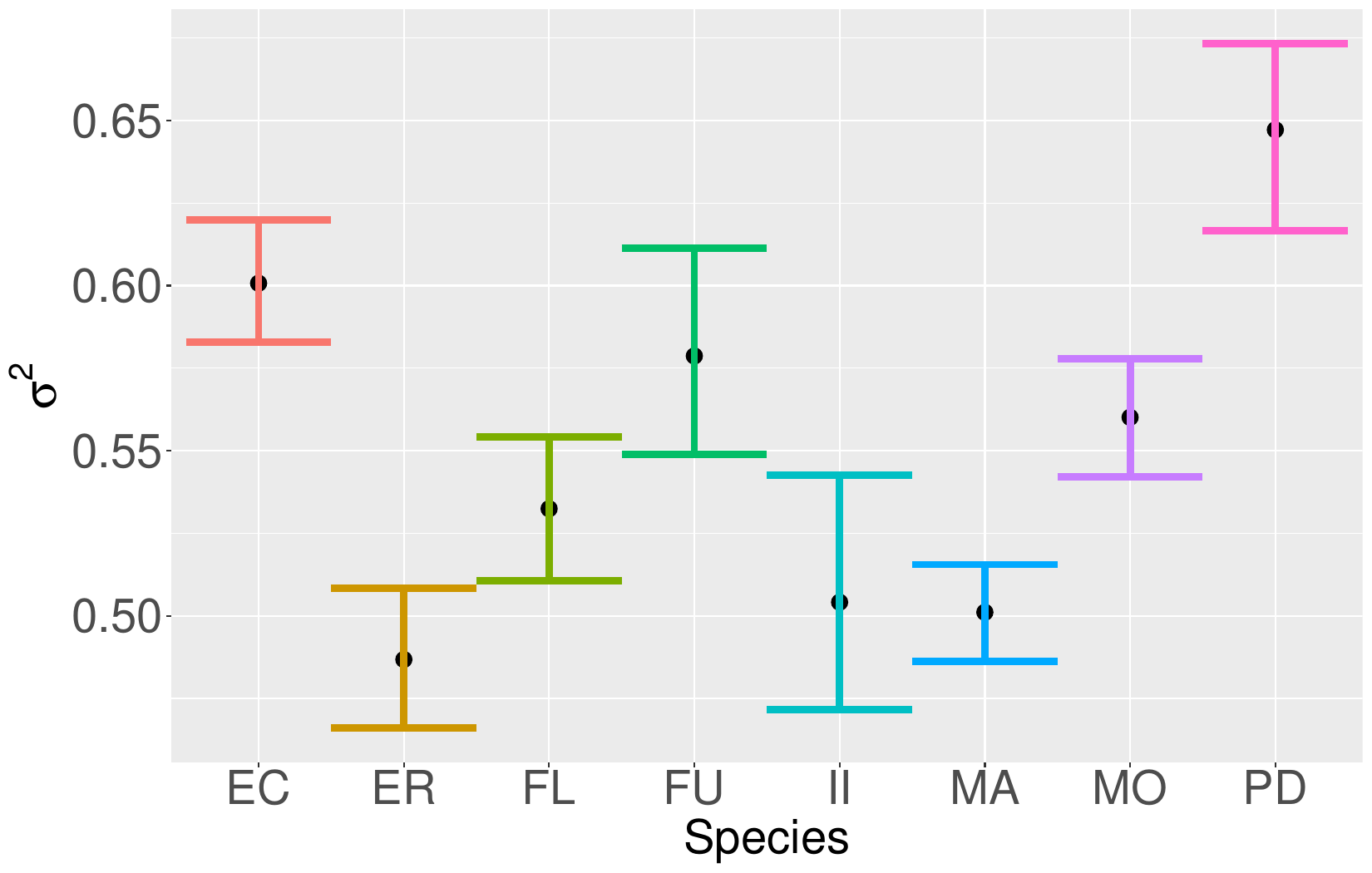}}} 
{\subfloat[$\lambda$]{\includegraphics[scale=0.16]{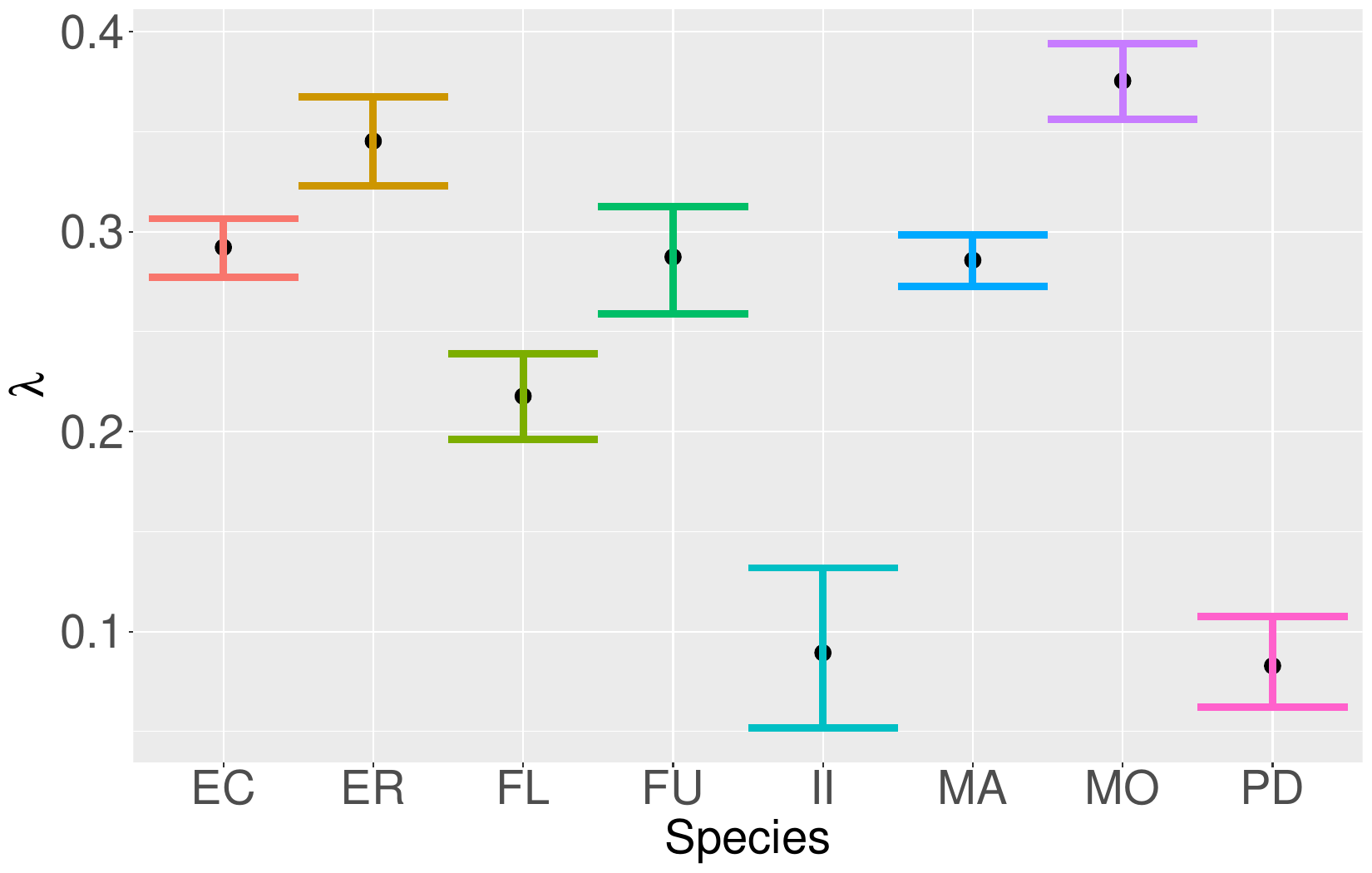}}} 
{\subfloat[$\gamma$]{\includegraphics[scale=0.16]{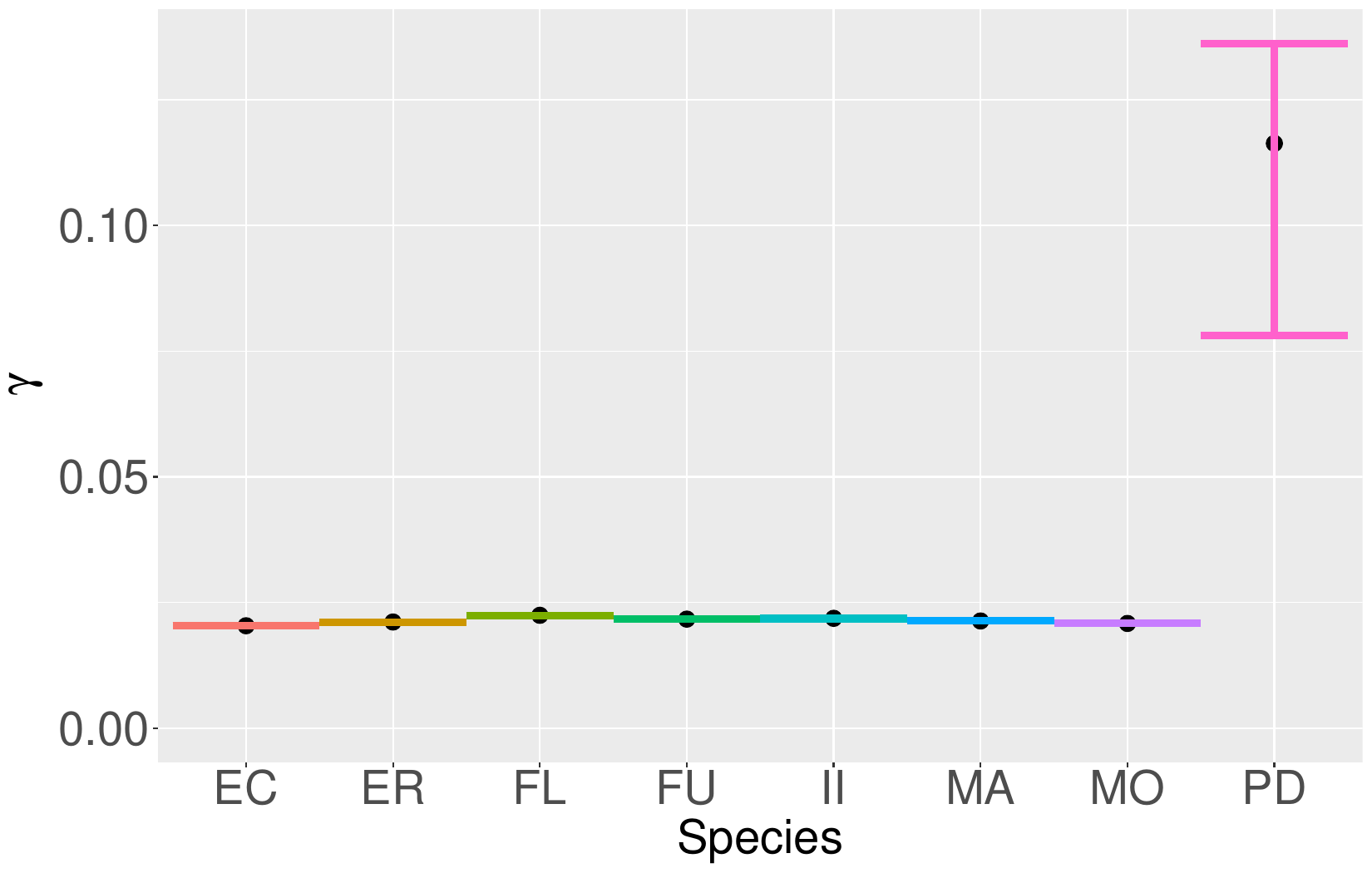}}} \\
{\subfloat[practical range $\phi_d$]{\includegraphics[scale=0.16]{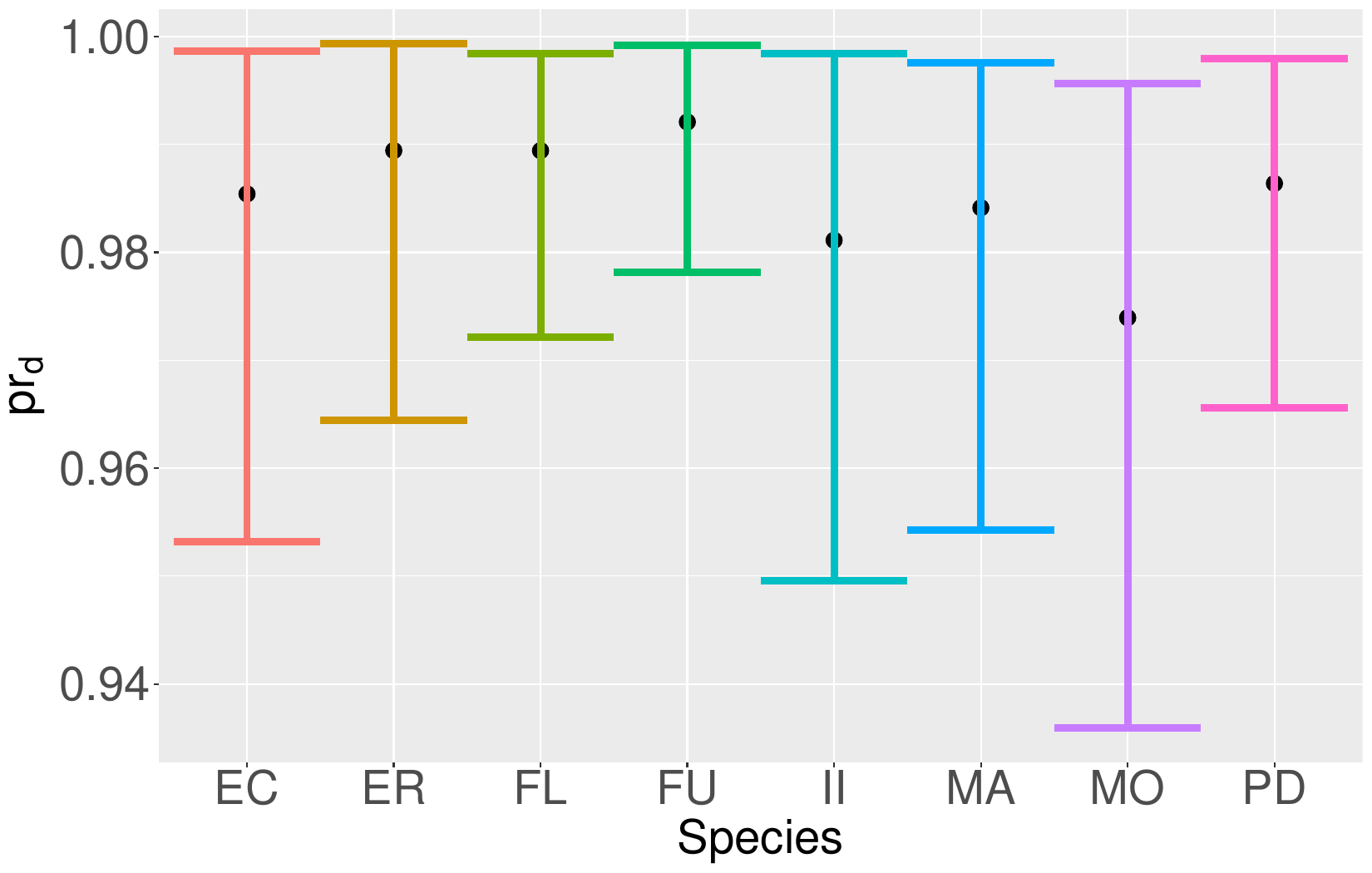}}} 
{\subfloat[practical range $\phi_h$]{\includegraphics[scale=0.16]{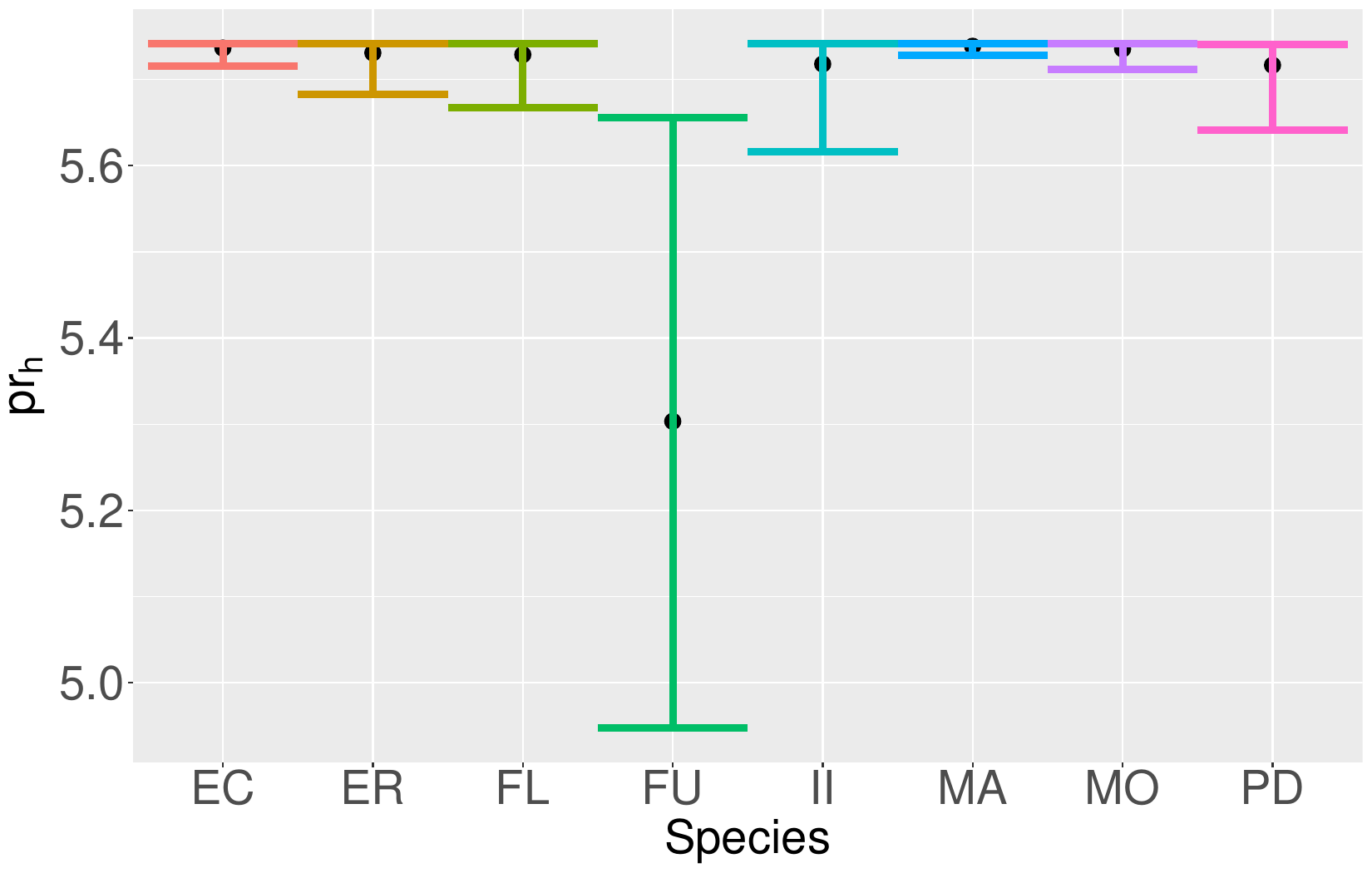}}} 
{\subfloat[practical range $\phi_c$]{\includegraphics[scale=0.16]{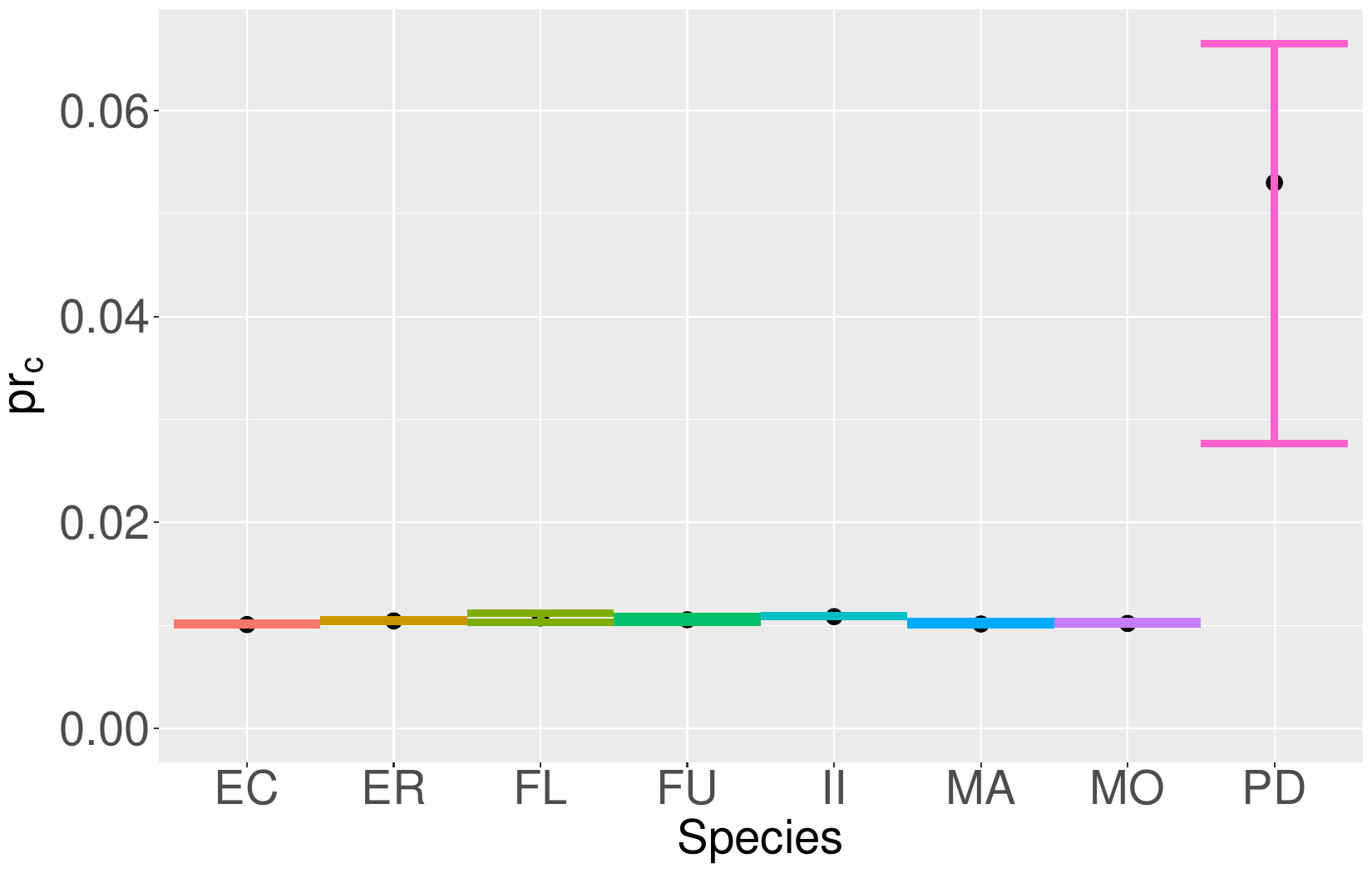}}} \\
{\subfloat[$\rho$]{\includegraphics[scale=0.16]{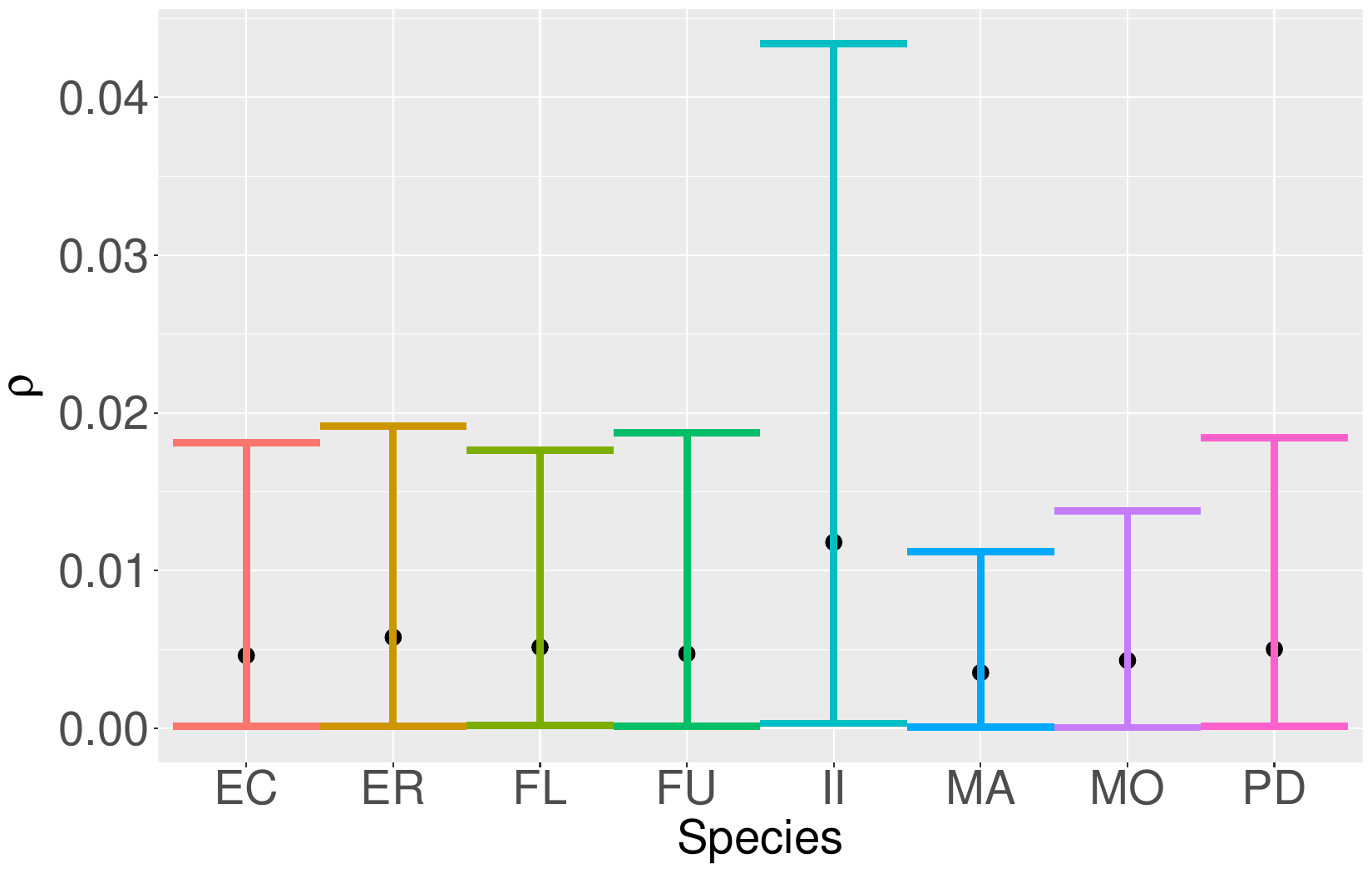}}}
\caption{Credible intervals (segments) and posterior means (dots) for parameters $\sigma^2$, $\lambda$, $\gamma$, $\rho$ and the three practical ranges.}
\label{fig:real_theta}
\end{figure} 

\begin{figure}[t]
{\subfloat[$\alpha$]{\includegraphics[scale=0.16]{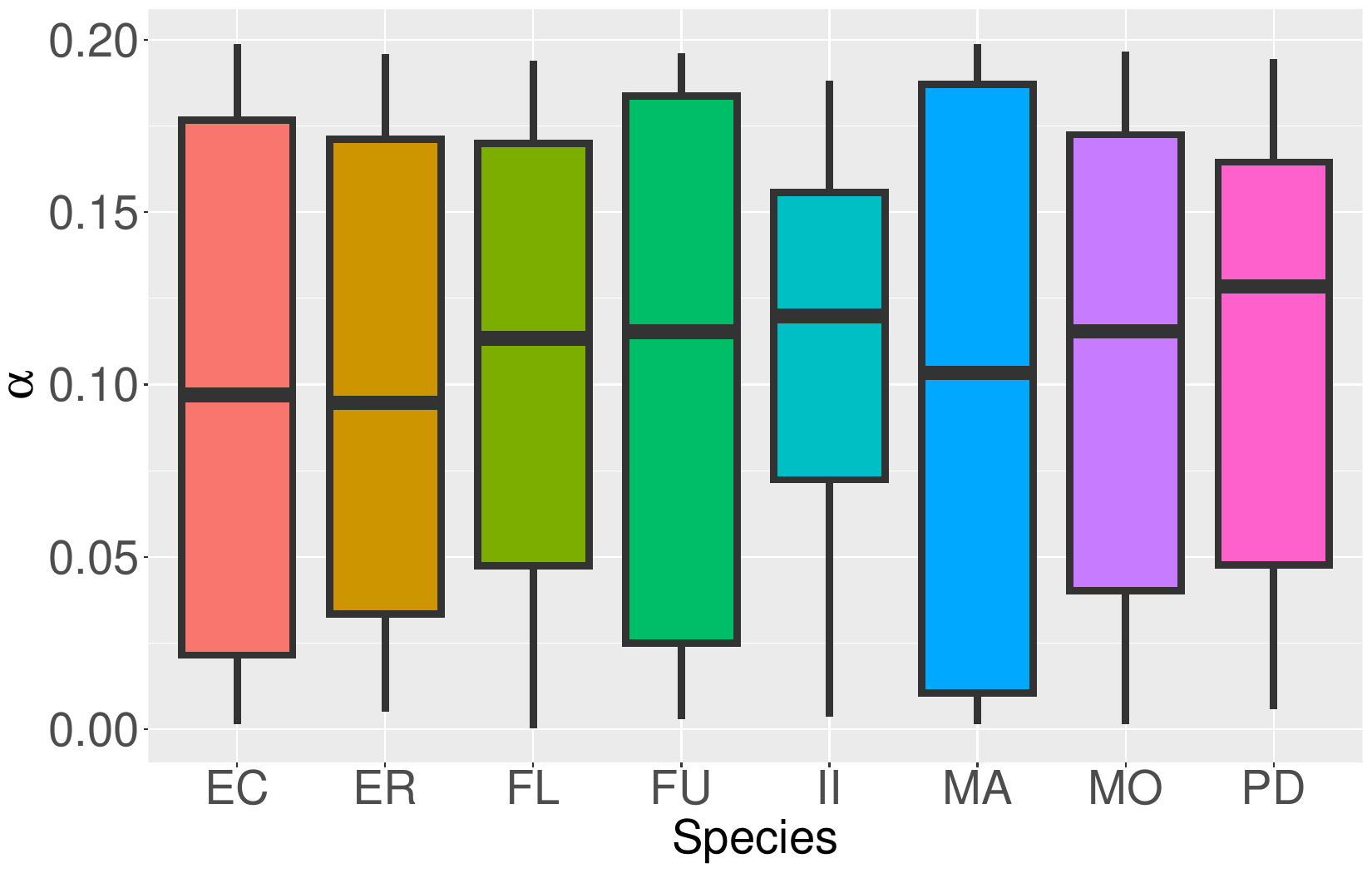}}} 
{\subfloat[$\beta$]{\includegraphics[scale=0.16]{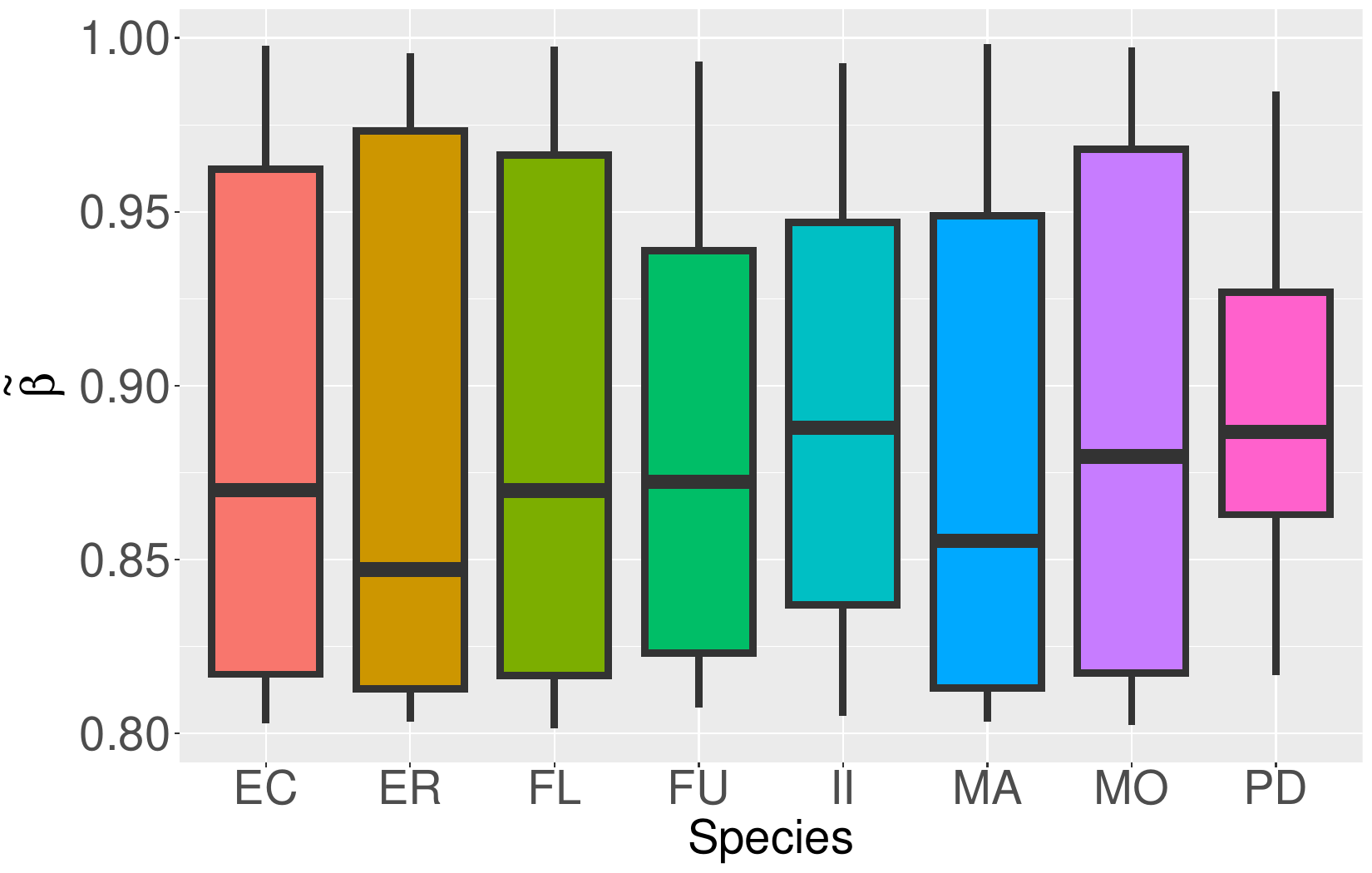}}} \\
{\subfloat[$\mu$]{\includegraphics[scale=0.16]{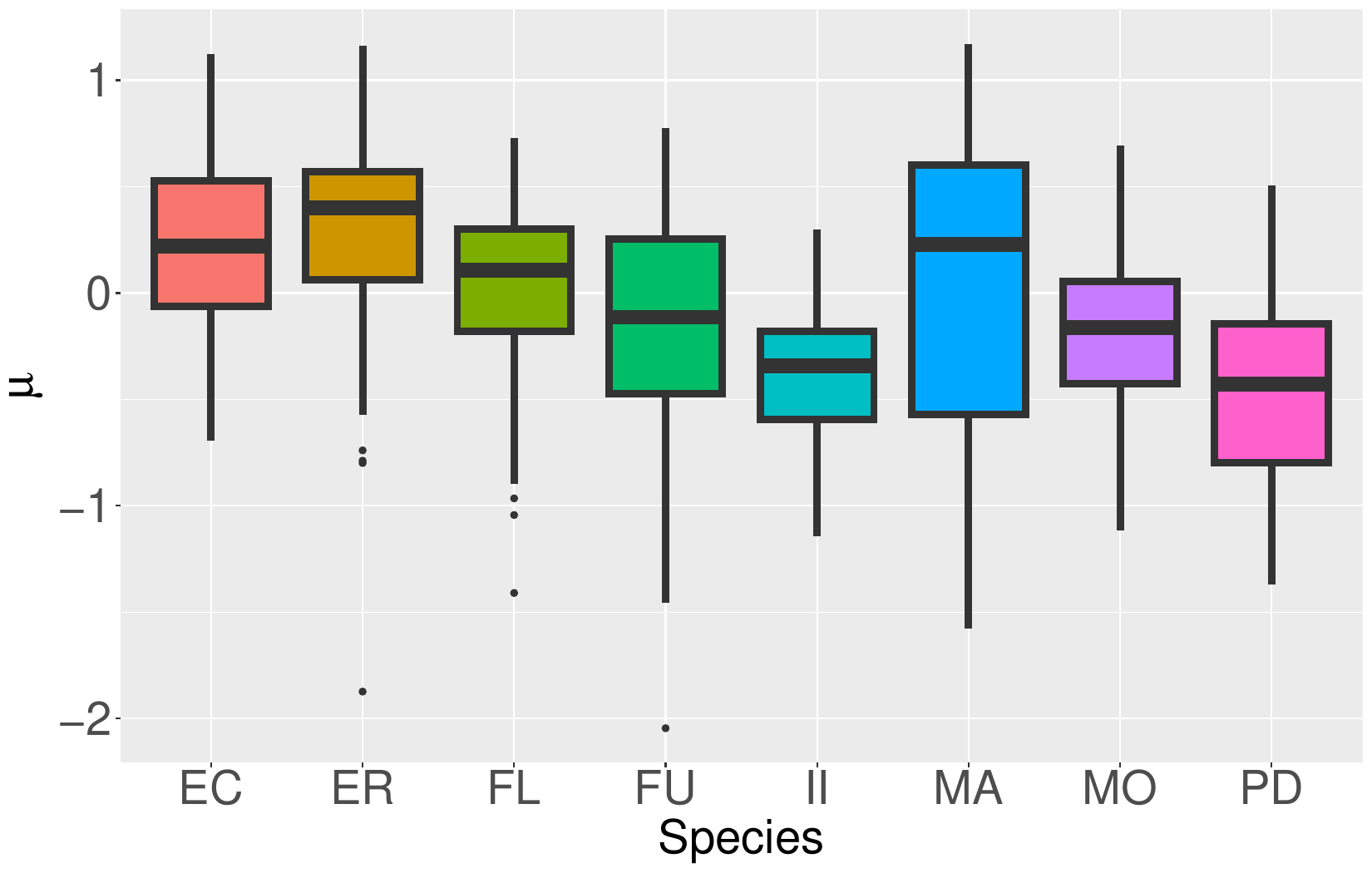}}} 
{\subfloat[$\tau^2$]{\includegraphics[scale=0.16]{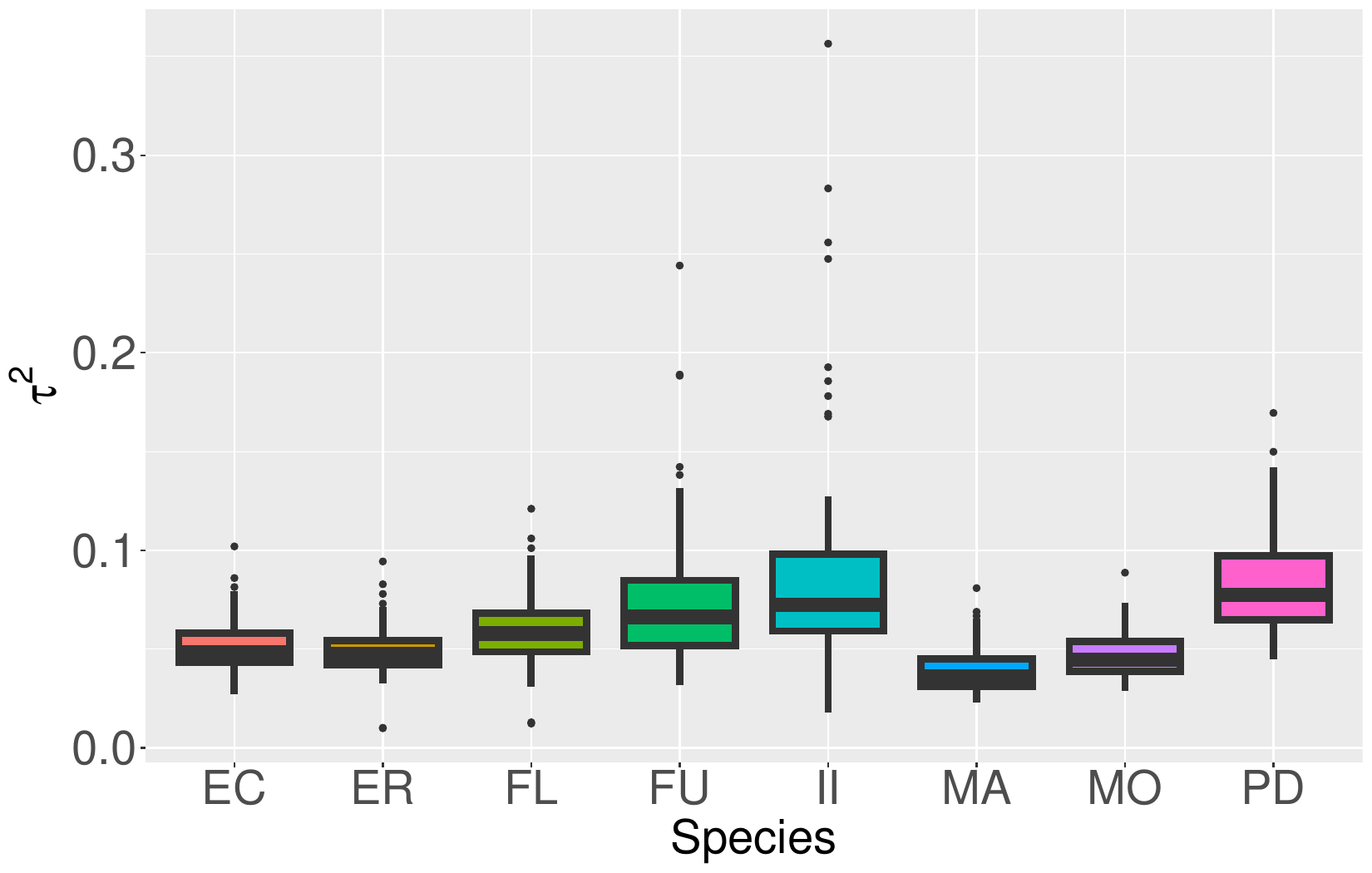}}} 
\caption{Box plots of the posterior means of $\alpha_i$, $\tilde{\beta}_i$, $\mu_i$ and $\tau^2_i$ Across individuals.}
\label{fig:real_eta}
\end{figure} 

\begin{figure}[t]
{\subfloat[EC]{\includegraphics[scale=0.14]{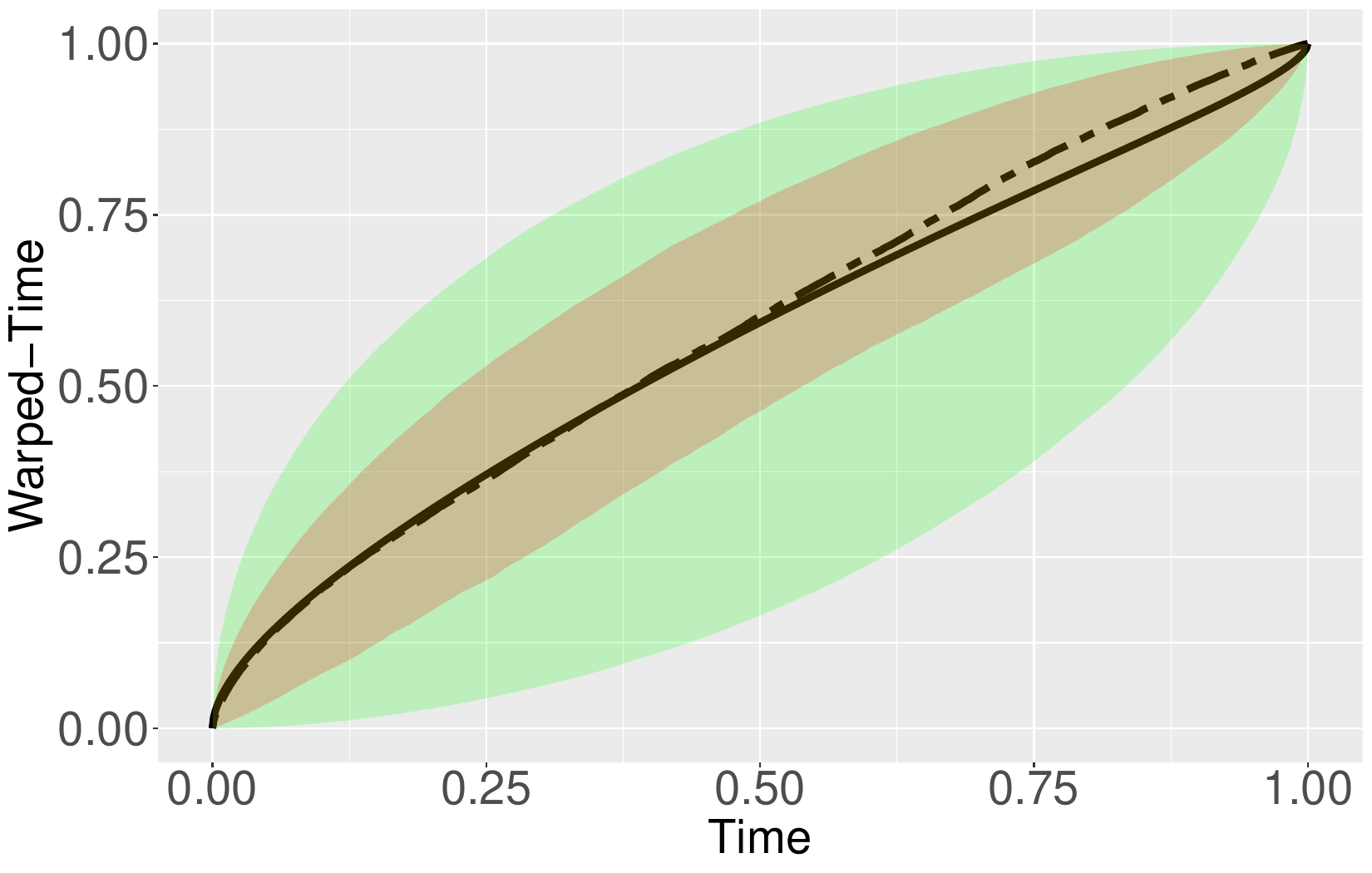}}} 
{\subfloat[ER]{\includegraphics[scale=0.14]{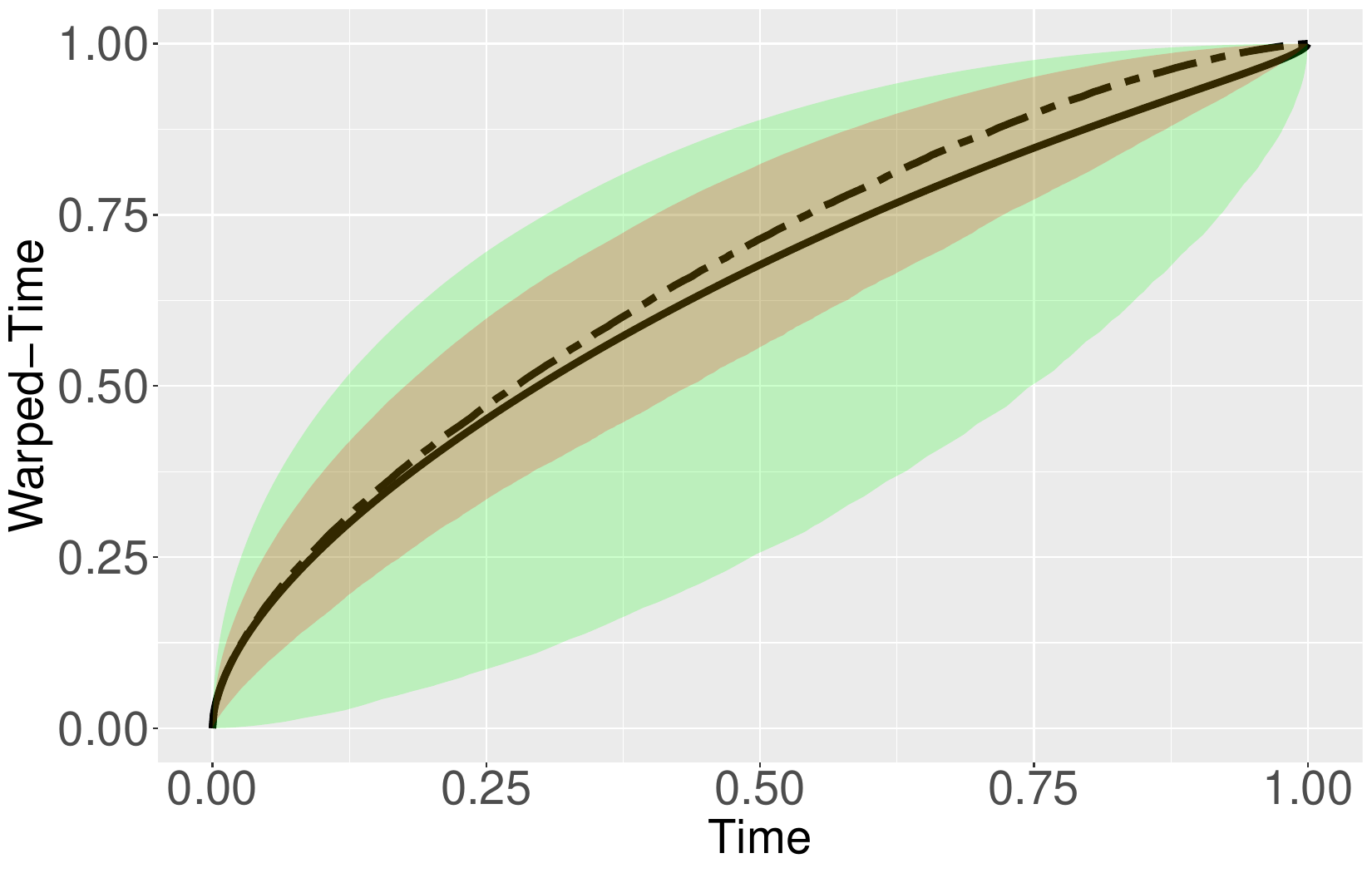}}} 
{\subfloat[FL]{\includegraphics[scale=0.14]{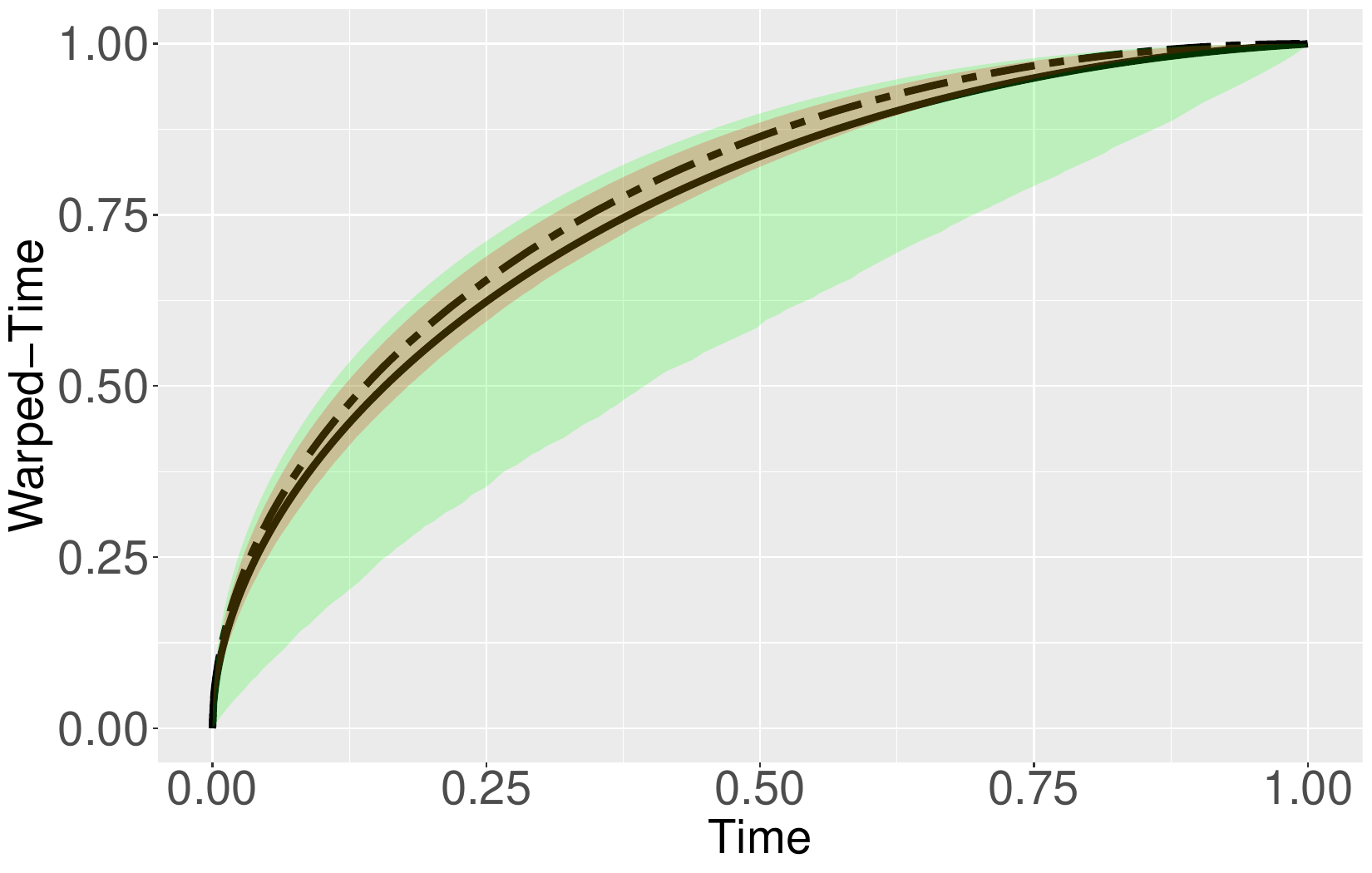}}} \\
{\subfloat[FU]{\includegraphics[scale=0.14]{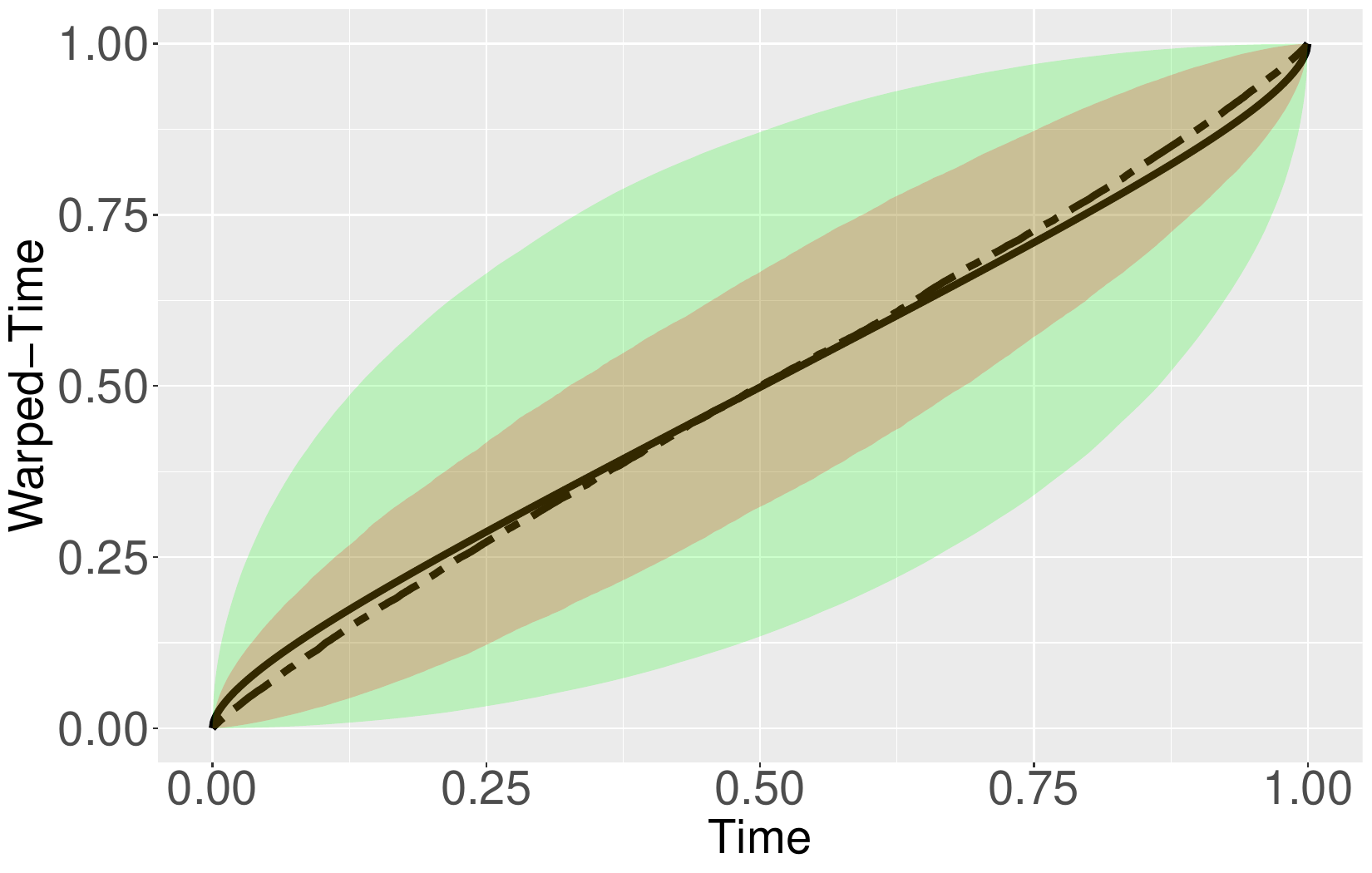}}} 
{\subfloat[II]{\includegraphics[scale=0.14]{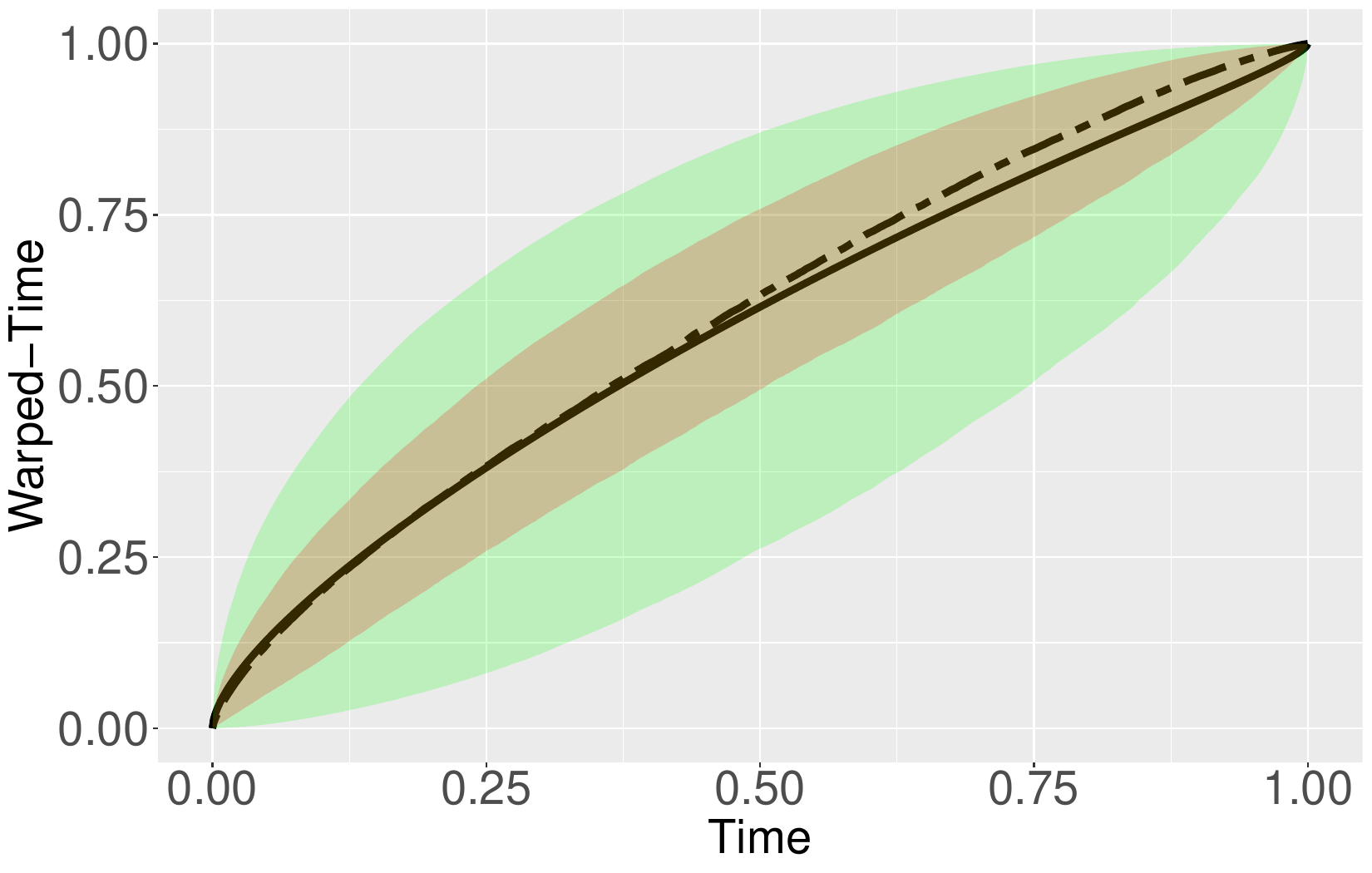}}} 
{\subfloat[MA]{\includegraphics[scale=0.14]{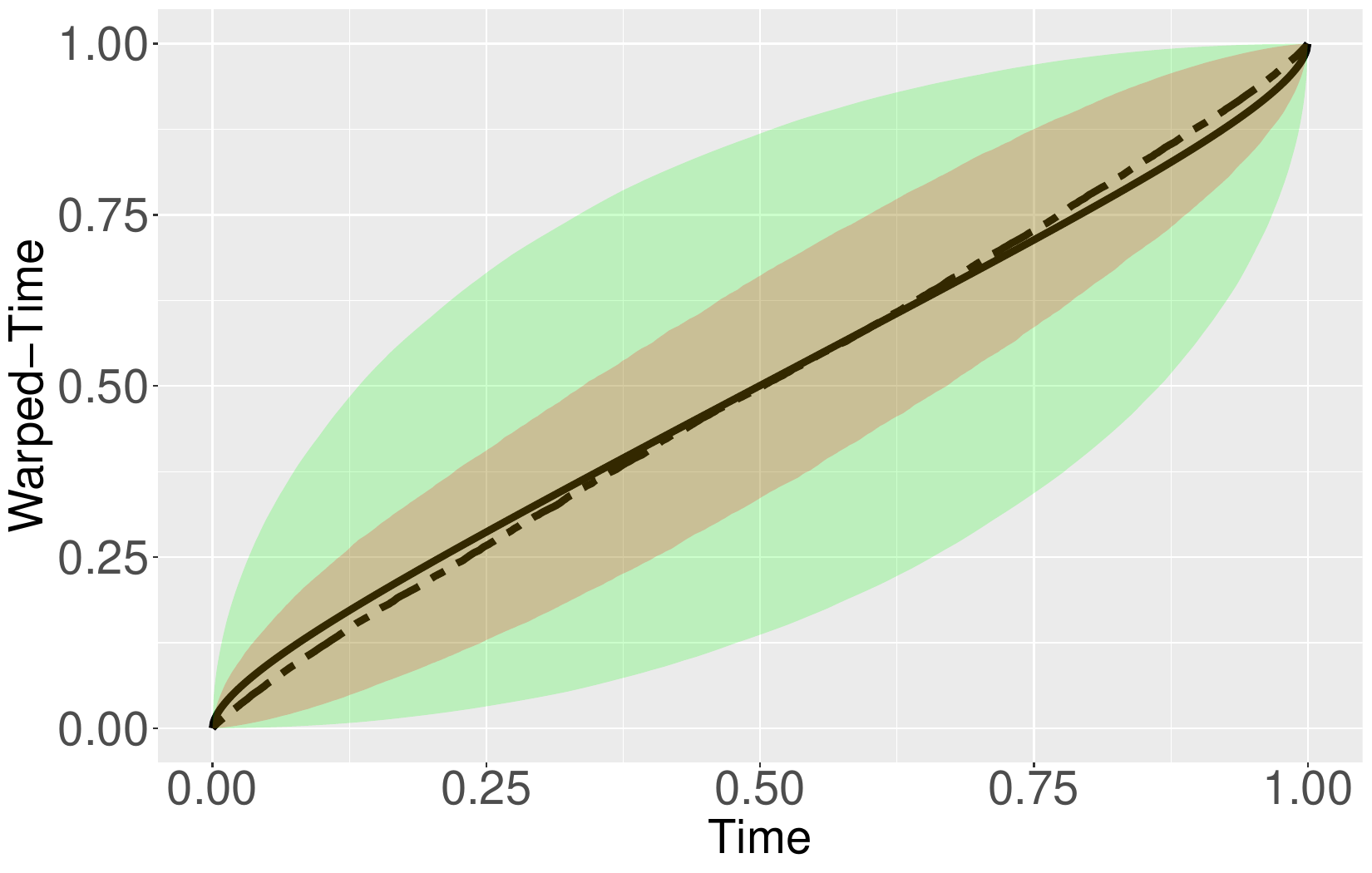}}} \\
{\subfloat[MO]{\includegraphics[scale=0.14]{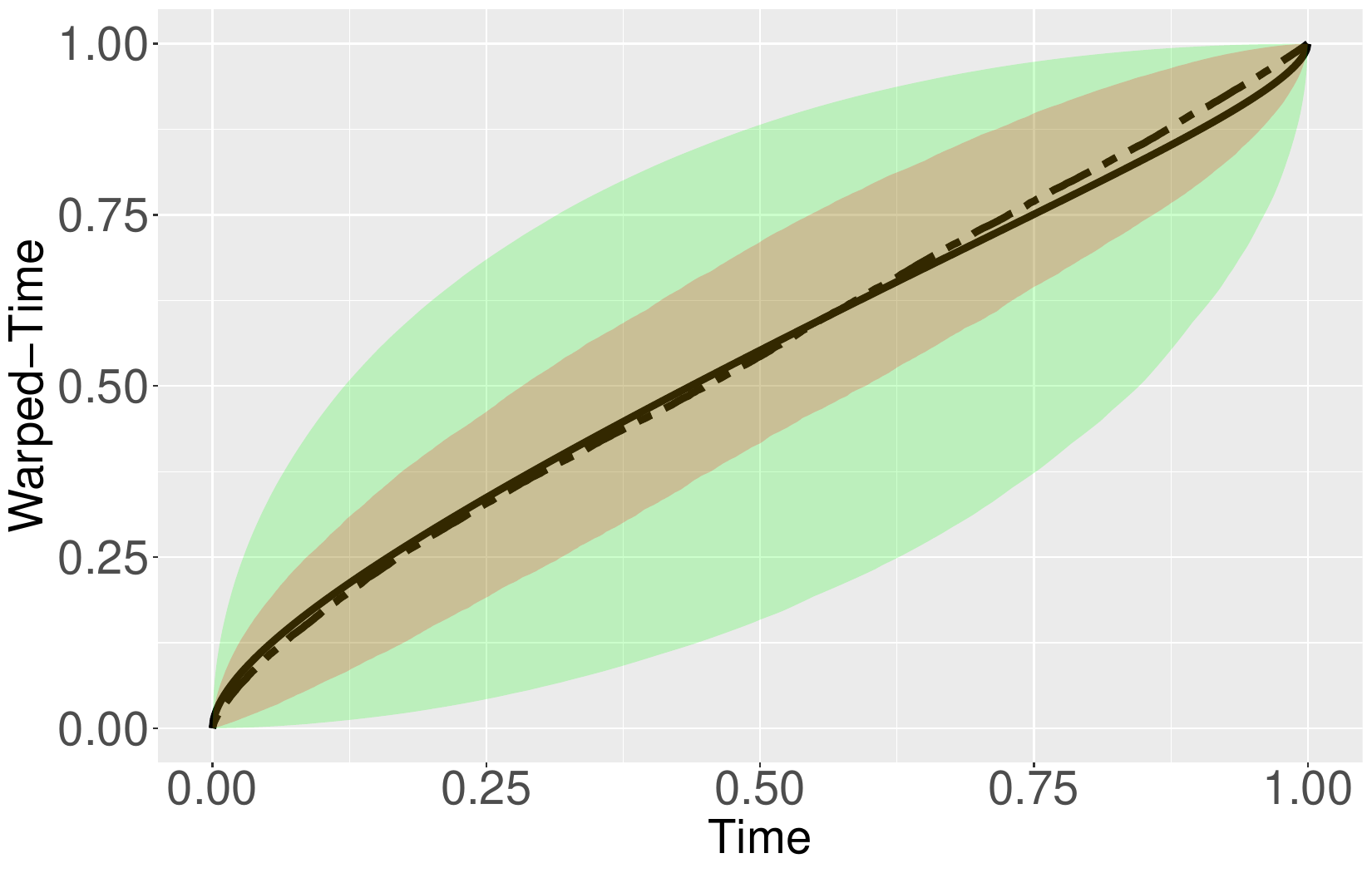}}} 
{\subfloat[PD]{\includegraphics[scale=0.14]{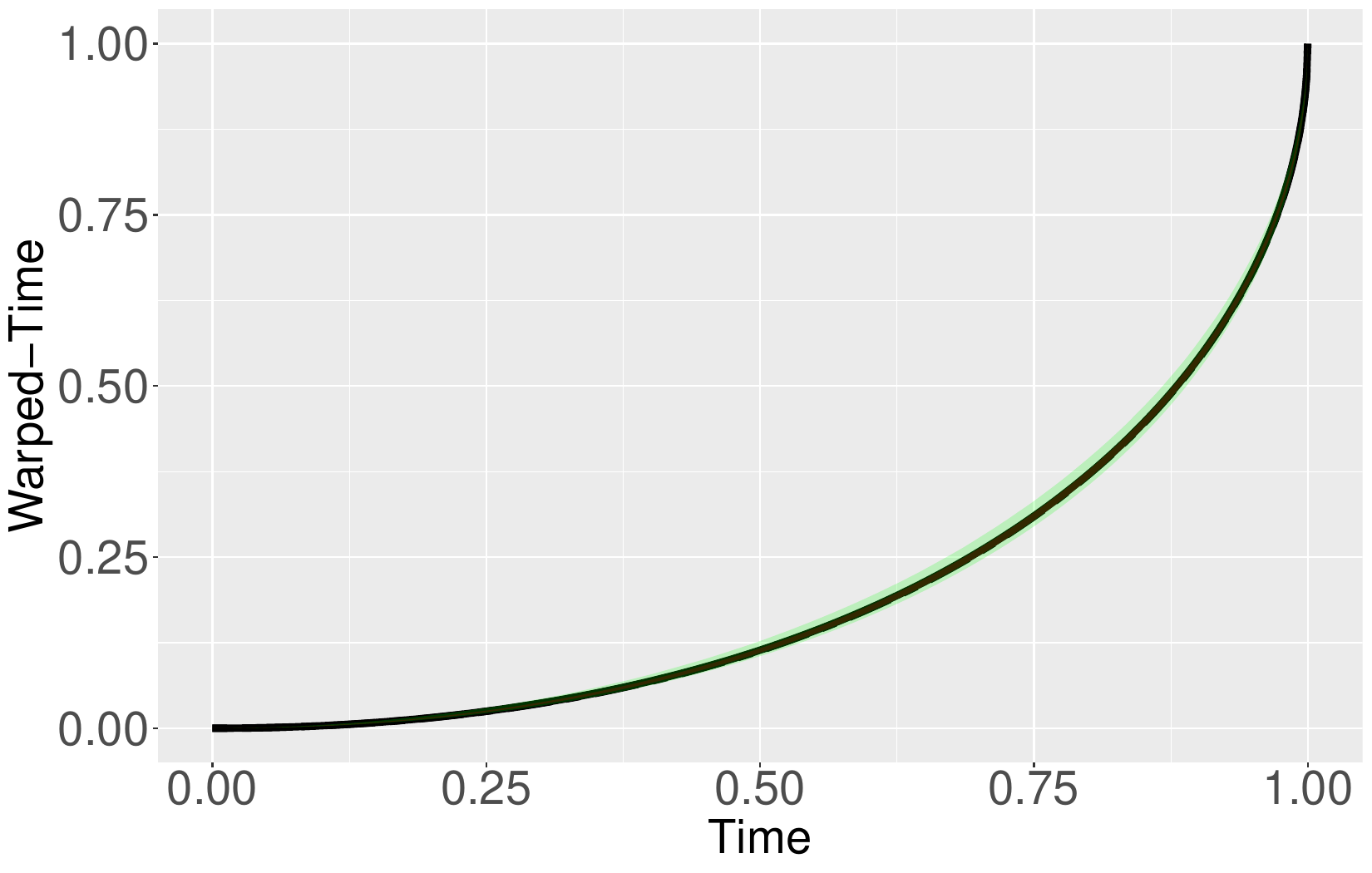}}} 
\caption{The representation illustrates the posterior value of the time-warping function for each species. The solid line denotes the posterior mean, the dashed line represents the posterior median, while the two shaded areas depict the 50\% and 95\% CIs.}
\label{fig:real_warp}
\end{figure}

\begin{figure}[t]
{\subfloat[EC]{\includegraphics[scale=0.15]{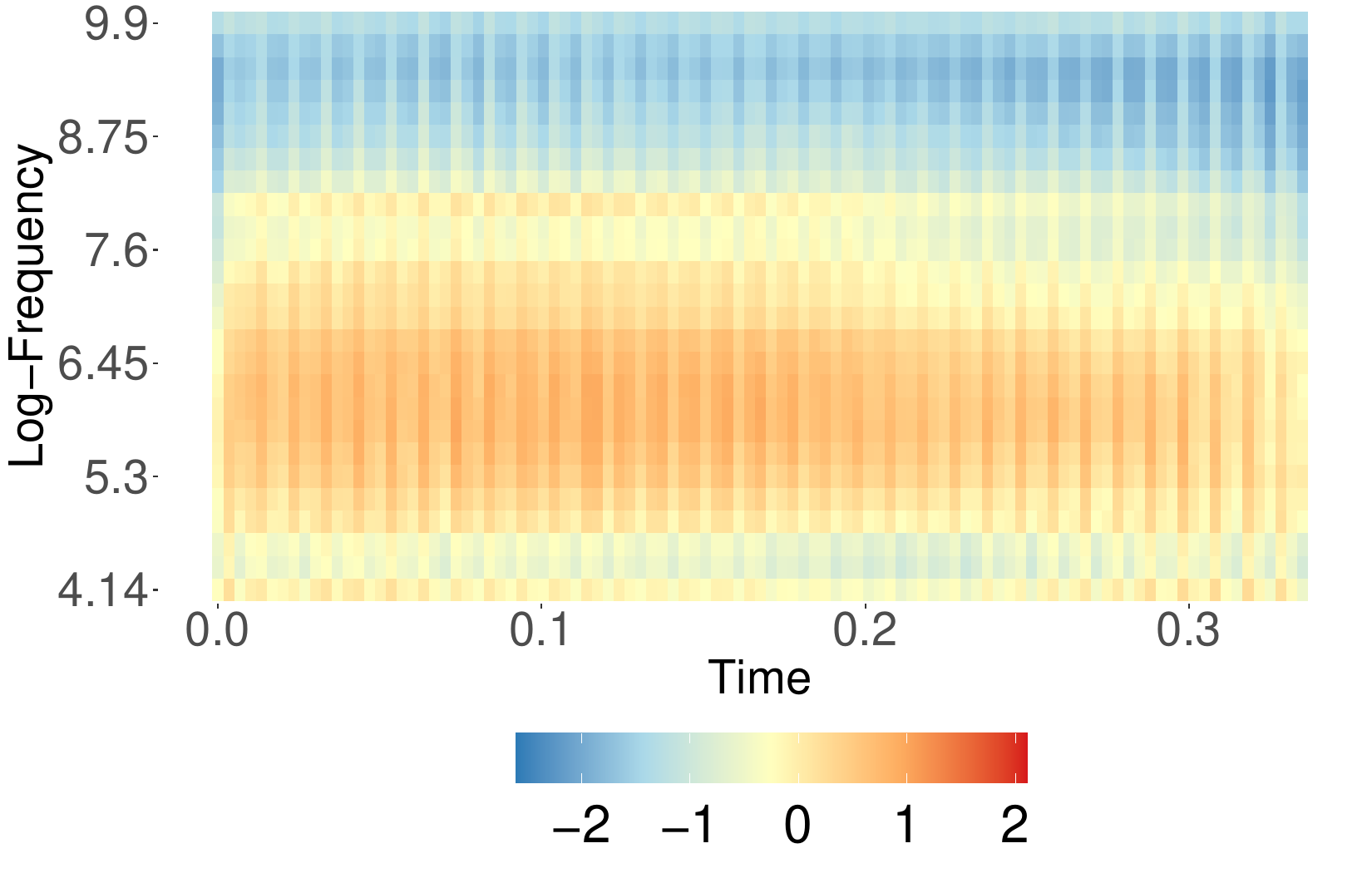}}} 
{\subfloat[ER]{\includegraphics[scale=0.15]{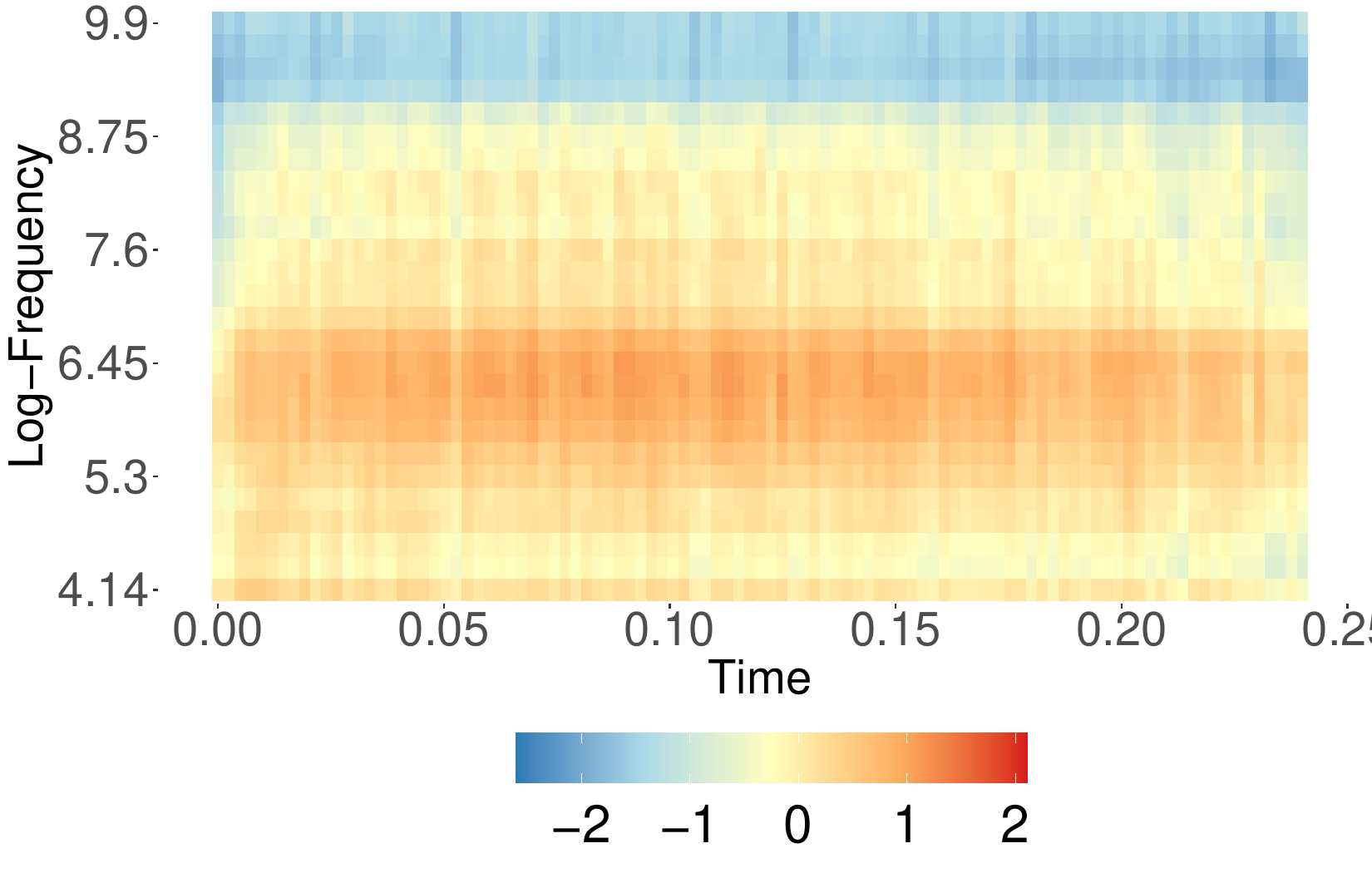}}} 
{\subfloat[FL]{\includegraphics[scale=0.15]{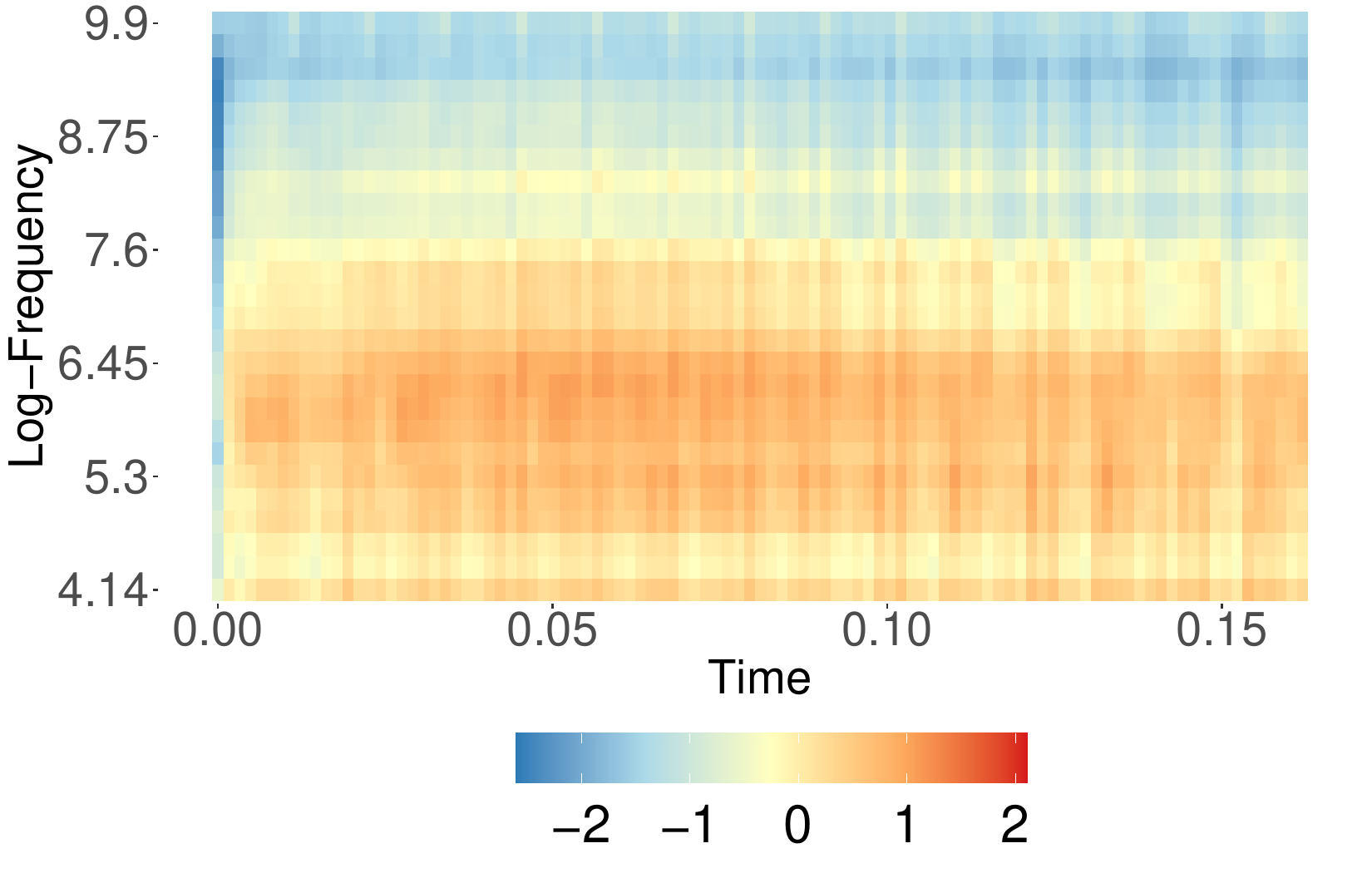}}} \\
{\subfloat[FU]{\includegraphics[scale=0.15]{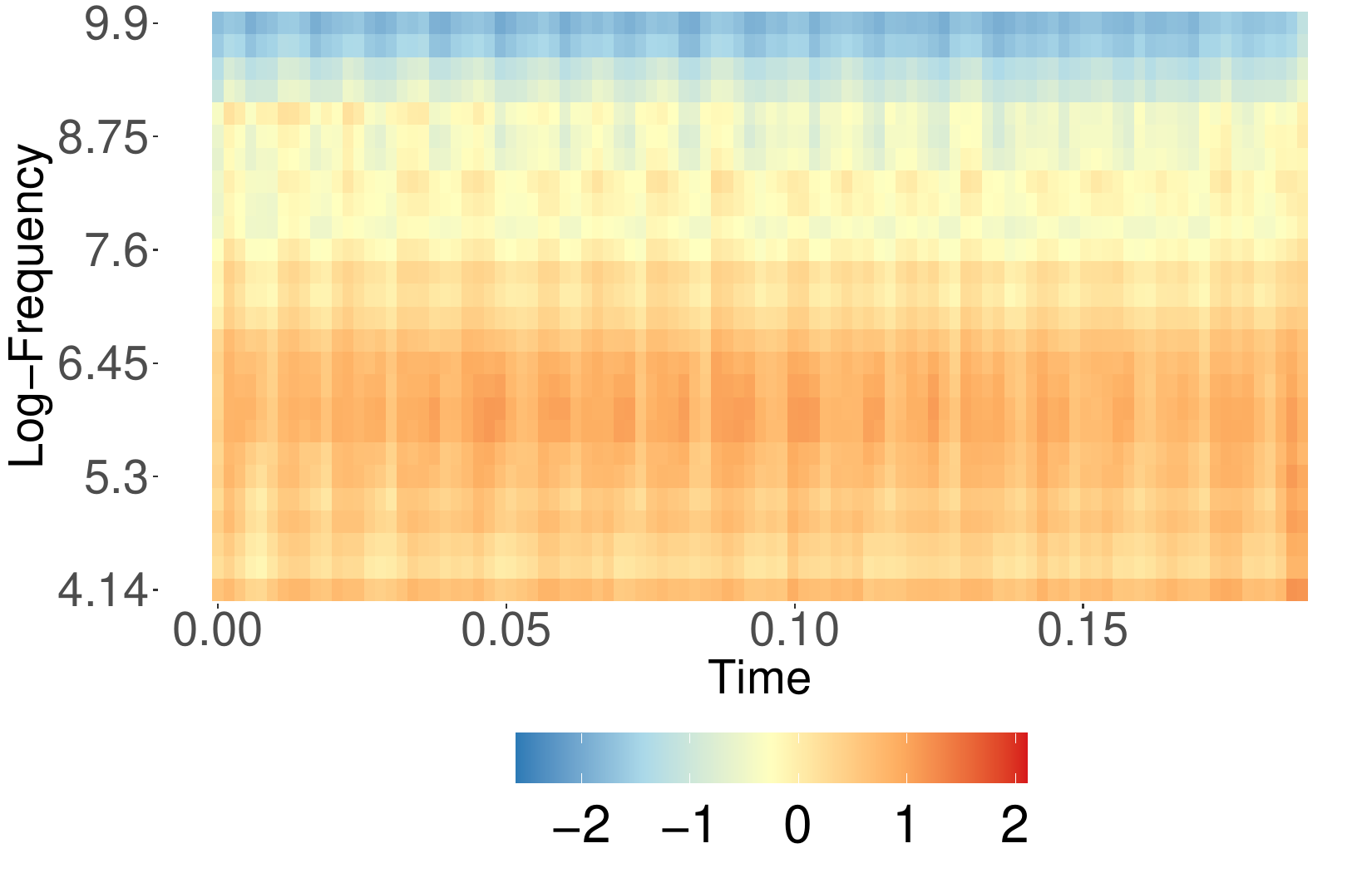}}} 
{\subfloat[II]{\includegraphics[scale=0.15]{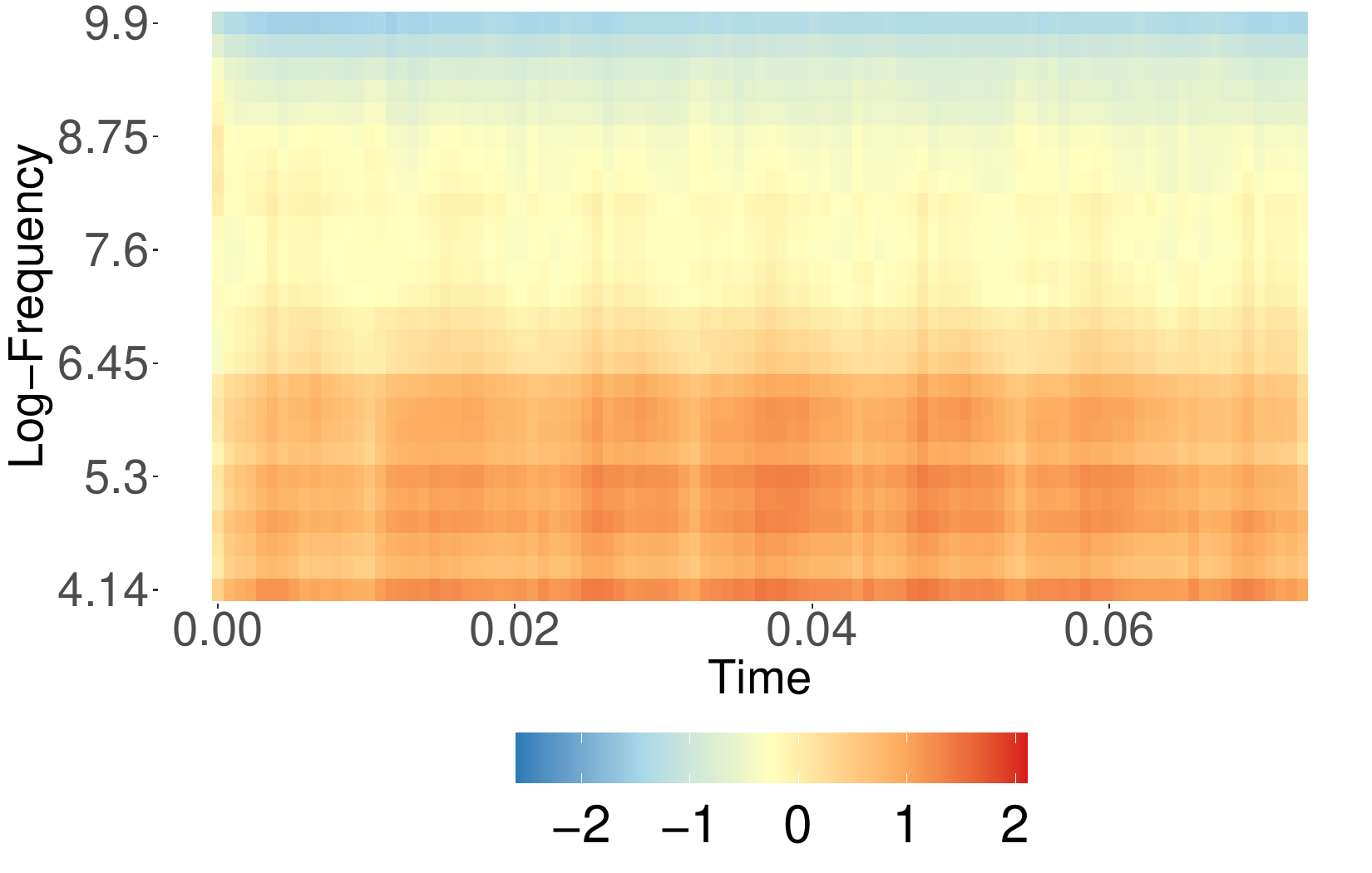}}}  
{\subfloat[MA]{\includegraphics[scale=0.15]{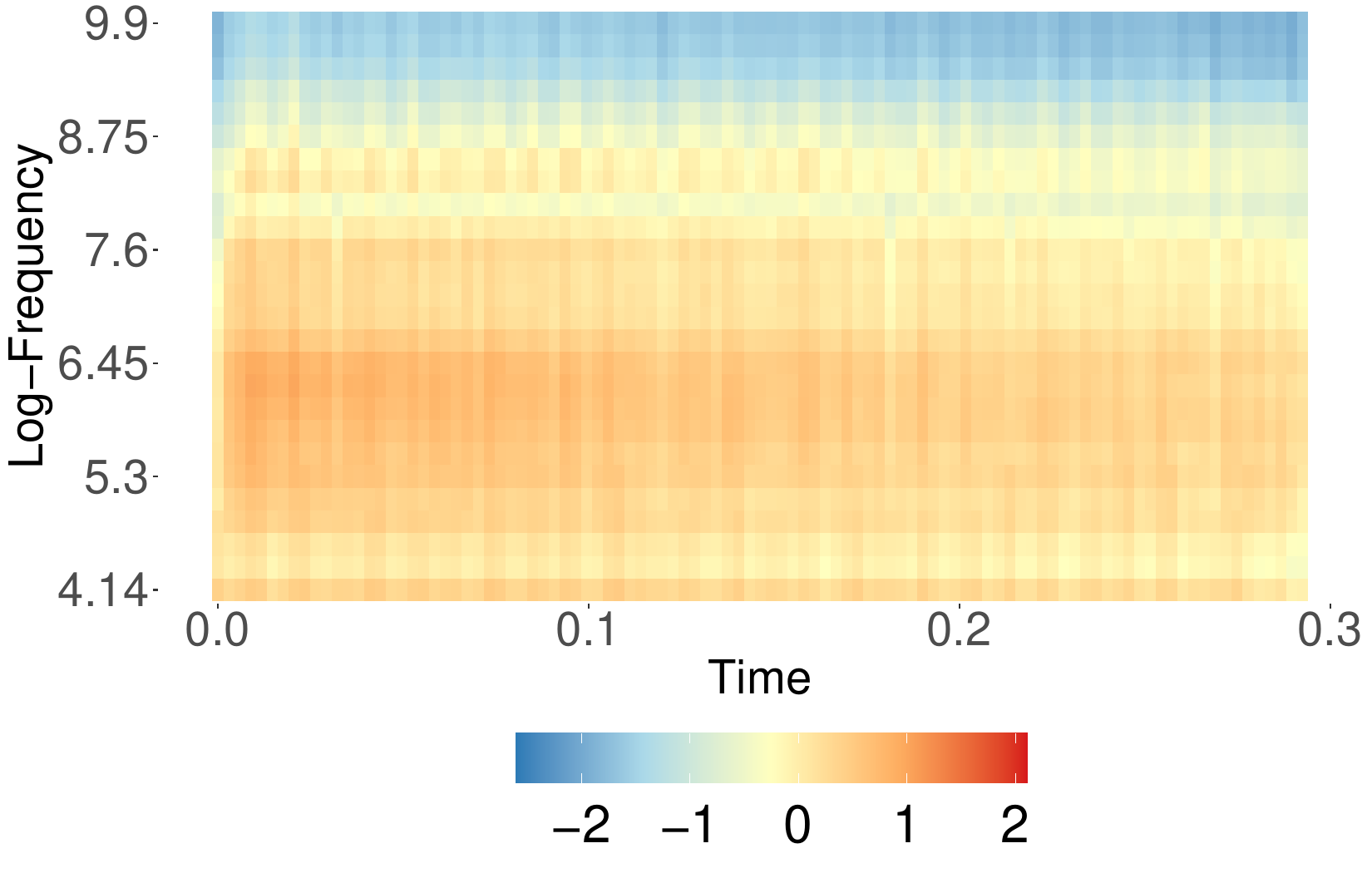}}} \\
{\subfloat[MO]{\includegraphics[scale=0.15]{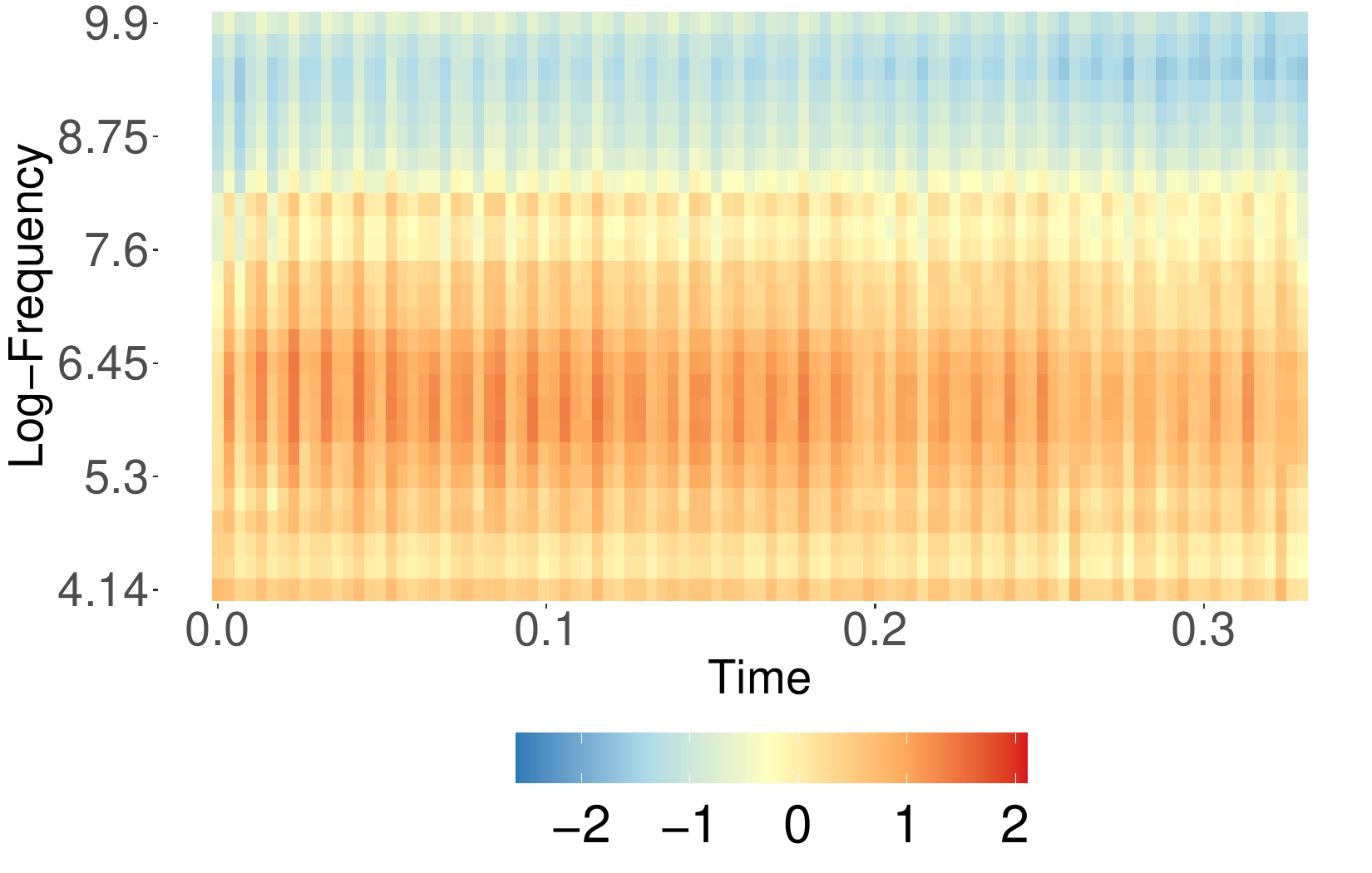}}} 
{\subfloat[PD]{\includegraphics[scale=0.15]{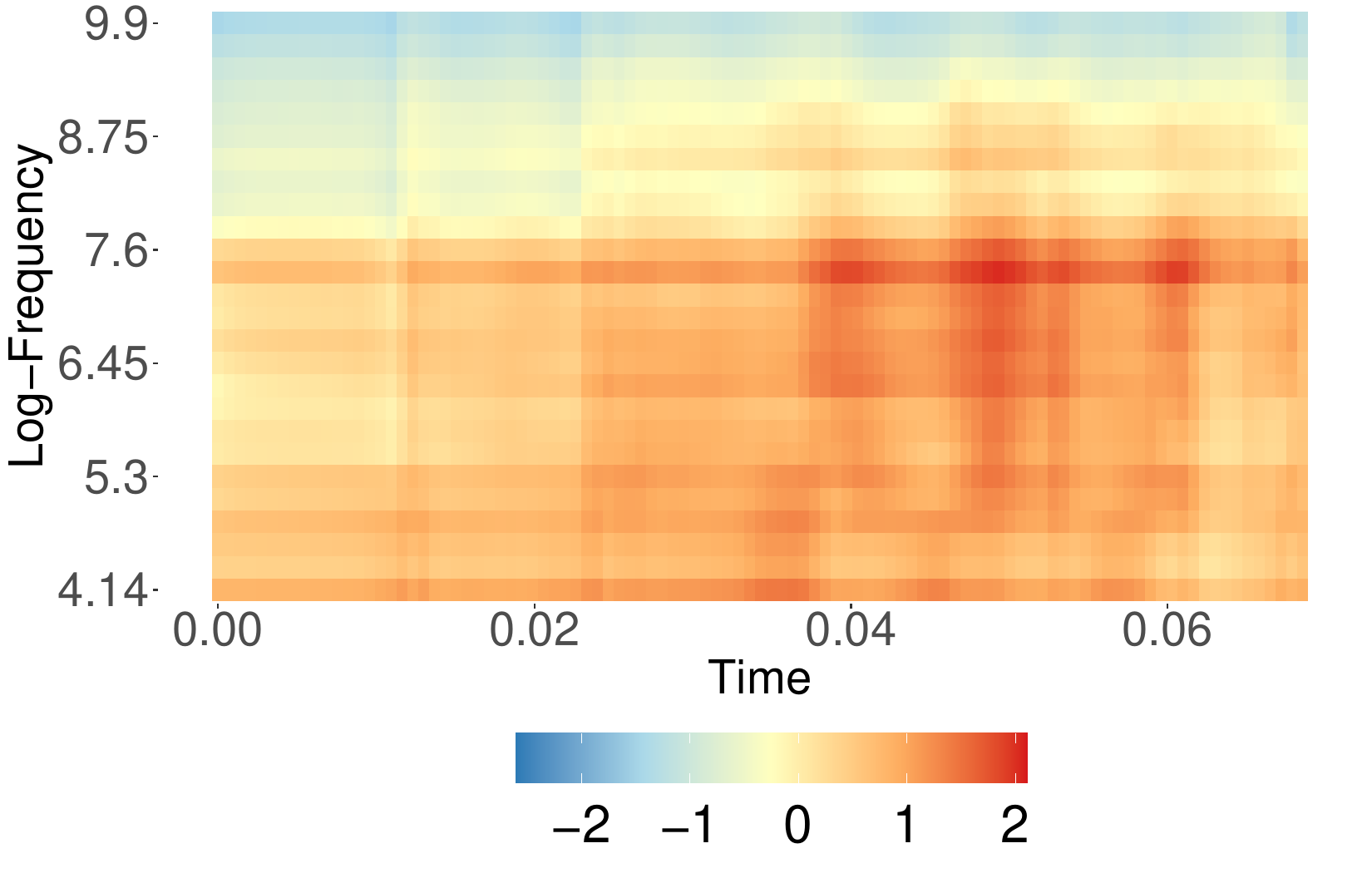}}} 
\caption{Posterior means of $\mathcal{A}_{\ell}(t,h)$. The color scale is shared across  pictures.}
\label{fig:real_wy1_mean}
\end{figure}

\begin{figure}[t]
{\subfloat[EC]{\includegraphics[scale=0.15]{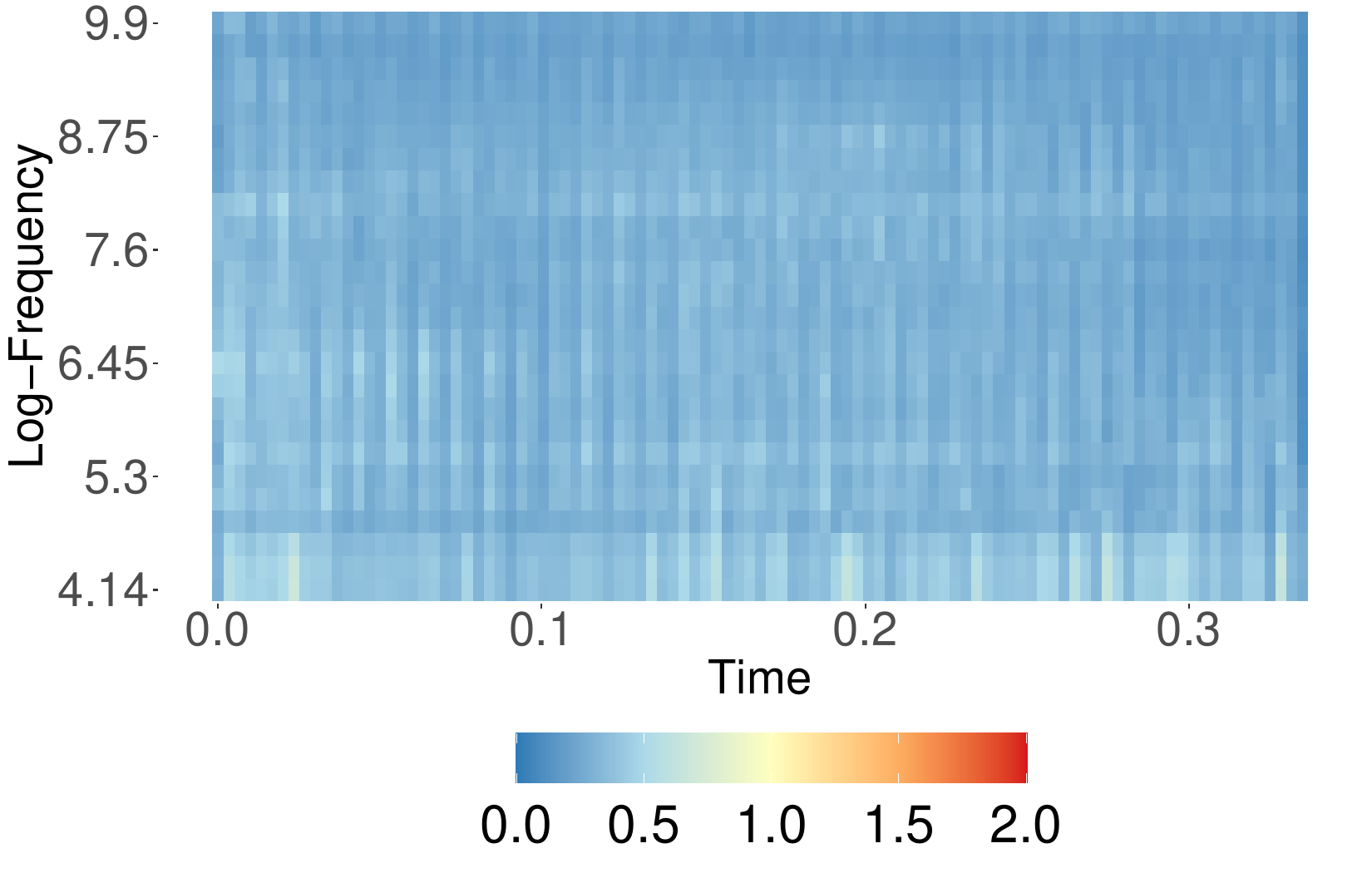}}} 
{\subfloat[ER]{\includegraphics[scale=0.15]{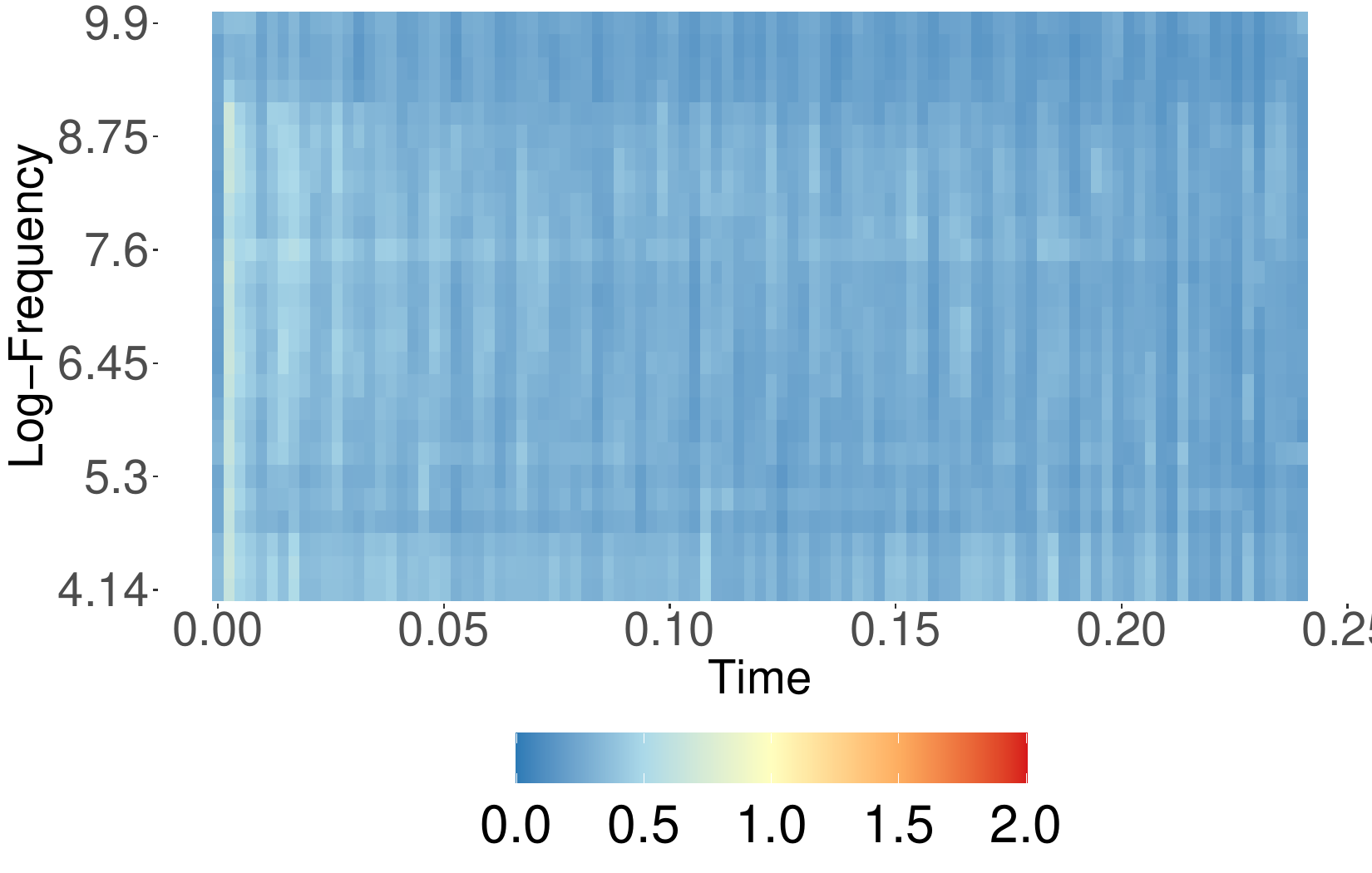}}} 
{\subfloat[FL]{\includegraphics[scale=0.15]{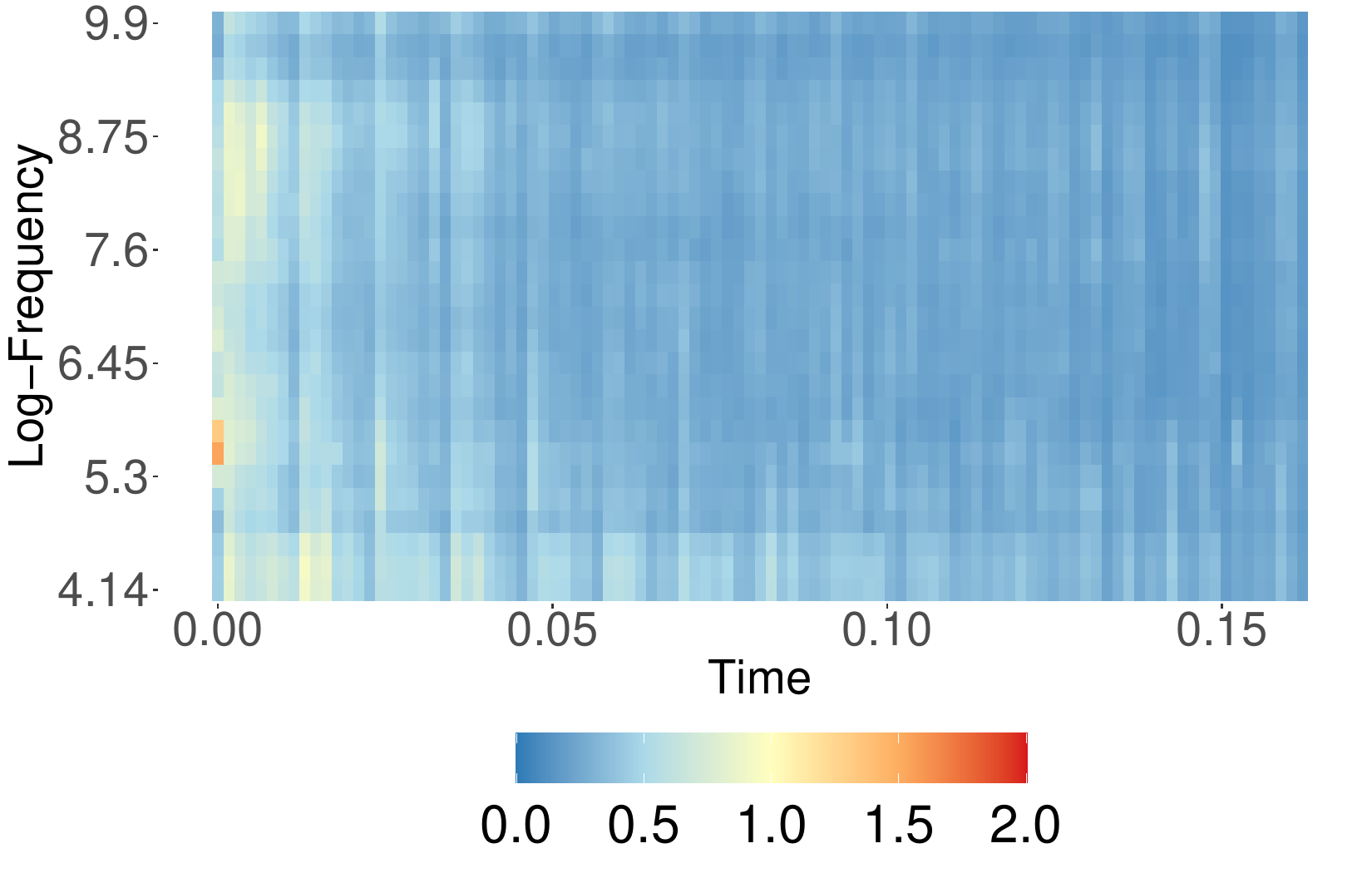}}} \\
{\subfloat[FU]{\includegraphics[scale=0.15]{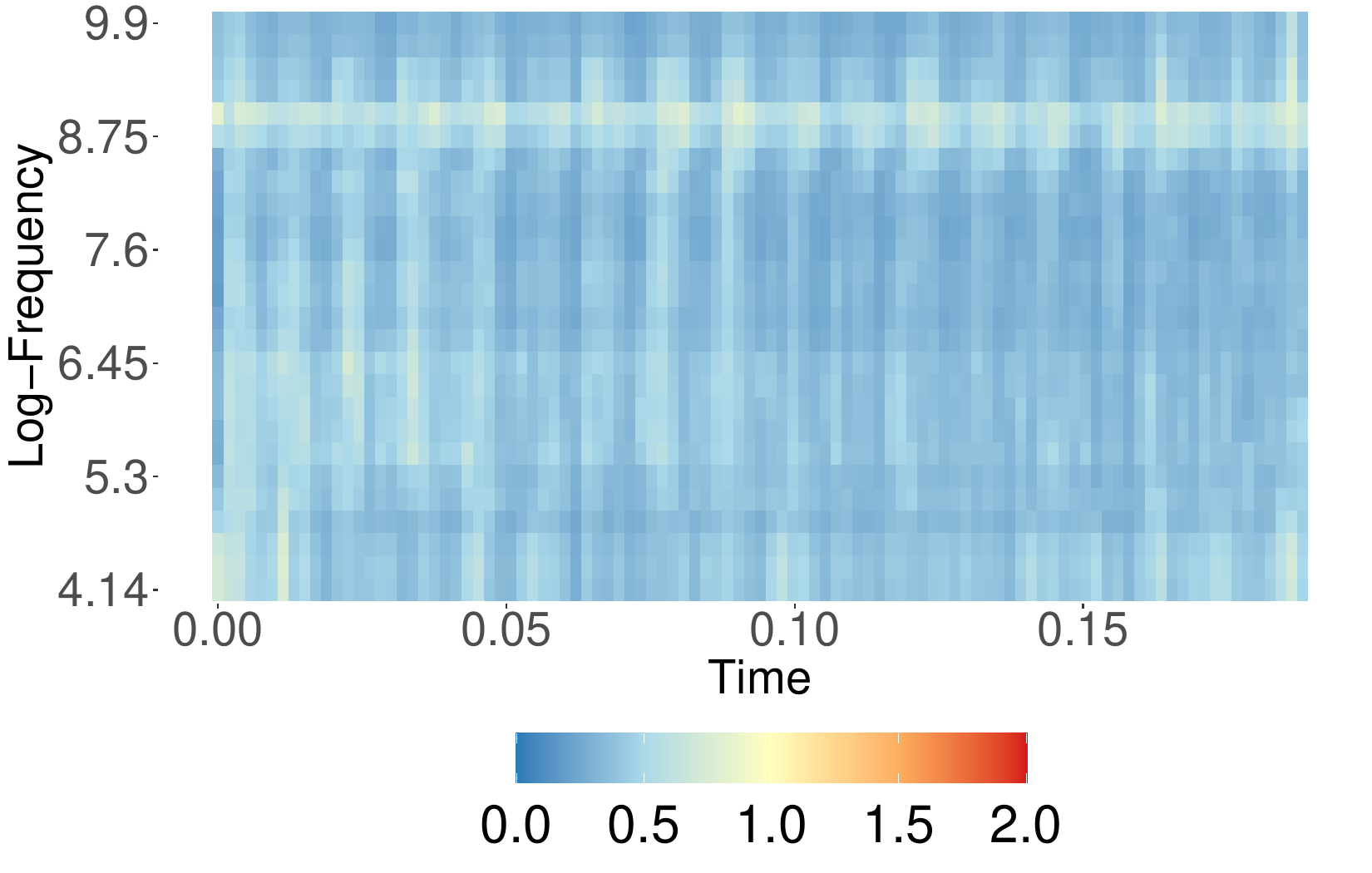}}} 
{\subfloat[II]{\includegraphics[scale=0.15]{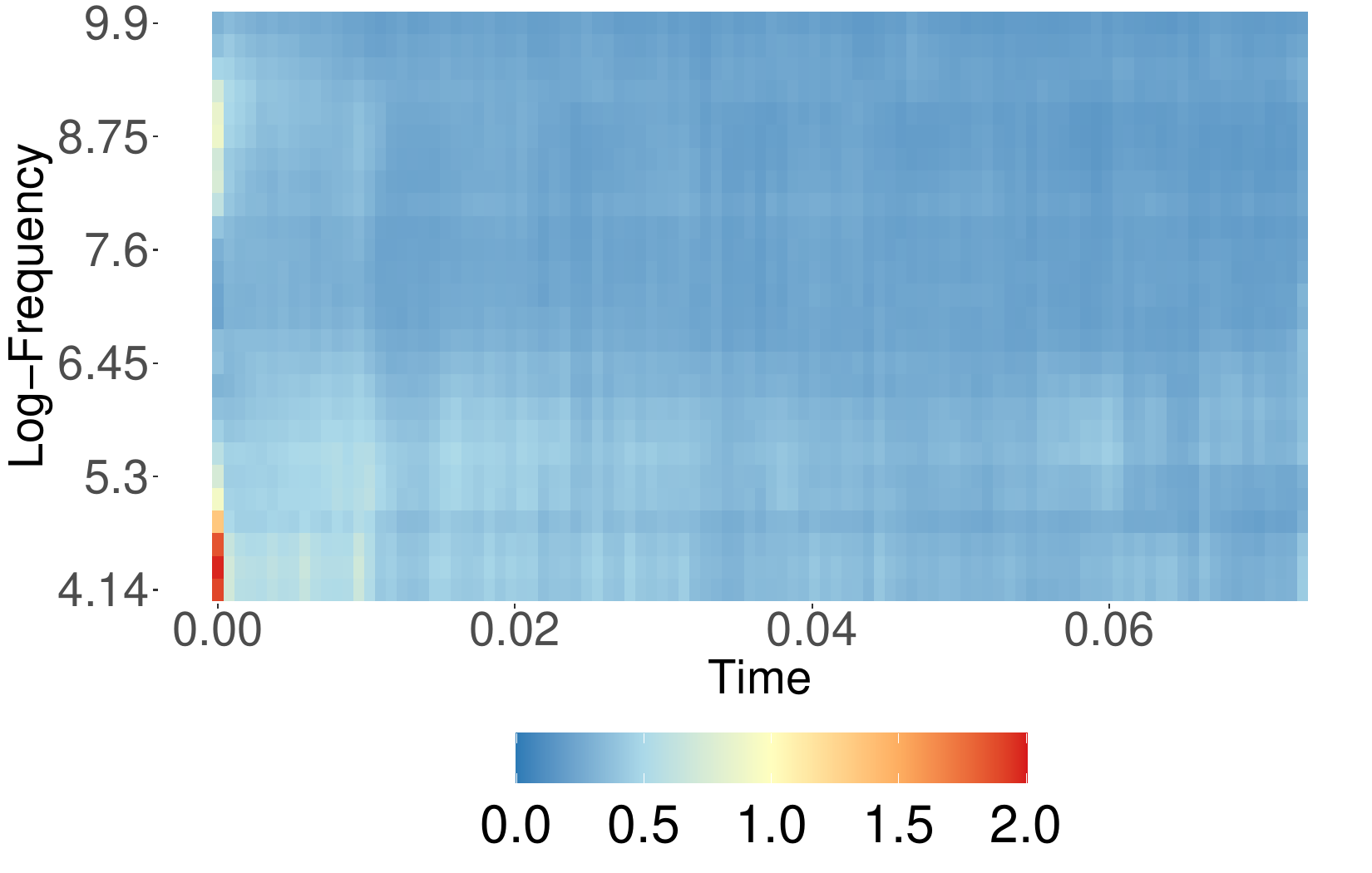}}}  
{\subfloat[MA]{\includegraphics[scale=0.15]{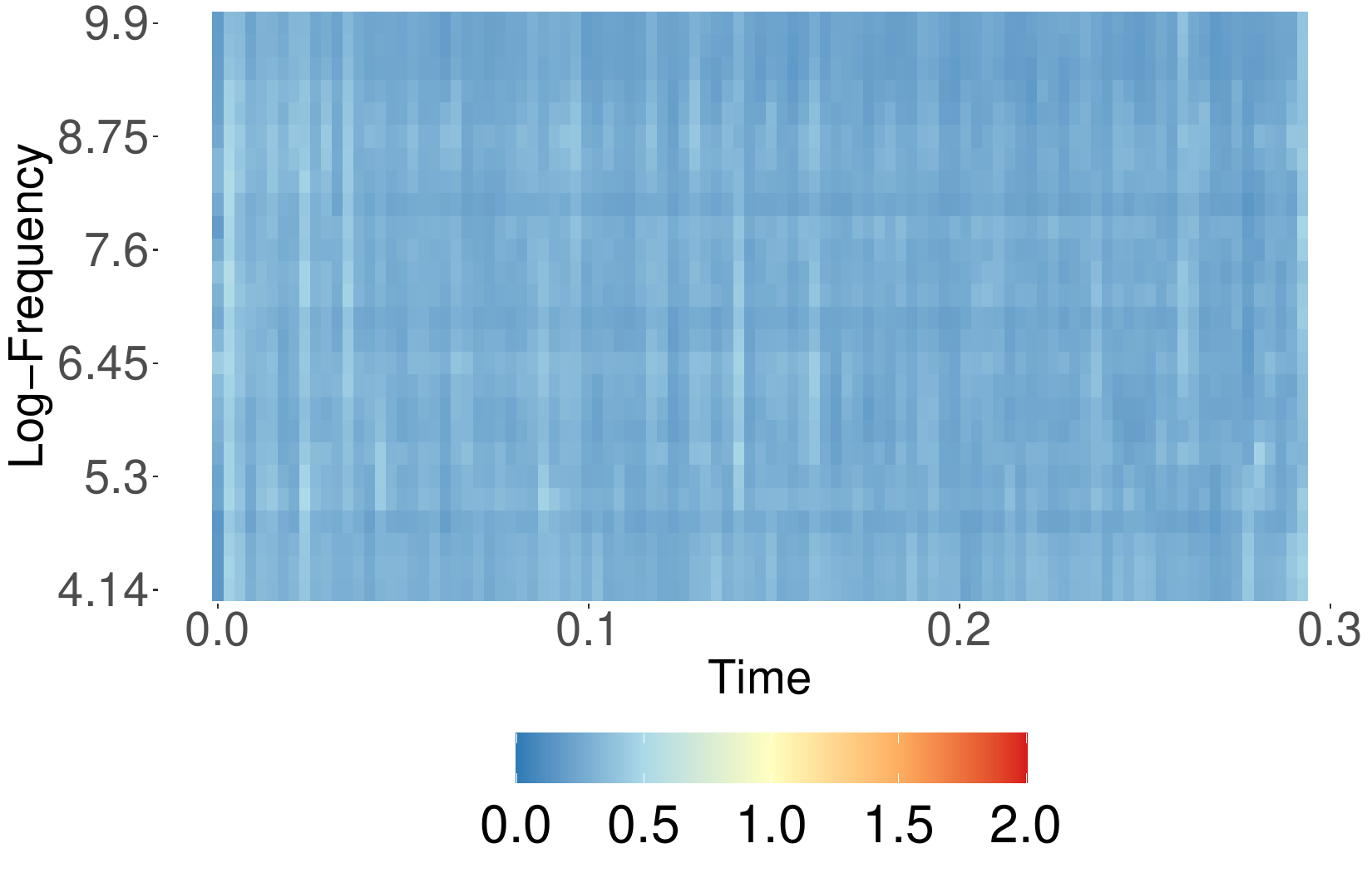}}} \\
{\subfloat[MO]{\includegraphics[scale=0.15]{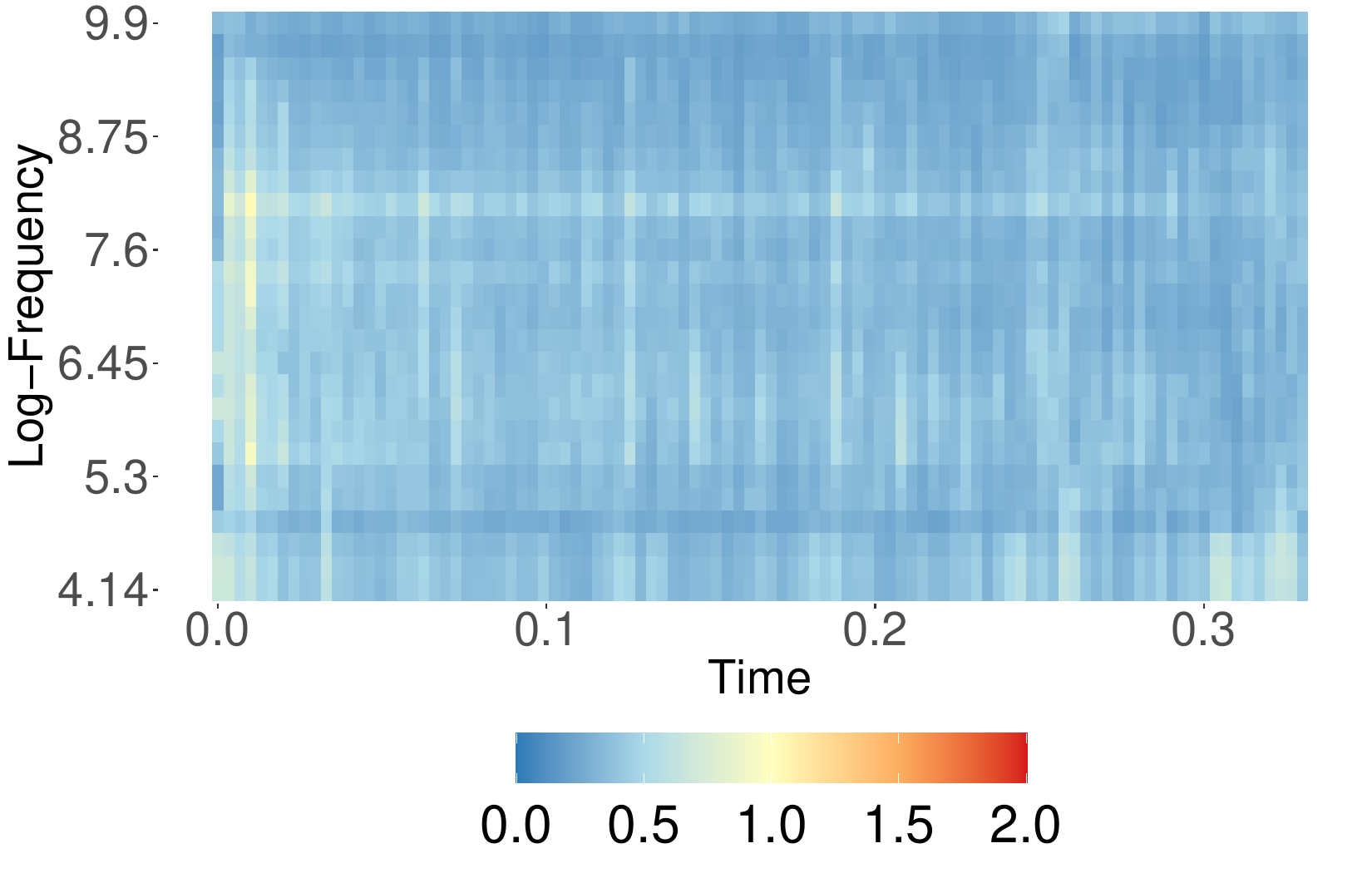}}} 
{\subfloat[PD]{\includegraphics[scale=0.15]{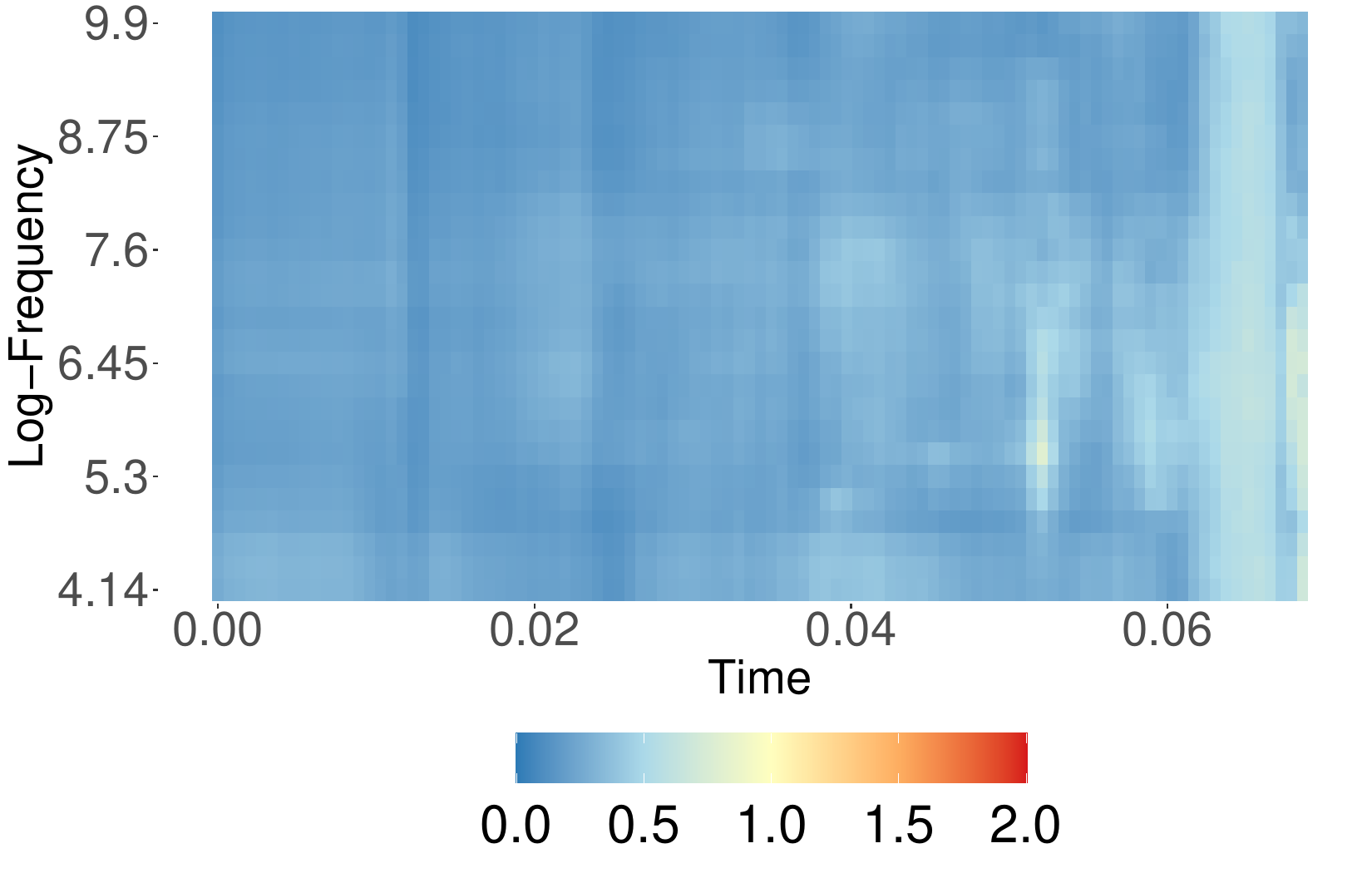}}} 
\caption{Posterior variance of $\mathcal{A}_{\ell}(t,h)$. The color scale is shared across  pictures.}
\label{fig:real_wy1_var}
\end{figure}

In this section, we present the results of the model applied to the data outlined in Section \ref{sec:data}. We estimate eight models, one for each species, with 60,000 iterations, discarding the first 4,800 as burn-in, and retaining every sixth iteration, resulting in  2,000 posterior samples for inference. The maximum number of neighbors in the NNGP is fixed at $4k = 40$. The total computation time, employing a parallel implementation in Julia \citep{bezanson2017julia} on 32 cores, ranged from a minimum of 4 days for the PD to 15 days for the EC.

For the distribution of the warping parameters, we assume $b_{\zeta} = b_{\delta} = 0.75$, with
\begin{align}
    m_{\zeta} &\sim N_{(-5,5)}(0, 0.75),\\
    m_{\delta} &\sim N_{(-5,5)}(0, 0.75),\\
    v_{\zeta} &\sim \text{IG}_{< 0.75}(0.01,0.01),\\
    v_{\delta} &\sim \text{IG}_{< 0.75}(0.01, 0.01).
\end{align}
Additionally, we assume ${\alpha}_i\sim U(0,0.2)$, $\tilde{\beta}_i\sim U(0.75,1)$, $\sigma^2$, $\tau_i^2 \sim \text{IG}(1.0, 1.0)$, and $\mu_i \sim N(0,100000)$.

%Figure \ref{fig:real_theta} displays the 95\% credible intervals (CIs) of the parameters in $\boldsymbol{\theta}$, with the ranges depicted in terms of practical range (see equations \eqref{eq:pr_h} and \eqref{eq:pf_cd}). In Figure \ref{fig:real_eta}, we compute the species-specific posterior means for each of the parameters in $\alpha_i$, $\tilde{\beta}_i$, $\mu_i$, and $\tau_i^2$, and the boxplot represents the distribution of the posterior means across $i=1,\dots,N$ for each of the eight species.
%Credible intervals (CIs) describing the posterior distribution of the parameters in $\boldsymbol{\theta}$ are shown in Figure \ref{fig:real_theta}, while Figure \ref{fig:real_eta} contains the boxplots of the posterior means across the spectrogram of the parameters $(\alpha_i, \beta_i, \mu_i, \tau_i^2)$. In Figure \ref{fig:real_warp}, each line represents the warping function of a single spectrogram computed using the posterior means of $(\zeta_i, \delta_i)$. 
%To describe the warping function, we display in Figure \ref{fig:real_warp} the posterior predictive distribution of the warping functions.
%Moreover, we derive the posterior mean and variance from the posterior distribution in \eqref{bho} of the representative sound $\mathcal{A}_{\ell}(t,h)$, assuming $l_{\ell}$ to be equal to the median length of the observed time lengths, with time points separated by 0.01, as the data observed,  and the same number and positions of log-frequency points: results as hown in \ref{fig:real_wy1} and \ref{fig:real_wy1a}.
Figure \ref{fig:real_theta} displays the 95\% credible intervals (CIs) of the parameters in $\boldsymbol{\theta}$, with the ranges depicted in terms of practical range (see equations \eqref{eq:pr_h} and \eqref{eq:pf_cd}). In Figure \ref{fig:real_eta}, we compute the species-specific posterior means for each of the parameters in $\alpha_i$, $\tilde{\beta}_i$, $\mu_i$, and $\tau_i^2$, and the boxplot represents the distribution of the posterior means across $i=1,\dots,N$ for each of the eight species.
Credible intervals (CIs) describing the posterior distribution of the parameters in $\boldsymbol{\theta}$ are shown in Figure \ref{fig:real_theta}, while Figure \ref{fig:real_eta} contains the boxplots of the posterior means across the spectrogram of the parameters $(\alpha_i, \beta_i, \mu_i, \tau_i^2)$. 
To describe the warping function, we display in Figure \ref{fig:real_warp} the posterior predictive distribution of the warping functions. Moreover, we derive the posterior mean and variance from the posterior distribution in \eqref{bho} of the representative sound $\mathcal{A}_{\ell}(t,h)$, assuming $l_{\ell}$ to be equal to the median length of the observed time  of the given species, with time points separated by 0.01, as the data observed, and the number and positions of log-frequency points equal to the one of the data: posterior means are in Figure \ref{fig:real_wy1_mean} while posterior variances are in Figure \ref{fig:real_wy1_var}.

\begin{figure}[t]
	{\subfloat[Quadratic distance]{\includegraphics[scale=0.24]{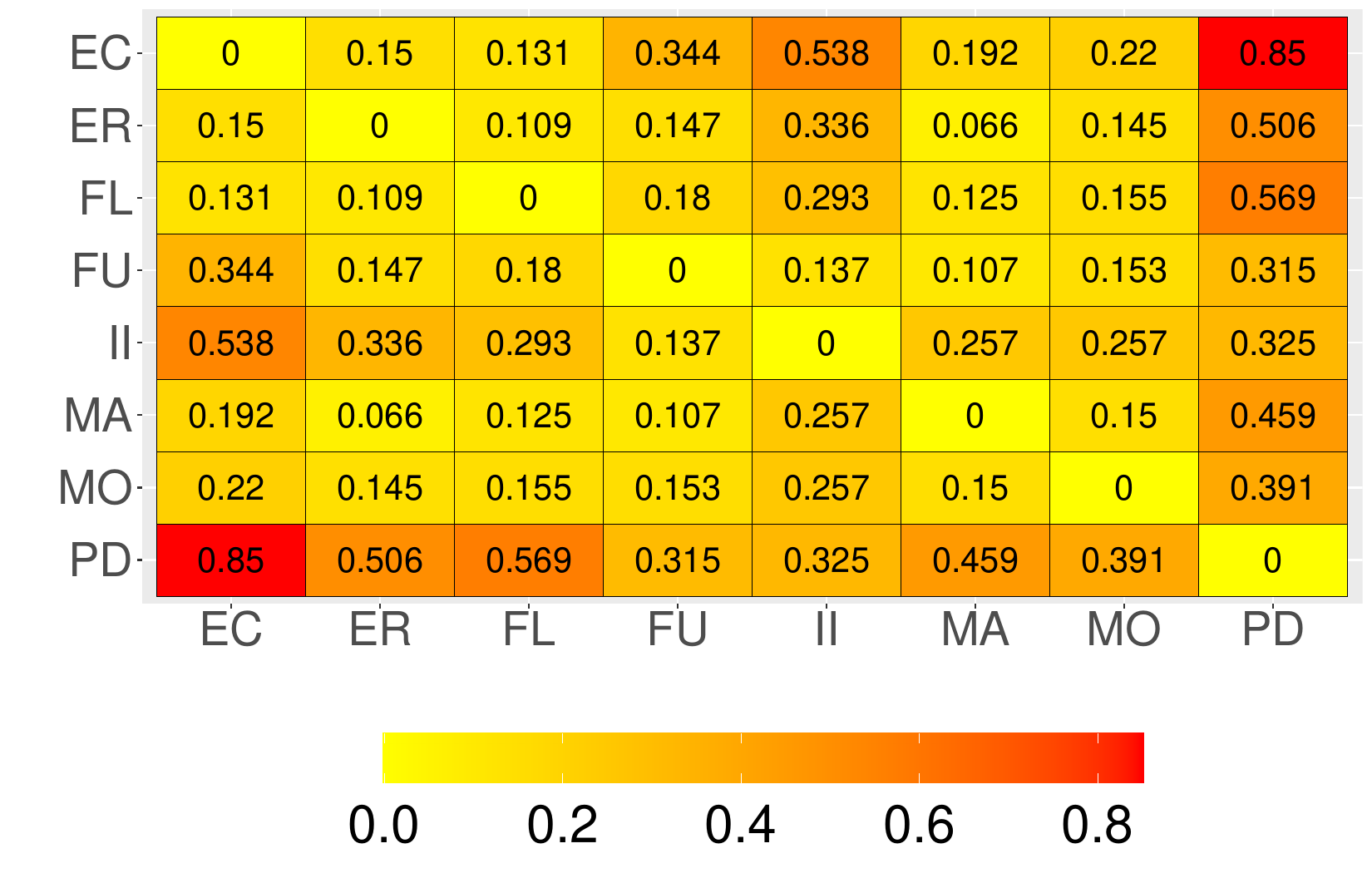}}} 
	{\subfloat[Phylogenetic tree]{\includegraphics[scale=0.24]{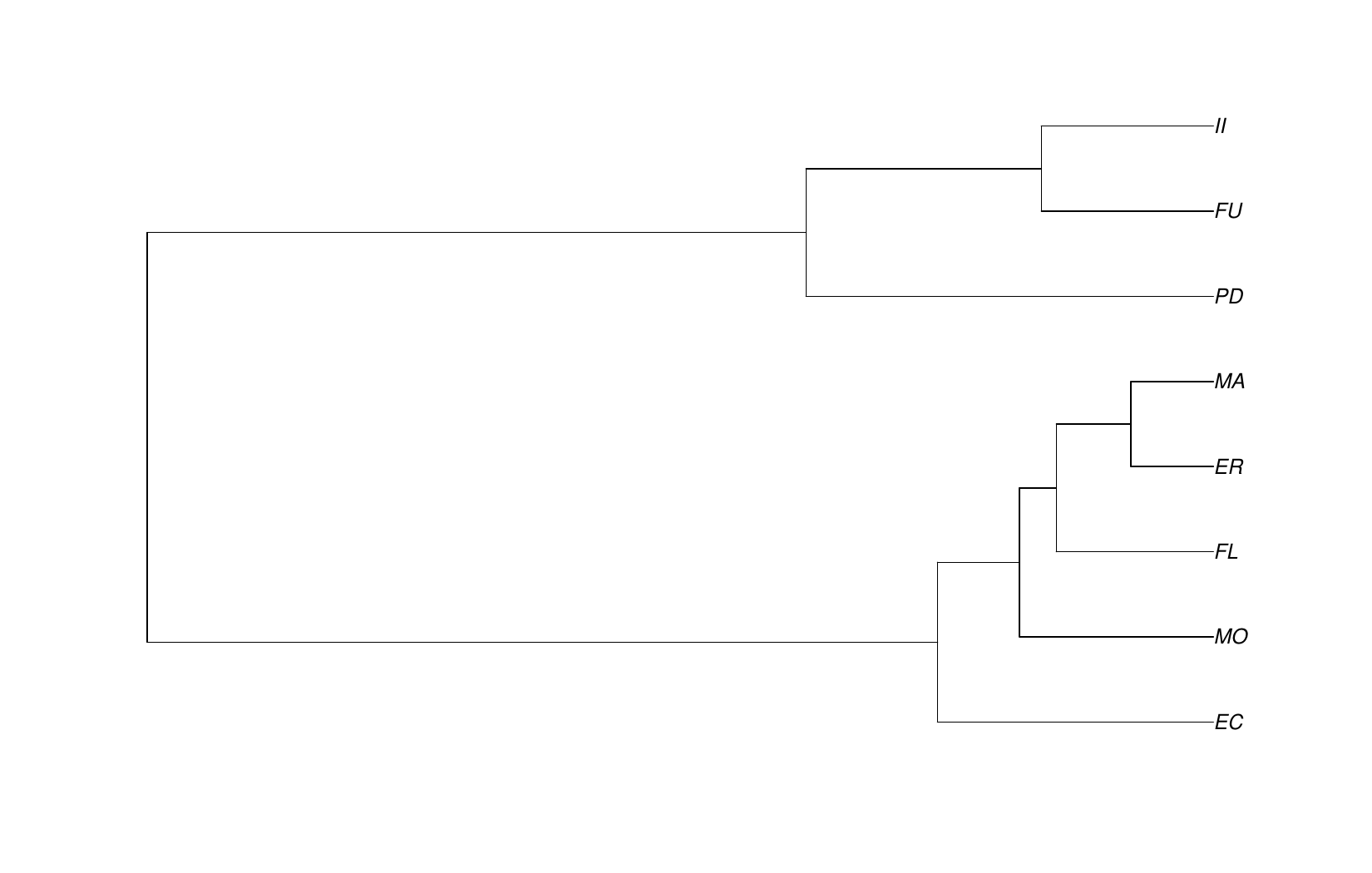}}} 
	\caption{Quadratic distance between the posterior means of $\mathcal{A}_{\ell}(t,h)$ and associated Phylogenetic tree}
	\label{fig:real_distance1}
\end{figure}

As a means to compare the representative sounds of each species, we decided to compute a distance between the posterior means of $\mathcal{A}_{\ell}(t,h)$. We can do this in different ways, but the easiest one is to evaluate all $\mathcal{A}_{\ell}(t,h)$ on a shared set of frequency and temporal points. Hence, for all species, we assume
$l_{\ell}$ to be equal to the median length of the observed time across all species, with time points separated by 0.01,  and the number and positions of log-frequency being the set $\mathcal{H}$.

Then, the distance between species A and B is the sample mean over all points in the spectrogram of the square differences between the intensity of the spectrogram at the same time-frequency points of the two species. The results are shown in Figure \ref{fig:real_distance1} (a).
From the distance matrix, by using the R library ``phylogram'' \citep{phylogram}, we compute a phylogenetic tree, which is shown in Figure \ref{fig:real_distance1} (b).
%\textcolor{red}{LA distanza di Bhattacharyya la levo, che i risultati sono simili, ma produce un albero peggiore e computazionalmente \`e' un disastro, se vogliamo possiamo dire di averlo fatto, ma i risultati sono simili}.

It is interesting to note that, for all species, the contribution of the cyclic component's contribution is smaller compared to the non-cyclical one, as evidenced by the values of $\lambda$, consistently lower than 0.5, as shown in Figure \ref{fig:real_theta} (b). However, for all species except PD, the cyclic components exhibit a period $\gamma \approx 0.025$, as illustrated in Figure \ref{fig:real_theta} (c). This observation is unsurprising, given that the spectrograms are all derived from the same time-window, and the frequency-range of the data is similar across species.
The different result obtained in the PD species is likely due to the very short time-length of their spectrograms (see Table \ref{tab1}), which may have hindered the learning of these parameters.
Moreover, for all species, there is a strong dependence in both shared-time and log-frequency distance, highlighted by the large value of practical ranges in Figure \ref{fig:real_theta} (d) and (e). Figure \ref{fig:real_theta} (g) emphasizes that there is a small, almost vanishing non-separability in the covariance.

The warping functions depicted in Figure \ref{fig:real_warp} reveal that for species ER, FL, II, and PD, there exists an expected shape distinct from the absence of warping (a 45-degree line). However, the results for the PD species, considering the aforementioned limitations, may not be entirely reliable. For the other species, it appears that the warping functions are primarily asymmetric fluctuations around the non-warped one.
In terms of error terms variability and mean value, we can see from the boxplots of Figure \ref{fig:real_eta} (c) and (d) that there are consistent across species. The same is true for parameters $\alpha_i$ and $\tilde{\beta}_i$.

In terms of posterior means of $\mathcal{A}_{\ell}(t,h)$, as shown in Figure \ref{fig:real_wy1_mean}, there are important differences, especially in the log-frequencies 4.14, which is the lower one, and between 6.9 and 8.75. In the frequency range (5.06, 6.65), all species have strong intensity, which is generally quite constant across time, with the exception of species EC and MA, which decreases along the time axis.
The variance, shown in Figure \ref{fig:real_wy1_var}, is approximately constant, with larger values in the first time points, which can be attributed to the synchronization function, since these points may be estimated with fewer observations, while the apparent cyclic pattern may be caused by the cyclic process.

\subsection{Biological interpretation}

The log-frequency distribution reflects the formant (local maxima of the spectrum) distribution of the respective species, showing that frequency variation in ER, FU, and MA  is higher than in the other species at higher frequencies. This result agrees with \cite{gamba2012} findings concerning formant distribution and variation across lemur species, where the fourth formant in FU, MA and ER was estimated and measured at around 6 kHz  (see also Figure 1 in the same paper). Variation in frequency distribution across species also reflects previous findings across Eulemur species, where distinctive species-specific traits were found in low-pitched grunts \cite[see][]{gamba2012,gamba2016}.

Reliable reconstructions of the phylogenesis have been prevented by the absence of fossils for lemurs, the high degree of parallelism among lemurs, and the inconsistency of reconstructions regarding relationships within Lemuridae. Concerning this last point, understanding relationships between the Eulemur species is the most daunting task, as demonstrated by the many available reconstructions that rarely agree. %Distances emerging from our resultis show that, unlike what we could expect, MA and FL never appear closer
%than the other taxa. 
According to the dissimilarity matrices in Figure \ref{fig:real_distance1} (a), the distances reveal that MA and FL are not closer than the other taxa, which is surprising considering they were earlier considered subspecies of Eulemur macaco. However, a study by \cite{gamba23008} alredy showed differences in grunt vocalizations among individuals of what were then two subspecies of the species Eulemur macaco, suggesting that divergence across vocalizations may only partially map differences at the phylogenetic level (in agreement with findings of \cite{Macedonia94}). These differences were mainly attributed to formants and fundamental frequency and were interpreted as informative of
morphological differences between the two species \citep{gamba23008}. Vocalisations in MA and FL may have diverged to prevent hybridisation, which has been observed in the wild. This is supported by research on African Cichlids, particularly of the genus Pseudotropheus, which reside in Lake Malawi. These fish have undergone adaptive divergence in genes that contribute to odour-mediated mate selection, showing how communicative signals can play a critical role in species-specific interactions. Although phylogenetic reconstructions obtained from the analyses of the lemurid data sets are often based on very different datasets, and those on communication data are rare, the trees generated
during this study show a relatively different picture from previous investigations. We must also consider that DNA research data must still establish a clear connection in some clades \citep{DelPero2006}. 
By looking at Figure \ref{fig:real_distance1} (b), we can see that the analysis of low-pitched calls we performed led to identification 
of II and PD as outgroups of the Lemuridae in this study, but FU surprising clustered together with II, possibly due to higher energy in low frequencies of their calls. On the other hand, as it has been found in all of our newly generated trees,  MA and ER are sister groups. 
We found inconsistent positioning of the Indridae species in this study, as PD and II did not show lower dissimilarity than the Lemuridae. This is probably related to an inherent problem with the sounds we considered for PD. The short duration of the PD sounds prevents proper construction of our models and mapping of differences with other species. Thus, limiting our analysis to the position of II, we can observe how this presents greater
distances than what we measured among Lemuridae species, particularly. 
%Finally, a closer look at the MAP distribution allows us to see how FU and MO appear relatively similar in agreement with the observations of Stoessel et al. (2023) concerning the evolution of the average auditory threshold. 

%As a way to compare the representative sounds of each species, and to measure how different they are, we use the Bhattacharyya distance \cite{Bhattacharyya} between the posteriors of $\mathcal{A}_i(t,h)$ and a quadratic distance between the posterior means of $\mathcal{A}_i(t,h)$. TO be able to compare sound from different species, we must evaluate them on a shared set of frequency and temporal points. We then decide to evaluate all $\mathcal{A}_i(t,h)$ on the same set of log-frequencies and to use a general $l_{\ell} = 0.25$ with 100 equally spaced time points.

%For the quadratic distance, we sum all the differences squared between the same time-frequency point of the posterior means of $\mathcal{A}_i(t,h)$ (see Figure \ref{fig:real_distance1}). On the other hand, 
%the Bhattacharyya distance, which measures the similarity of two probability distributions, $P()$ and $Q()$, with density function $p()$ and $q()$, it is defined as 
%$$
%D_B(P,Q) = - \log (\text{BC}(P,Q))
%$$
%where $\text{BC}(P,Q)$ is the Bhattacharyya coefficient  defined as
%\begin{equation} \label{eq:bc}
%	\text{BC}(P,Q) = \int_{\mathcal{X}} \sqrt{p(x)q(x)} dx  = \int_{\mathcal{X}} \sqrt{p(x)/q(x)} q(x)dx 
%\end{equation}
%All the integrals can be computed as Monte Carlo integrations and the results are in
% Figure \ref{fig:real_distance2}

\subsection{Cross-validation}
\begin{table}[t]
	\centering
	\begin{tabular}{r|rrrr}
	  \hline
	 & Our & NoWarp & NoCirc & NoAl \\ 
	  \hline
	  EC & \textbf{0.95} & 0.95 & 1.22 & 0.97 \\ 
	  ER & \textbf{0.91} & 0.92 & 1.17 & 0.98 \\ 
	  FL & \textbf{0.92} & 0.96 & 1.31 & 0.94 \\ 
	  FU & 0.91 & 1.02 & 1.33 &  \textbf{0.90} \\ 
	  II &  \textbf{0.93} & 1.06 & 1.38 & 0.94 \\ 
	  MA & 0.82 & 0.87 & 1.18 &  \textbf{0.81} \\ 
	  MO &  \textbf{0.92} & 0.91 & 1.29 & 0.93 \\ 
	  PD & 1.17 & 1.35 & 1.43 &  \textbf{0.71} \\ 
	   \hline
	\end{tabular}\caption{CRPS index. In bold it is shown the best model for each species.}\label{tab:crps}
\end{table}

We opt for cross-validation to validate the model along with all its components. Specifically, for each species and spectrogram, we select, randomly,  5\% of the time-frequency points and remove them from the data. Utilizing only the remaining 95\% of the data, we fit the model to all species under different settings: i) our proposed model (Prop.); ii) a model with no warping (NoWarp); iii) a model without the cyclic component (NoCirc); and iv) a model where $\alpha_i=0$ and $\beta_i=0$, representing a model without alignment (NoAl).
Subsequently, using the holdout sample, we compute the Continuous Ranked Probability Score (CRPS), and the results are presented in Table \ref{tab:crps}.

In 5 out of the 8 species, our proposed model shows the lowest CRPS. For species FU and MA, the NoAl model has a slightly lower CRPS, but our model's CRPS is very close to it.
Interestingly, removing the cyclic component results in a worsened CRPS for all species, with a significant increase in margin. Conversely, alignment and warping generally produce smaller CRPS values. However, since the differences are not substantial, in future applications or when data are not highly informative, it may be better to estimate a model without these components.
For species PD, the best model is the one without alignment. This  can be attributed to the short observations present in this dataset, which do not allow for alignment.

%%%%%%%%%%%%%%%%%%%%%%%%%
% Section 6.0 		Conclusion & Future development  %
%%%%%%%%%%%%%%%%%%%%%%%%%
\section{Conclusion and Future development}\label{sec:conclude}

This work has presented a spatial-temporal model for bioacoustic data with non-stationary temporal patterns and periodic artifacts. The model combines novel ideas with several techniques, including the NNGP, to describe the complex structures that are intrinsic to bioacoustic data and to handle the sheer size of the available dataset.
The model was applied to a motivating example, where a large dataset of recorded spectrograms was analyzed. From the model output, we were able to estimate a representative sound for each species, which we then compared in terms of quadratic distance and phylogenetic tree. 
We also performed cross-validation to evaluate if all the model components were necessary to describe the data. Although this was not always true, with the exception of one species, our model was always the best or very close to the best.

We are currently working on the application of this methodology to a larger dataset comprising different calls. This will provide a better description of the distance between species in relation with a phylogenetic tree.

%The future will find the authors looking for a proper a metric that can properly measure the distance between the posterior predictive distributions of the species-specific latent ``MC'' so as to provide a better understanding on the role of acoustics in the evolutionary distinctiveness of the various species. Another direction of future work is to extend the current model to become a predictive model for the call-type and the species labels.

%%%%%%%%%%% 
% Acknowledgment  %
%%%%%%%%%%% 
\section*{Acknowledgment}\label{sec:acknowledge}

The work of the E. Bibbona was partially funded under the National Plan for Complementary Investments to the NRRP, project ‘‘D34H–Digital Driven Diagnostics, prognostics and therapeutics for sustainable Health care’’ (project code: PNC0000001), Spoke 4, funded by the Italian Ministry of University and Research.\\
The work of G. Mastrantonio was partially carried out within the FAIR - Future Artificial Intelligence Research and received funding from the European Union Next-GenerationEU (PIANO NAZIONALE DI RIPRESA E RESILIENZA (PNRR) – MISSIONE 4 COMPONENTE 2, INVESTIMENTO 1.3 – D.D.1555 11/10/2022, PE00000013).\\
We aknowledge the support of the SmartData@PoliTO center for Big Data and Machine Learning technologies. Computational resources were provided by HPC@POLITO (http://www.hpc.polito.it).

\appendix   

\section{Simulated study}\label{subsec:applysim}

\begin{table}[t]
\centering
\scriptsize
\begin{tabular}{rrrrrrr}
\hline
& $\tilde{\alpha}_i$ & $\tilde{\beta}_i$& $\zeta_i$ & $\delta_i$ & $\mu_i$ & $\tau_i^2$ \\ [2pt]
\hline
  1 & 0.00 & 1.00 & 0.75 & 0.88 & -1.12 & 1.16 \\ 
  2 & 0.07 & 0.78 & 0.34 & 0.42 & -0.08 & 1.27 \\ 
  3 & 0.18 & 0.84 & 0.19 & -0.26 & 3.86 & 1.39 \\ 
  4 & 0.08 & 0.92 & 0.94 & 0.42 & 7.13 & 1.53 \\ 
  5 & 0.16 & 0.82 & -0.14 & 0.12 & 2.00 & 0.96 \\ 
  6 & 0.12 & 0.88 & 0.85 & 0.60 & -1.95 & 0.87 \\ 
  7 & 0.09 & 0.91 & 0.13 & 0.78 & -6.12 & 0.85 \\ 
  8 & 0.10 & 0.77 & 0.60 & 0.50 & -3.76 & 0.64 \\ 
  9 & 0.17 & 0.88 & 0.44 & -0.19 & -0.38 & 0.70 \\ 
  10 & 0.07 & 0.90 & 0.83 & 0.68 & 5.34 & 0.60 \\ 
  11 & 0.07 & 0.76 & 0.66 & 0.90 & -1.46 & 1.31 \\ 
  12 & 0.08 & 0.79 & 0.61 & -0.06 & -6.44 & 0.82 \\ 
  13 & 0.19 & 0.85 & 0.15 & 0.11 & -2.81 & 1.20 \\ 
  14 & 0.06 & 0.78 & 0.82 & 0.81 & -0.32 & 0.67 \\ 
  15 & 0.07 & 0.85 & 0.13 & 0.82 & -2.14 & 1.06 \\ 
\hline
\end{tabular}\caption{Simulated values of the data-specific parameters used to generate the synthetic data set.}
\label{tab:simwarp}
\end{table}

\begin{table}[t]
\centering
\scriptsize
\begin{tabular}{ r|llllll }
	\hline
	  					 	& ${\alpha}_i$ & $\tilde{\beta}_i$& $\zeta_i$ & $\delta_i$ & $\mu_i$ & $\tau_i^2$ 	\\ [2pt]
	\hline 
	\multirow{3}{2em}{1}	& 0.00 & 1.00 & 0.75 & 0.88 & -1.12 & 1.16 \\ 
							& 0.083 & 0.932  & 0.507 & 0.696 & -0.217 & 0.764 	\\
							& (0.00 \ 0.218) & (0.759 \ 1.00) & (0.112 \ 0.858) & (0.232 \ 1.062) & (-1.689 \ 1.207) & (0.516 \ 1.066) 	\\ [3pt] 
	\multirow{3}{2em}{2}    & 0.07 & 0.78 & 0.34 & 0.42 & -0.08 & 1.27 \\ 
							& 0.113 & 0.695 & 0.596 & $0.911$ & 1.024 & 1.41 	\\ 
							& (0.00 \ 0.228) & (0.561 \ 0.808) & (0.109 \ 1.034) & (0.56 \ 1.198) & (-0.482 \ 2.503) & (1.025 \ 1.826) 		\\ [3pt]
	\multirow{3}{2em}{3} 	& 0.18 & 0.84 & 0.19 & -0.26 & 3.86 & 1.39 \\ 
							& 0.174 & 0.873 & -0.001 & -0.214 & 3.843 & 1.529	 \\ 
							& (0.104 \ 0.226) & (0.767 \ 0.964) & (-0.427 \ 0.372) & (-0.568 \ 0.126) & (2.144 \ 5.737) & (1.233 \ 1.846)	 \\ [3pt]
	\multirow{3}{2em}{4} 	& 0.08 & 0.92 & 0.94 & 0.42 & 7.13 & 1.53 \\
							& 0.082 & 0.761 & 0.879 & 0.482 & 7.777 & 1.935 \\ 
							& (0.033 \ 0.156) & (0.57 \ 1) & (0.531 \ 1.163) & (-0.016 \ 0.831) & (6.165 \ 9.284) & (1.471 \ 2.509) \\ [3pt]
	\multirow{3}{2em}{5} 	& 0.16 & 0.82 & -0.14 & 0.12 & 2.00 & 0.96 \\ 
							& 0.151 & 0.96 & -0.014 & 0.368 & 2.643 & 0.866 \\ 
							& (0.098 \ 0.234) & (0.878 \ 1) & (-0.278 \ 0.25) & (0.015 \ 0.707) & (1.125 \ 4.196) & (0.572 \ 1.201) \\ [3pt]
	\multirow{3}{2em}{6} 	& 0.12 & 0.88 & 0.85 & 0.60 & -1.95 & 0.87 \\ 
							& 0.069 & 0.913 & 0.753 & 0.64 & -0.735 & 1.127 \\ 
   							& (0.01 \ 0.143) & (0.815 \ 1) & (0.419 \ 0.942) & (0.287 \ 0.884) & (-2.157 \ 0.623) & (0.871 \ 1.415) \\ [3pt]
	\multirow{3}{2em}{7} 	& 0.09 & 0.91 & 0.13 & 0.78 & -6.12 & 0.85 \\ 
							& 0.089 & 0.964 & 0.202 & 0.886 & -5.289 & 0.952 \\ 
   							& (0.04 \ 0.171) & (0.905 \ 1) & (0.018 \ 0.383) & (0.621 \ 1.117) & (-6.716 \ -3.862) & (0.731 \ 1.215) \\ [3pt]
	\multirow{3}{2em}{8} 	& 0.10 & 0.77 & 0.60 & 0.50 & -3.76 & 0.64 \\ 
							& 0.034 & 0.853 & 0.401 & 0.436 & -2.861 & 0.641 \\ 
   							& (0 \ 0.108) & (0.781 \ 0.908) & (0.068 \ 0.625) & (0.198 \ 0.646) & (-4.206 \ -1.569) & (0.467 \ 0.855) \\ [3pt]
	\multirow{3}{2em}{9} 	& 0.17 & 0.88 & 0.44 & -0.19 & -0.38 & 0.70 \\ 
							& 0.185 & 0.962 & $0.721$ & 0.242 & 0.131 & 0.578 \\ 
  							& (0.14 \ 0.26) & (0.895 \ 1) & (0.508 \ 0.979) & (-0.02 \ 0.547) & (-1.492 \ 1.761) & (0.326 \ 0.894) \\ [3pt]
	\multirow{3}{2em}{10}  	 & 0.07 & 0.90 & 0.83 & 0.68 & 5.34 & 0.60 \\ 
							 & 0.03 & 0.959 & 0.777 & 0.69 & 6.014 & 0.657 \\ 
   							 & (0 \ 0.122) & (0.907 \ 0.995) & (0.489 \ 1.022) & (0.418 \ 0.935) & (4.467 \ 7.618) & (0.446 \ 0.917) \\ [3pt]
	\multirow{3}{2em}{11}    & 0.07 & 0.76 & 0.66 & 0.90 & -1.46 & 1.31 \\ 
							 & 0.072 & 0.893 & 0.327 & 0.568 & -1.379 & 1.06 \\ 
   							 & (0 \ 0.21) & (0.657 \ 1) & (-0.104 \ 0.87) & (0.093 \ 1.07) & (-2.896 \ 0.098) & (0.695 \ 1.498) \\ [3pt]
	\multirow{3}{2em}{12} 	 & 0.08 & 0.79 & 0.61 & -0.06 & -6.44 & 0.82 \\ 
							 & 0.067 & 0.855 & 0.798 & 0.258 & -6.604 & 1.074 \\ 
   							 & (0 \ 0.172) & (0.735 \ 0.958) & (0.504 \ 1.13) & (-0.053 \ 0.658) & (-8.022 \ -4.992) & (0.807 \ 1.344) \\ [3pt]
	\multirow{3}{2em}{13}   & 0.19 & 0.85 & 0.15 & 0.11 & -2.81 & 1.20 \\ 
							 & 0.196 & 0.946 & 0.197 & 0.293 & -2.575 & 1.329 \\ 
							 & (0.15 \ 0.262) & (0.816 \ 1) & (-0.038 \ 0.501) & (0.009 \ 0.696) & (-4.026 \ -0.83) & (1.055 \ 1.629) \\ [3pt]
	\multirow{3}{2em}{14}   & 0.06 & 0.78 & 0.82 & 0.81 & -0.32 & 0.67 \\ 
							 & 0.027 & 0.835 & 0.574 & 0.69 & 0.548 & 0.637 \\ 
   						 	 & (0 \ 0.109) & (0.75 \ 0.984) & (0.126 \ 0.89) & (0.201 \ 1.005) & (-0.897 \ 2.041) & (0.372 \ 0.956) \\ [3pt]
	\multirow{3}{2em}{15}    & 0.07 & 0.85 & 0.13 & 0.82 & -2.14 & 1.06 \\ 
							 & 0.07 & 0.875 & 0.06 & 0.918 & -2.476 & 1.276 \\ 
   							 & (0 \ 0.173) & (0.78 \ 0.966) & (-0.172 \ 0.298) & (0.58 \ 1.187) & (-4.388 \ -0.411) & (1.009 \ 1.603) \\ 
	\hline 
\end{tabular}\caption{Results of the simulated study - The true simulated values, posterior means and 95\% credible intervals (from top to bottom of each row) of the data-specific parameters for all $15$ synthetic sounds.}\label{tab:simwarpresults}
\end{table}

\begin{table}[t]
\centering
\scriptsize
\begin{tabular}{rlllllll}
	\hline
	 & $\sigma^2$ & $\lambda$ & $\gamma$ & $\rho$ & $\phi_d$ & $\phi_h$ & $\phi_c$ \\ [2pt]
	\hline
	& 10 & 0.5 & 0.06 & 0.85 & 206 & 0.69 & 766 \\ 
	 & 9.702 & 0.476 & 0.06 & 0.805 & 167.821 & 0.038 & 1021.255 \\ 
	 & (8.637 \ 10.763) & (0.434 \ 0.519) & (0.06 \ 0.06) & (0.394 \ 0.992) & (123.967 \ 228.794) & (0.034 \ 0.044) & (730.176 \ 1246.129) \\ 
	\hline
\end{tabular}\caption{Results of the simulated study - The true simulated values, posterior means and 95\% credible intervals (from top to bottom of each row) of the general parameters for the synthetic latent ``MC''.}\label{tab:simresults}
\end{table}

\begin{figure}[t]
{\subfloat[$\mathbf{y}_1$]{\includegraphics[scale=0.25]{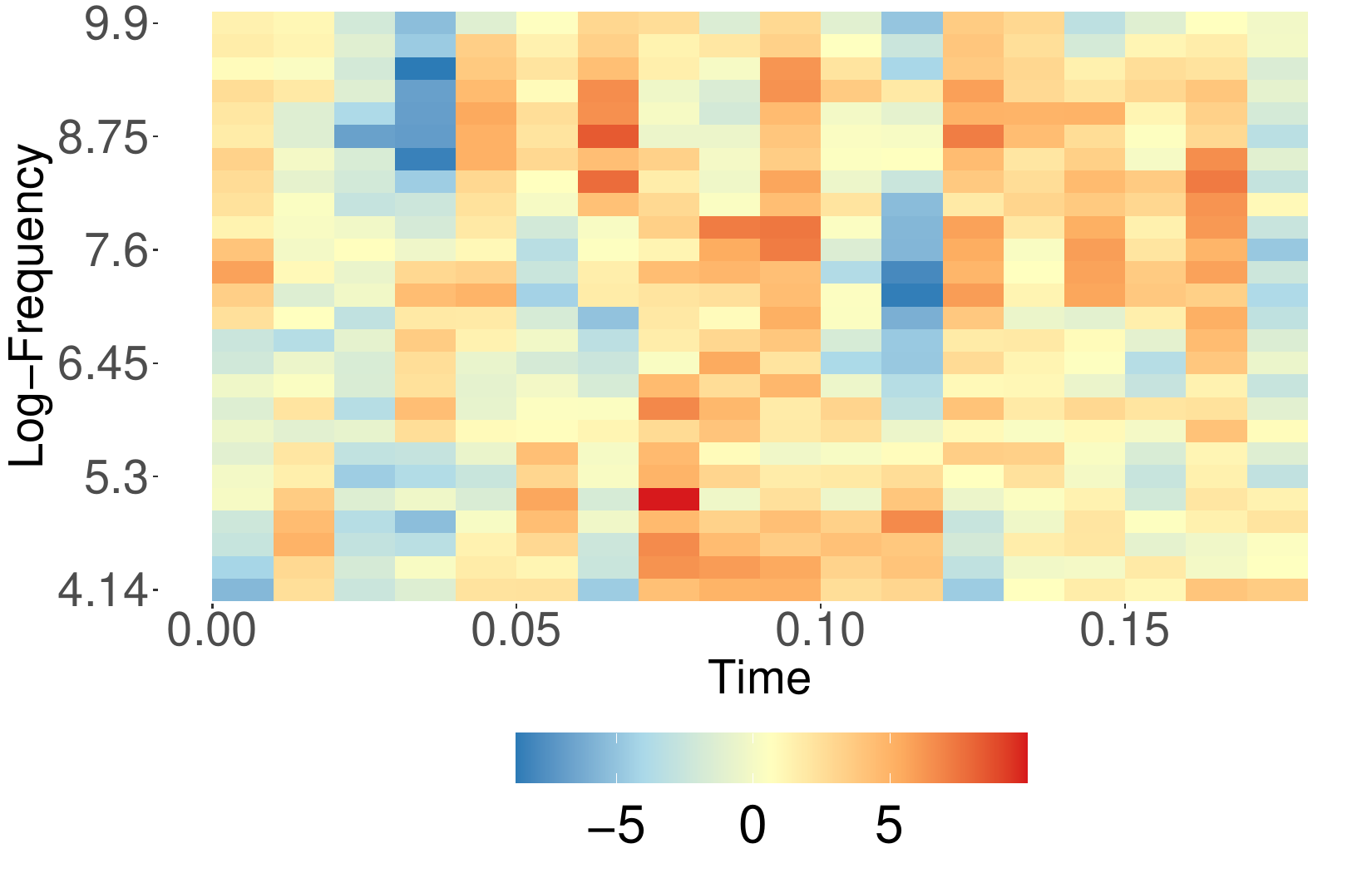}}} 
{\subfloat[$\mathbf{y}_1$]{\includegraphics[scale=0.25]{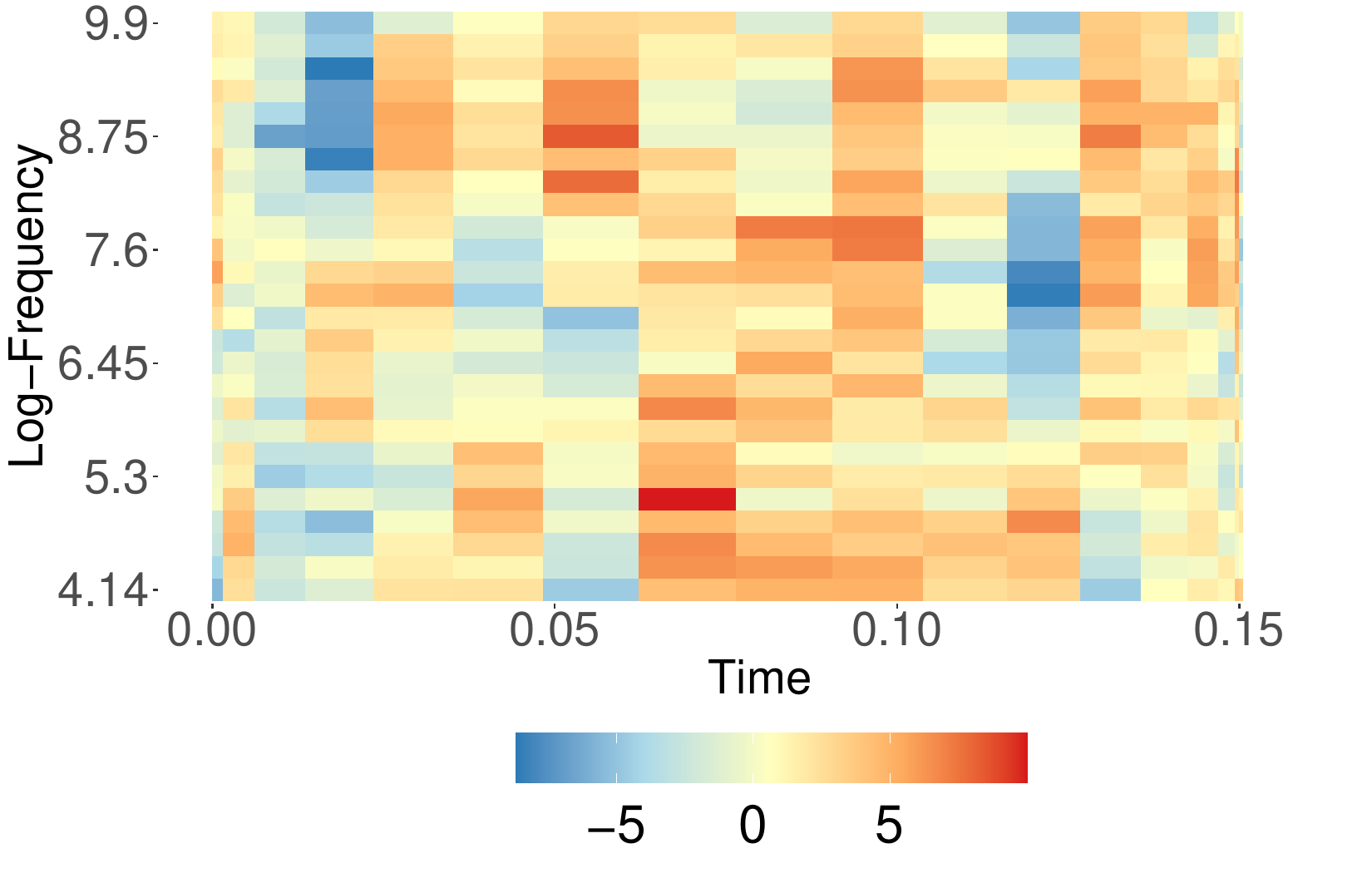}}} \\
{\subfloat[$\mathbf{y}_3$]{\includegraphics[scale=0.25]{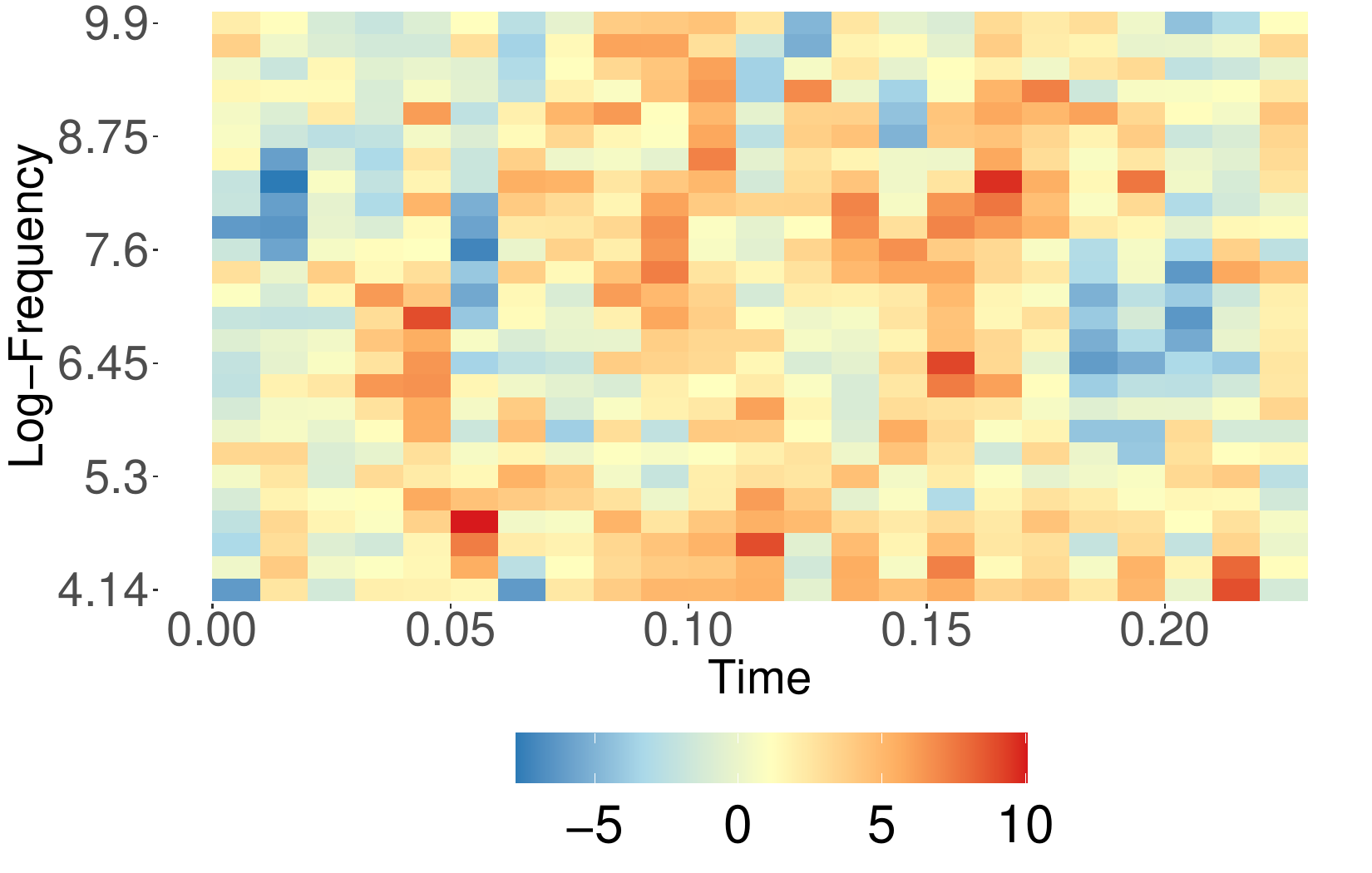}}} 
{\subfloat[$\mathbf{y}_3$]{\includegraphics[scale=0.25]{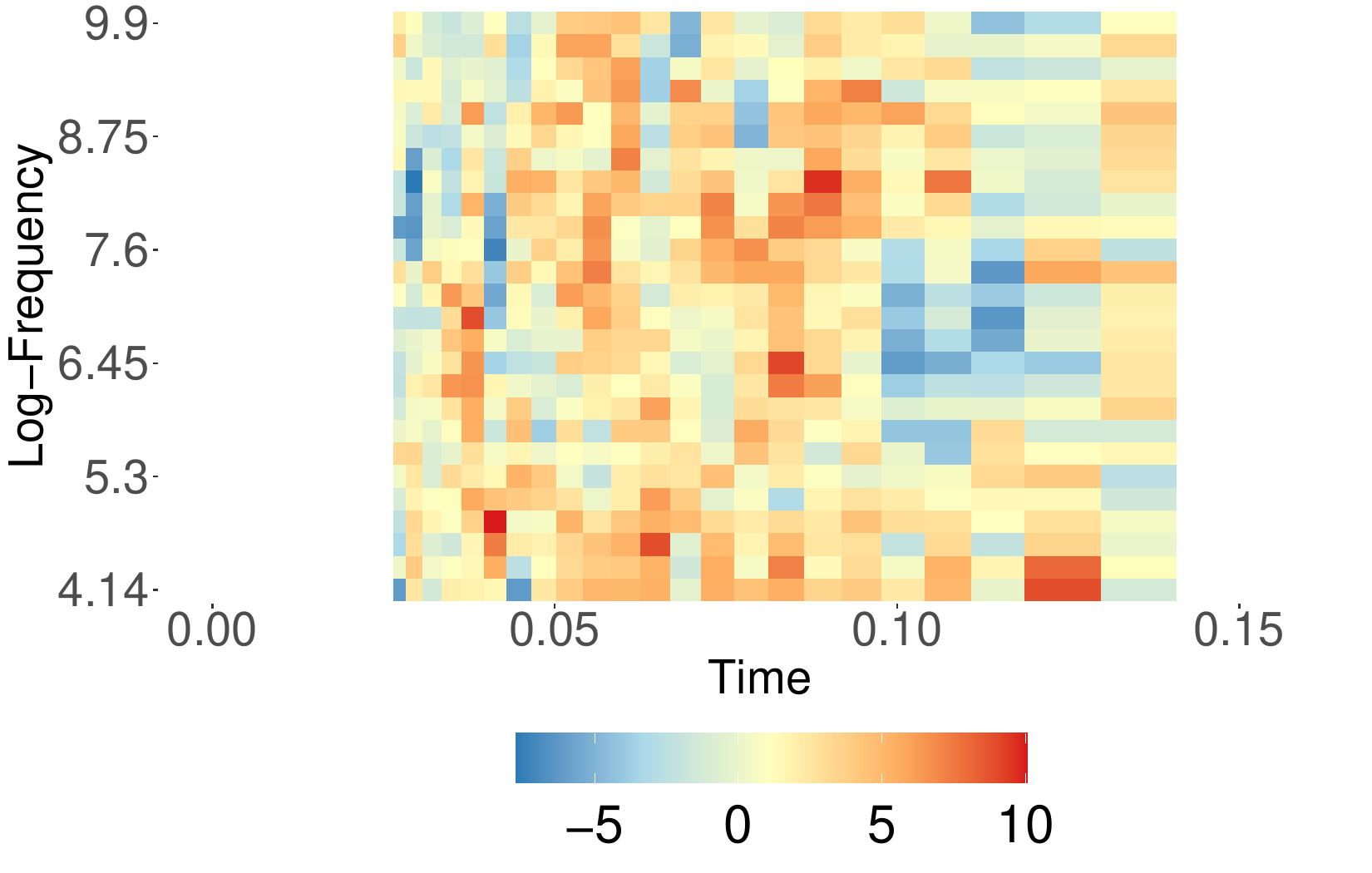}}} \\
{\subfloat[$\mathbf{y}_9$]{\includegraphics[scale=0.25]{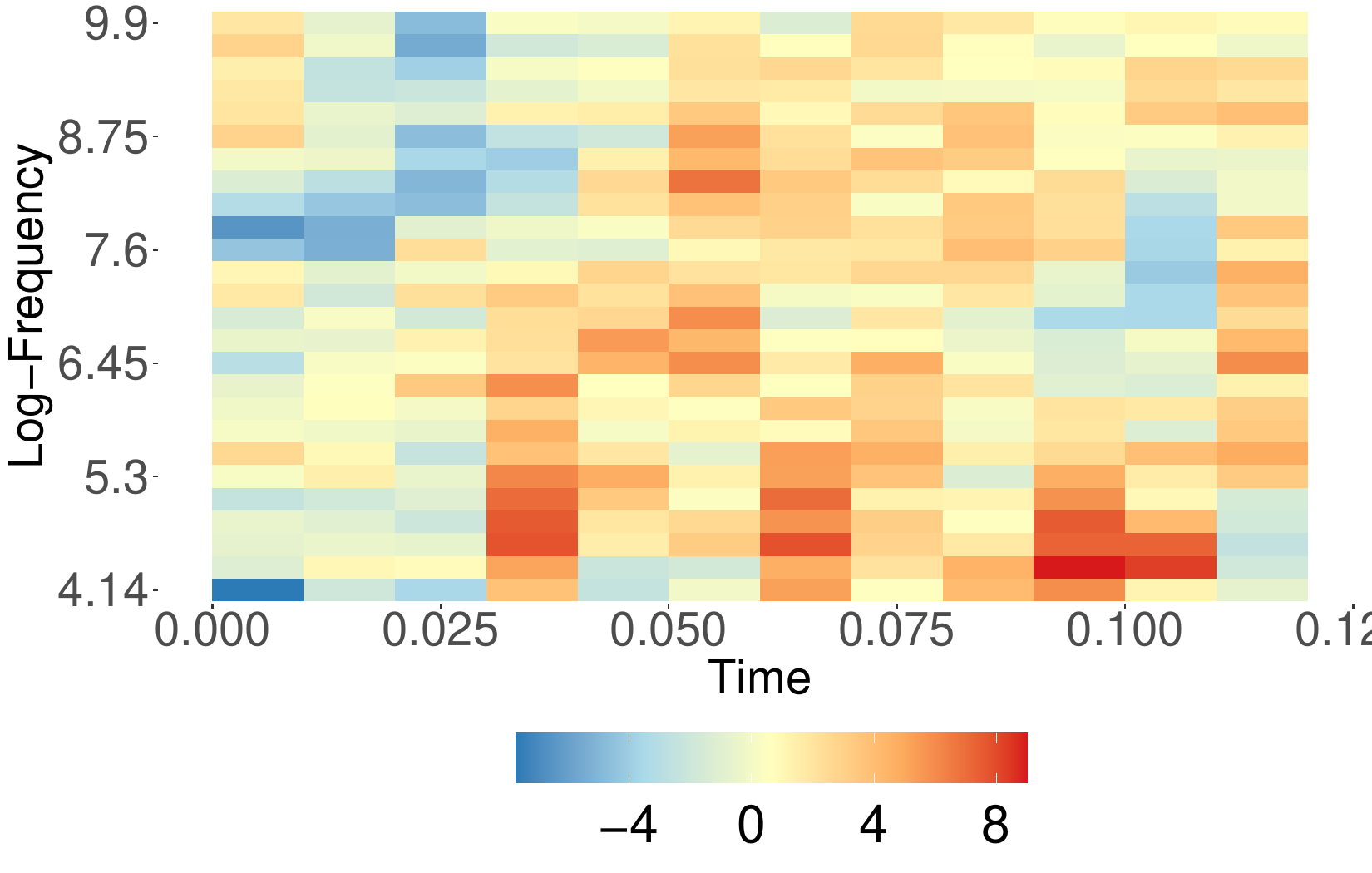}}} 
{\subfloat[$\mathbf{y}_9$]{\includegraphics[scale=0.25]{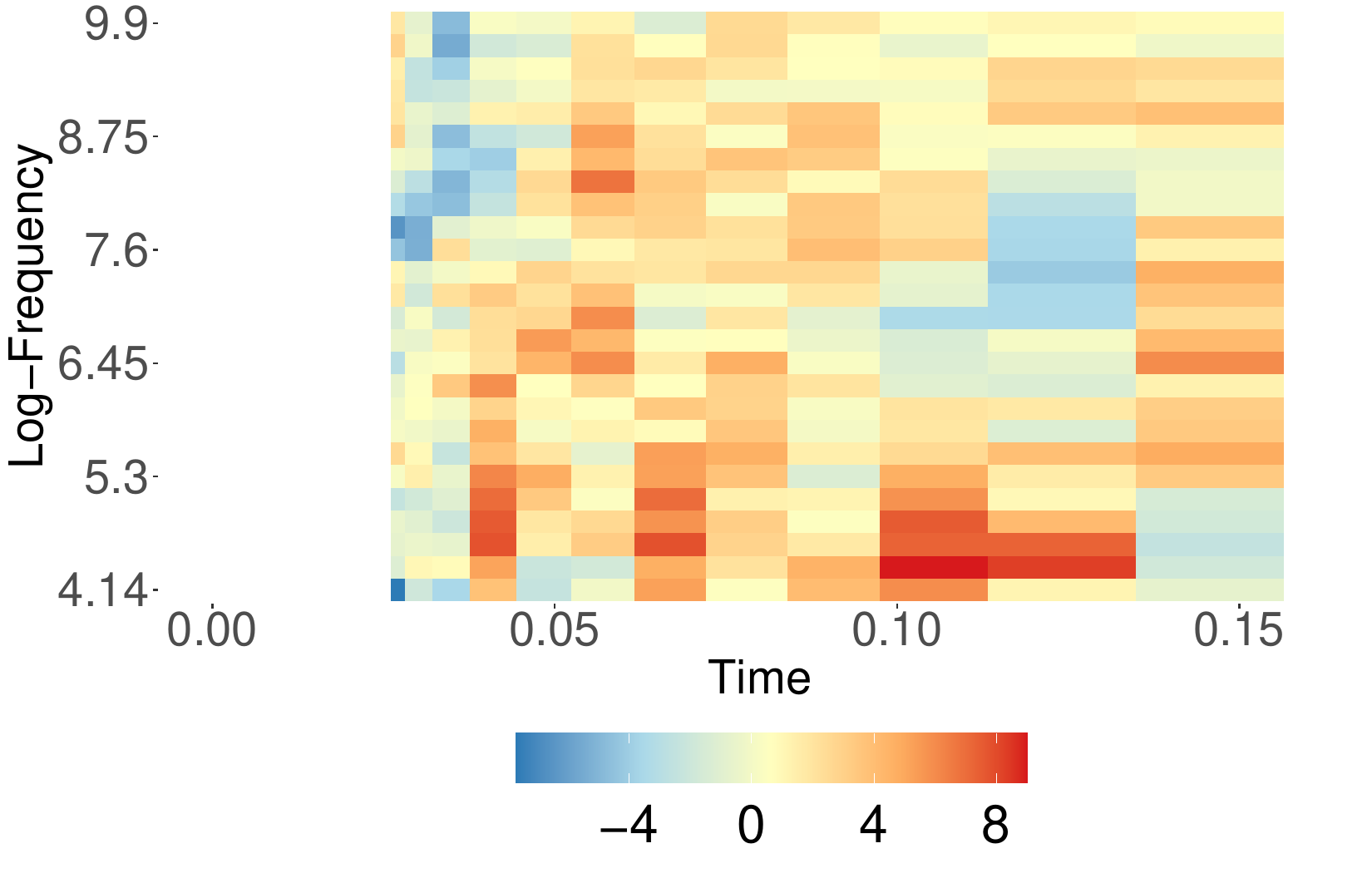}}} \\
{\subfloat[$\mathbf{y}_{14}$]{\includegraphics[scale=0.25]{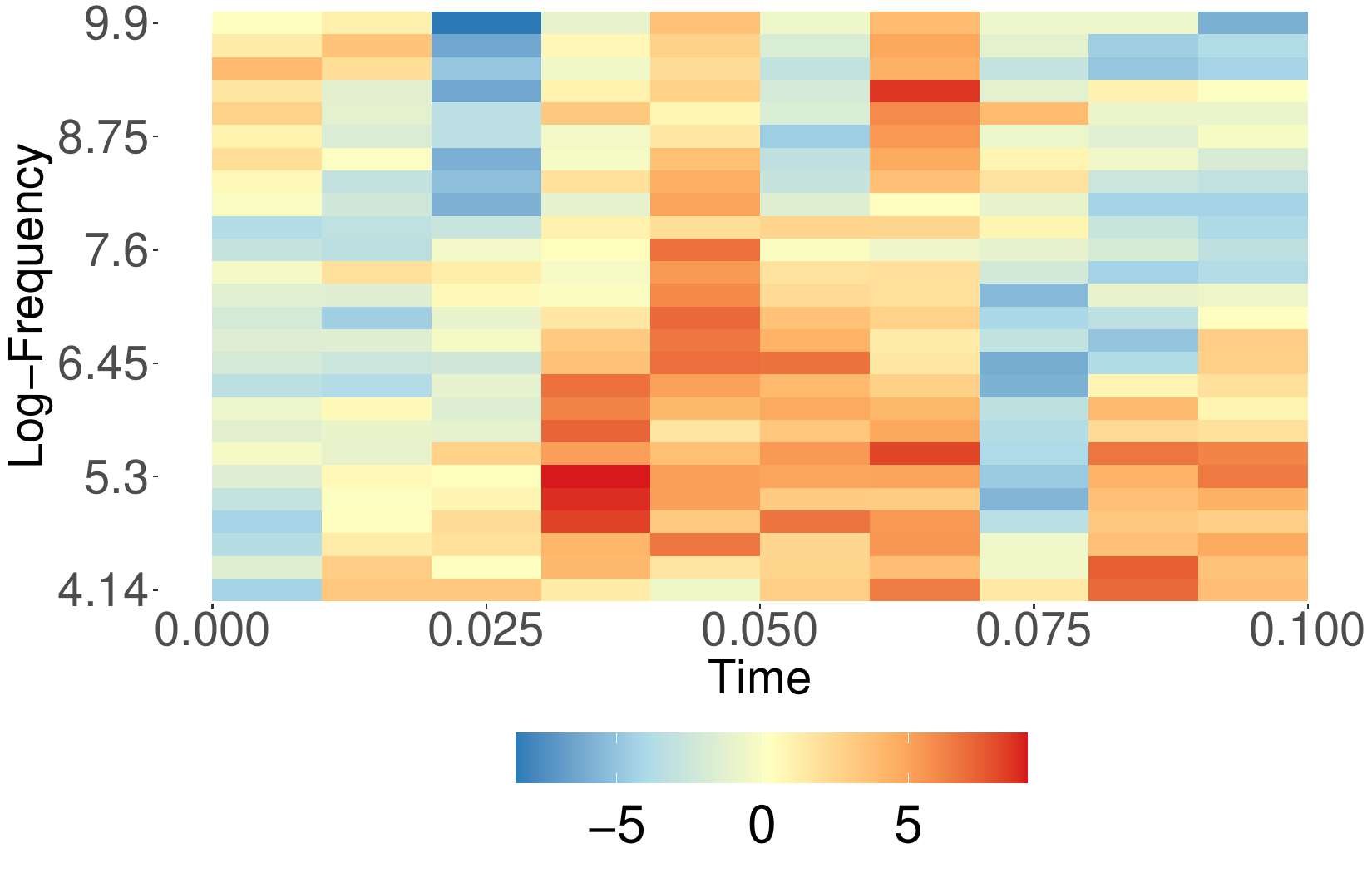}}} 
{\subfloat[$\mathbf{y}_{14}$]{\includegraphics[scale=0.25]{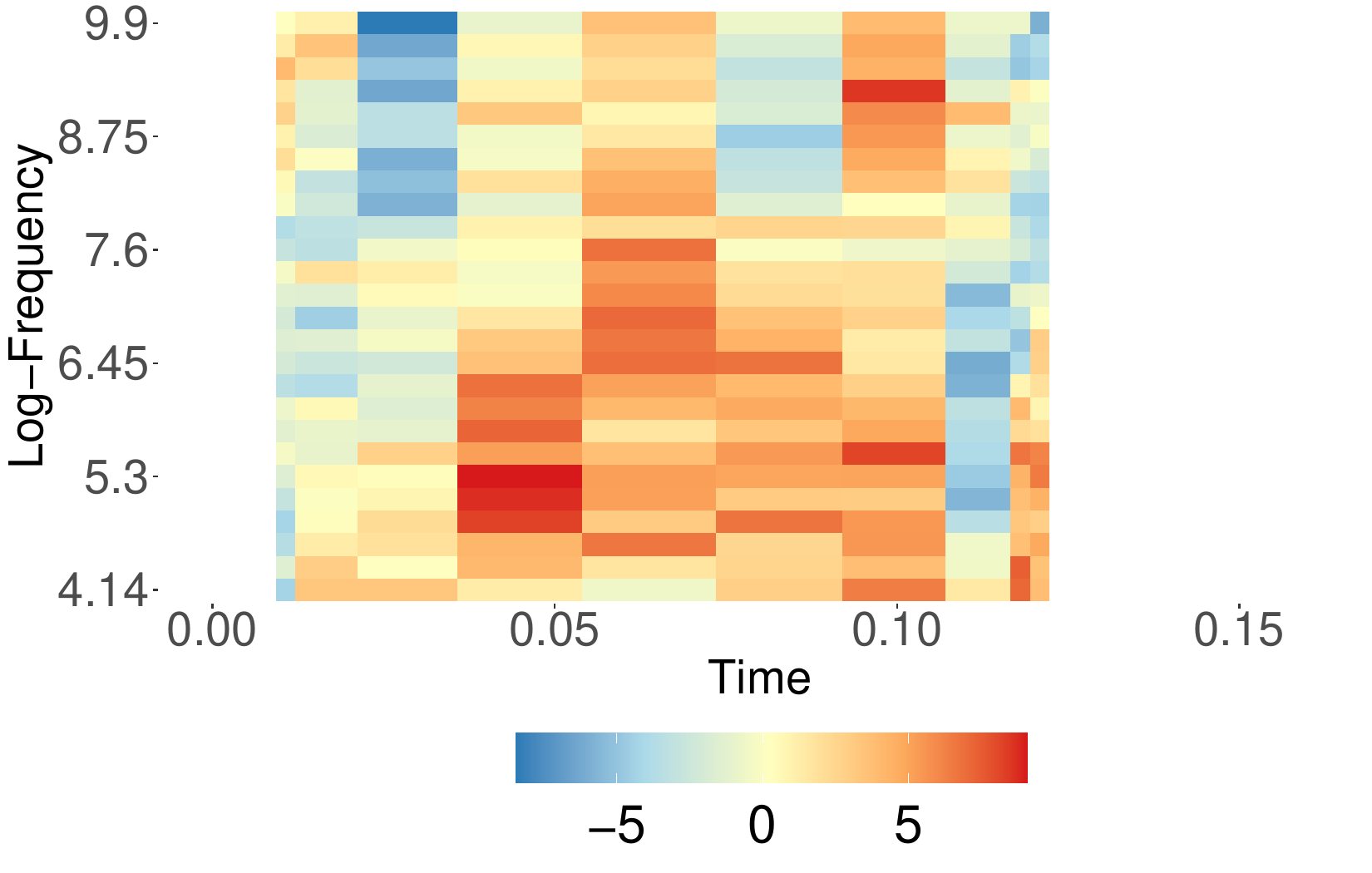}}} 
\caption{Plots of $4$ synthetic sounds. Each row contains two different plots of the same synthetic sound. The $4$ plots on the left column are the spectrogram representation with the $x$-axis being the real-time axis given by the simulated $T_i$ and the constant $0.01$ time-step. The $4$ plots on the right column are the same spectrogram representations with the $x$-axis being given by $\psi(t|\boldsymbol{\chi}_i)$ under the simulated parametrizations in Table \ref{tab:simwarp}. Color graduation is the same across all plots.}
\label{fig:simdata}
\end{figure} 

%\begin{figure}[t]
%{\subfloat[Posterior Mean $\boldsymbol{w}^0$]{\includegraphics[scale=0.25]{wmean}}} 
%{\subfloat[Posterior Variance $\boldsymbol{w}^0$]{\includegraphics[scale=0.25]{wvar}}} \\
%{\subfloat[Posterior Mean $\boldsymbol{y}^0$]{\includegraphics[scale=0.25]{ymean}}} 
%{\subfloat[Posterior Variance $\boldsymbol{y}^0$]{\includegraphics[scale=0.25]{yvar}}} 
%\caption{Results of the simulated study.}
%\label{fig:simdata_res1}
%\end{figure} 

In this section, the model is estimated using a set of synthetic data to see if the model is able to retrieve the parameters used to simulate the data. A total of 15 synthetic sounds are generated over predefined time-frequency grids designed to closely resemble the real data. The number of log-frequency bins for all synthetic sounds is $H = 26$, identical to the available real data, while the number of time coordinates for each synthetic sound, $T_i$, is simulated uniformly from an interval of integers in the range $[10, 25]$. The temporal distance between two consecutive time coordinates is set at $0.01$ seconds, precisely matching that of the real data.

The parameters $\boldsymbol{\theta} = (\phi_h, \phi_d, \phi_c, \gamma, \rho, \sigma, \lambda)$ are deliberately set to make the estimation challenging, providing a level of assurance regarding the model's performance in real data applications. Specifically, the decays are $\phi_h = 0.69$, $\phi_d = 206$, and $\phi_c = 766$, resulting in practical ranges of $4.3$, $0.092$, and $0.024$, respectively. The periodicity is set to $\gamma = 0.06$. Notably, the practical range for the circular decay implies that the circular correlation between any two time coordinates separated by three steps is nearly zero, thereby increasing the difficulty of its estimation. 
The non-separable parameter, variance, and weight between the two latent processes are set to $\rho = 0.85$, $\sigma^2 = 10$, and $\lambda = 0.5$, respectively. 
The values of the data-specific parameters $\boldsymbol{\chi}_i = (\alpha_i,\beta_i,\boldsymbol{\xi}_i)$ used for the simulation are summarized in Table \ref{tab:simwarp} for reference, while we simulate ${\alpha}_i$ and $\tilde{\beta}_i $ from, respectively, $ {\alpha}_i\sim U(0.05,0.2)$ and $\tilde{\beta}_i \sim U(0.75, 0.95)$, with the exception of $\tilde{\alpha}_1$ and $\tilde{\beta}_1$, which are set to $0$ and $1$. The warping parameters $\boldsymbol{\xi}_i = (\zeta_i,\delta_i)$ are simulated from their random effect distributions under the assumption that $a_{\zeta} = a_{\delta} = -1.5$, $b_{\zeta} = b_{\delta} = 1.5$, $\mu_{\zeta} = \mu_{\delta} = 0.7$, and $v_{\zeta}= v_{\delta} = 0.3$. A selected few of the synthetic sounds are depicted in Figure \ref{fig:simdata}.

The model is implemented using the same number of iterations, thinning, burn-in, and priors as the real data. Table \ref{tab:simwarpresults} and Table \ref{tab:simresults} present the estimation summary for the data-specific parameters and the general parameters, respectively. The estimation of the frequency decay $\phi_h$, and the warping parameters $\delta_2$ and $\zeta_9$ yields posterior means with relatively small $95$\% CIs that do not include the true values. However, the inference has largely demonstrated good recovery of the other $94$ out of all $97$ simulated parameters.

\bibliographystyle{imsart-nameyear} % Style BST file
\bibliography{anabibli.bib}  % Bibliography file (usually '*.bib')

\begin{thebibliography}{33}
% BibTex style file: imsart-nameyear.bst, 2017-11-03
% Default style options (sort=1,type=nameyear).
% Used options (sort=1,type=nameyear).

\bibitem[\protect\citeauthoryear{Banerjee, Gelfand and Carlin}{2014}]{Banerjee2014}
\begin{bbook}[author]
\bauthor{\bsnm{Banerjee},~\bfnm{Sudipto}\binits{S.}}, \bauthor{\bsnm{Gelfand},~\bfnm{Alan~E.}\binits{A.~E.}} \AND \bauthor{\bsnm{Carlin},~\bfnm{Bradley~P.}\binits{B.~P.}}
(\byear{2014}).
\btitle{Hierarchical Modeling and Analysis for Spatial Data},
\bedition{Second} ed.
\bpublisher{Chapman and Hall/CRC}, \baddress{New York}.
\end{bbook}
\endbibitem

\bibitem[\protect\citeauthoryear{Bezanson et~al.}{2017}]{bezanson2017julia}
\begin{barticle}[author]
\bauthor{\bsnm{Bezanson},~\bfnm{Jeff}\binits{J.}}, \bauthor{\bsnm{Edelman},~\bfnm{Alan}\binits{A.}}, \bauthor{\bsnm{Karpinski},~\bfnm{Stefan}\binits{S.}} \AND \bauthor{\bsnm{Shah},~\bfnm{Viral~B}\binits{V.~B.}}
(\byear{2017}).
\btitle{Julia: A fresh approach to numerical computing}.
\bjournal{SIAM review}
\bvolume{59}
\bpages{65--98}.
\end{barticle}
\endbibitem

\bibitem[\protect\citeauthoryear{Boersma}{2001}]{praat}
\begin{barticle}[author]
\bauthor{\bsnm{Boersma},~\bfnm{Paul}\binits{P.}}
(\byear{2001}).
\btitle{Praat, a system for doing phonetics by computer}.
\bjournal{Glot International}
\bvolume{5}
\bpages{341--345}.
\end{barticle}
\endbibitem

\bibitem[\protect\citeauthoryear{Datta et~al.}{2016a}]{datta}
\begin{barticle}[author]
\bauthor{\bsnm{Datta},~\bfnm{Abhirup}\binits{A.}}, \bauthor{\bsnm{Banerjee},~\bfnm{Sudipto}\binits{S.}}, \bauthor{\bsnm{Finley},~\bfnm{Andrew~O.}\binits{A.~O.}} \AND \bauthor{\bsnm{Gelfand},~\bfnm{Alan~E.}\binits{A.~E.}}
(\byear{2016}a).
\btitle{Hierarchical nearest-neighbor Gaussian process models for large geostatistical datasets}.
\bjournal{Journal of the American Statistical Association}
\bvolume{111}
\bpages{800-812}.
\bnote{PMID: 29720777}.
\bdoi{10.1080/01621459.2015.1044091}
\end{barticle}
\endbibitem

\bibitem[\protect\citeauthoryear{Datta et~al.}{2016b}]{datta2}
\begin{barticle}[author]
\bauthor{\bsnm{Datta},~\bfnm{Abhirup}\binits{A.}}, \bauthor{\bsnm{Banerjee},~\bfnm{Sudipto}\binits{S.}}, \bauthor{\bsnm{Finley},~\bfnm{Andrew~O.}\binits{A.~O.}}, \bauthor{\bsnm{Hamm},~\bfnm{Nicholas A.~S.}\binits{N.~A.~S.}} \AND \bauthor{\bsnm{Schaap},~\bfnm{Martijn}\binits{M.}}
(\byear{2016}b).
\btitle{Nonseparable dynamic nearest neighbor Gaussian process models for large spatio-temporal data with an application to particulate matter analysis}.
\bjournal{The Annals of Applied Statistics}
\bvolume{10}
\bpages{1286 -- 1316}.
\bdoi{10.1214/16-AOAS931}
\end{barticle}
\endbibitem

\bibitem[\protect\citeauthoryear{DelPero, Pozzi and Masters}{2006}]{DelPero2006}
\begin{barticle}[author]
\bauthor{\bsnm{DelPero},~\bfnm{Massimiliano}\binits{M.}}, \bauthor{\bsnm{Pozzi},~\bfnm{Luca}\binits{L.}} \AND \bauthor{\bsnm{Masters},~\bfnm{Judith~C.}\binits{J.~C.}}
(\byear{2006}).
\btitle{A Composite Molecular Phylogeny of Living Lemuroid Primates}.
\bjournal{Folia Primatologica}
\bvolume{77}
\bpages{434-445}.
\bdoi{10.1159/000095390}
\end{barticle}
\endbibitem

\bibitem[\protect\citeauthoryear{Gamba, Friard and Giacoma}{2012}]{gamba2012}
\begin{barticle}[author]
\bauthor{\bsnm{Gamba},~\bfnm{Marco}\binits{M.}}, \bauthor{\bsnm{Friard},~\bfnm{Olivier}\binits{O.}} \AND \bauthor{\bsnm{Giacoma},~\bfnm{Cristina}\binits{C.}}
(\byear{2012}).
\btitle{Vocal tract morphology determines species-specific features in vocal signals of lemurs (Eulemur)}.
\bjournal{International Journal of Primatology}
\bvolume{33}
\bpages{1453-1466}.
\bdoi{10.1007/s10764-012-9635-y}
\end{barticle}
\endbibitem

\bibitem[\protect\citeauthoryear{Gamba and Giacoma}{2007}]{gamba2007}
\begin{barticle}[author]
\bauthor{\bsnm{Gamba},~\bfnm{Marco}\binits{M.}} \AND \bauthor{\bsnm{Giacoma},~\bfnm{Cristina}\binits{C.}}
(\byear{2007}).
\btitle{Quantitative acoustic analysis of the vocal repertoire of the crowned lemur}.
\bjournal{Ethology Ecology \& Evolution}
\bvolume{19(4)}
\bpages{323-343}.
\bdoi{10.1080/08927014.2007.9522555}
\end{barticle}
\endbibitem

\bibitem[\protect\citeauthoryear{Gamba and Giacoma}{2008}]{gamba23008}
\begin{barticle}[author]
\bauthor{\bsnm{Gamba},~\bfnm{M.}\binits{M.}} \AND \bauthor{\bsnm{Giacoma},~\bfnm{C.}\binits{C.}}
(\byear{2008}).
\btitle{Subspecific Divergence in the Black Lemur’s Low-Pitched Vocalizations}.
\bjournal{The Open Acoustics Journal}
\bvolume{7}
\bpages{49-53}.
\end{barticle}
\endbibitem

\bibitem[\protect\citeauthoryear{Gamba et~al.}{2016}]{gamba2016}
\begin{barticle}[author]
\bauthor{\bsnm{Gamba},~\bfnm{Marco}\binits{M.}}, \bauthor{\bsnm{Torti},~\bfnm{Valeria}\binits{V.}}, \bauthor{\bsnm{Estienne},~\bfnm{Vittoria}\binits{V.}}, \bauthor{\bsnm{Randrianarison},~\bfnm{RoseM.}\binits{R.}}, \bauthor{\bsnm{Valente},~\bfnm{Daria}\binits{D.}}, \bauthor{\bsnm{Rovara},~\bfnm{Paolo}\binits{P.}}, \bauthor{\bsnm{Bonadonna},~\bfnm{Giovanna}\binits{G.}}, \bauthor{\bsnm{Friard},~\bfnm{Olivier}\binits{O.}} \AND \bauthor{\bsnm{Giacoma},~\bfnm{Cristina}\binits{C.}}
(\byear{2016}).
\btitle{The Indris have got rhythm! Timing and pitch variation of a primate song examined between sexes and age classes}.
\bjournal{Frontiers in Neuroscience}
\bvolume{10(249)}.
\bdoi{10.3389/fnins.2016.00249}
\end{barticle}
\endbibitem

\bibitem[\protect\citeauthoryear{Gelfand et~al.}{2010}]{gelfand2010}
\begin{bbook}[author]
\bauthor{\bsnm{Gelfand},~\bfnm{A.}\binits{A.}}, \bauthor{\bsnm{Diggle},~\bfnm{P.}\binits{P.}}, \bauthor{\bsnm{Fuentes},~\bfnm{M.}\binits{M.}} \AND \bauthor{\bsnm{Guttorp},~\bfnm{P.}\binits{P.}}
(\byear{2010}).
\btitle{Handbook of Spatial Statistics}.
\bpublisher{Chapman and Hall}.
\end{bbook}
\endbibitem

\bibitem[\protect\citeauthoryear{Gneiting}{2002}]{gneiting}
\begin{barticle}[author]
\bauthor{\bsnm{Gneiting},~\bfnm{Tilmann}\binits{T.}}
(\byear{2002}).
\btitle{Nonseparable, stationary covariance functions for space-time data}.
\bjournal{Journal of the American Statistical Association}
\bvolume{97(458)}
\bpages{590-600}.
\end{barticle}
\endbibitem

\bibitem[\protect\citeauthoryear{Graps}{1995}]{Graps1995}
\begin{barticle}[author]
\bauthor{\bsnm{Graps},~\bfnm{A.}\binits{A.}}
(\byear{1995}).
\btitle{An introduction to wavelets}.
\bjournal{IEEE Computational Science and Engineering}
\bvolume{2}
\bpages{50-61}.
\bdoi{10.1109/99.388960}
\end{barticle}
\endbibitem

\bibitem[\protect\citeauthoryear{{De Gregorio} et~al.}{2021}]{DEGREGORIO2021R1379}
\begin{barticle}[author]
\bauthor{\bsnm{{De Gregorio}},~\bfnm{Chiara}\binits{C.}}, \bauthor{\bsnm{Valente},~\bfnm{Daria}\binits{D.}}, \bauthor{\bsnm{Raimondi},~\bfnm{Teresa}\binits{T.}}, \bauthor{\bsnm{Torti},~\bfnm{Valeria}\binits{V.}}, \bauthor{\bsnm{Miaretsoa},~\bfnm{Longondraza}\binits{L.}}, \bauthor{\bsnm{Friard},~\bfnm{Olivier}\binits{O.}}, \bauthor{\bsnm{Giacoma},~\bfnm{Cristina}\binits{C.}}, \bauthor{\bsnm{Ravignani},~\bfnm{Andrea}\binits{A.}} \AND \bauthor{\bsnm{Gamba},~\bfnm{Marco}\binits{M.}}
(\byear{2021}).
\btitle{Categorical rhythms in a singing primate}.
\bjournal{Current Biology}
\bvolume{31}
\bpages{R1379-R1380}.
\end{barticle}
\endbibitem

\bibitem[\protect\citeauthoryear{Jona~Lasinio, Mastrantonio and Pollice}{2013}]{Jona2013b}
\begin{barticle}[author]
\bauthor{\bsnm{Jona~Lasinio},~\bfnm{G.}\binits{G.}}, \bauthor{\bsnm{Mastrantonio},~\bfnm{G.}\binits{G.}} \AND \bauthor{\bsnm{Pollice},~\bfnm{A.}\binits{A.}}
(\byear{2013}).
\btitle{Discussing the ``big n problem''}.
\bjournal{Statistical Methods and Applications}
\bvolume{22}
\bpages{97-112}.
\end{barticle}
\endbibitem

\bibitem[\protect\citeauthoryear{Kershenbaum et~al.}{2016}]{kershenbaum2016}
\begin{barticle}[author]
\bauthor{\bsnm{Kershenbaum},~\bfnm{Arik}\binits{A.}}, \bauthor{\bsnm{Blumstein},~\bfnm{Daniel~T.}\binits{D.~T.}}, \bauthor{\bsnm{Roch},~\bfnm{Marie~A.}\binits{M.~A.}}, \bauthor{\bsnm{Akcay},~\bfnm{Caglar}\binits{C.}} \AND \bauthor{\bparticle{et} \bsnm{al.}}
(\byear{2016}).
\btitle{Acoustic sequences in non-human animals: a tutorial review and prospectus}.
\bjournal{Biological Review}
\bvolume{91(1)}
\bpages{13–52}.
\end{barticle}
\endbibitem

\bibitem[\protect\citeauthoryear{Kumar and Foufoula-Georgiou}{1997}]{kumar1997}
\begin{barticle}[author]
\bauthor{\bsnm{Kumar},~\bfnm{Praveen}\binits{P.}} \AND \bauthor{\bsnm{Foufoula-Georgiou},~\bfnm{Efi}\binits{E.}}
(\byear{1997}).
\btitle{Wavelet analysis for geophysical applications}.
\bjournal{Reviews of geophysics}
\bvolume{35}
\bpages{385--412}.
\end{barticle}
\endbibitem

\bibitem[\protect\citeauthoryear{Macedonia and Stanger}{1994}]{Macedonia94}
\begin{barticle}[author]
\bauthor{\bsnm{Macedonia},~\bfnm{Joseph~M.}\binits{J.~M.}} \AND \bauthor{\bsnm{Stanger},~\bfnm{Kathrin~F.}\binits{K.~F.}}
(\byear{1994}).
\btitle{Phylogeny of the Lemuridae Revisited: Evidence from Communication Signals}.
\bjournal{Folia Primatologica}
\bvolume{63}
\bpages{1 - 43}.
\bdoi{10.1159/000156787}
\end{barticle}
\endbibitem

\bibitem[\protect\citeauthoryear{Maretti et~al.}{2010}]{Maretti2010}
\begin{barticle}[author]
\bauthor{\bsnm{Maretti},~\bfnm{Giovanna}\binits{G.}}, \bauthor{\bsnm{Sorrentino},~\bfnm{Viviana}\binits{V.}}, \bauthor{\bsnm{Finomana},~\bfnm{Andriamasitoly}\binits{A.}}, \bauthor{\bsnm{Gamba},~\bfnm{Marco}\binits{M.}} \AND \bauthor{\bsnm{Giacoma},~\bfnm{Cristina}\binits{C.}}
(\byear{2010}).
\btitle{Not just a pretty song: an overview of the vocal repertoire of Indri Indri}.
\bjournal{Journal of Anthropological Sciences}
\bvolume{88}
\bpages{151-165}.
\bnote{PMID: 20834055}.
\end{barticle}
\endbibitem

\bibitem[\protect\citeauthoryear{Mastrantonio, Gelfand and Jona~Lasinio}{2016}]{mastrantonio2015c}
\begin{barticle}[author]
\bauthor{\bsnm{Mastrantonio},~\bfnm{G.}\binits{G.}}, \bauthor{\bsnm{Gelfand},~\bfnm{A.~E.}\binits{A.~E.}} \AND \bauthor{\bsnm{Jona~Lasinio},~\bfnm{G.}\binits{G.}}
(\byear{2016}).
\btitle{The wrapped skew {G}aussian process for analyzing spatio-temporal data}.
\bjournal{Stochastic Environmental Research and Risk Assessment}
\bvolume{30}
\bpages{2231–2242}.
\bdoi{10.1007/s00477-015-1163-9}
\end{barticle}
\endbibitem

\bibitem[\protect\citeauthoryear{Mastrantonio, Jona~Lasinio and Gelfand}{2015}]{mastrantonio2015b}
\begin{barticle}[author]
\bauthor{\bsnm{Mastrantonio},~\bfnm{G.}\binits{G.}}, \bauthor{\bsnm{Jona~Lasinio},~\bfnm{G.}\binits{G.}} \AND \bauthor{\bsnm{Gelfand},~\bfnm{A.~E.}\binits{A.~E.}}
(\byear{2015}).
\btitle{Spatio-temporal circular models with non-separable covariance structure}.
\bjournal{TEST}
\bvolume{25}
\bpages{331--350}.
\bdoi{10.1007/s11749-015-0458-y}
\end{barticle}
\endbibitem

\bibitem[\protect\citeauthoryear{Mastrantonio et~al.}{2017}]{mastrantonio}
\begin{barticle}[author]
\bauthor{\bsnm{Mastrantonio},~\bfnm{Gianluca}\binits{G.}}, \bauthor{\bsnm{Lasinio},~\bfnm{Giovanna~Jona}\binits{G.~J.}}, \bauthor{\bsnm{Pollice},~\bfnm{Aleesio}\binits{A.}}, \bauthor{\bsnm{Capotorti},~\bfnm{Giulia}\binits{G.}}, \bauthor{\bsnm{Teodonio},~\bfnm{Lorenzo}\binits{L.}}, \bauthor{\bsnm{Genova},~\bfnm{Giulio}\binits{G.}} \AND \bauthor{\bsnm{Blasi},~\bfnm{Carlo}\binits{C.}}
(\byear{2017}).
\btitle{A hierarchical multivariate spatio-temporal model for large clustered climate data with annual cycles}.
\bjournal{Annals of Applied Statistics}
\bvolume{13(2)}
\bpages{797-823}.
\bdoi{10.1214/18-AOAS1212}
\end{barticle}
\endbibitem

\bibitem[\protect\citeauthoryear{Pfl{\"u}ger and Fichtel}{2012}]{Pfluger}
\begin{barticle}[author]
\bauthor{\bsnm{Pfl{\"u}ger},~\bfnm{Femke~J.}\binits{F.~J.}} \AND \bauthor{\bsnm{Fichtel},~\bfnm{Claudia}\binits{C.}}
(\byear{2012}).
\btitle{On the function of redfronted lemur's close calls}.
\bjournal{Animal Cognition}
\bvolume{15}
\bpages{823--831}.
\bdoi{10.1007/s10071-012-0507-9}
\end{barticle}
\endbibitem

\bibitem[\protect\citeauthoryear{Pozzi, Gamba and Giacoma}{2010}]{pozzi2010}
\begin{barticle}[author]
\bauthor{\bsnm{Pozzi},~\bfnm{Luca}\binits{L.}}, \bauthor{\bsnm{Gamba},~\bfnm{Marco}\binits{M.}} \AND \bauthor{\bsnm{Giacoma},~\bfnm{Cristina}\binits{C.}}
(\byear{2010}).
\btitle{The use of Artificial Neural Networks to classify primate vocalizations: a pilot study on black lemurs}.
\bjournal{American Journal of Primatology}
\bvolume{72}
\bpages{337-348}.
\bdoi{https://doi.org/10.1002/ajp.20786}
\end{barticle}
\endbibitem

\bibitem[\protect\citeauthoryear{Sainburg, Thielk and Gentner}{2020}]{sainburg}
\begin{barticle}[author]
\bauthor{\bsnm{Sainburg},~\bfnm{Tim}\binits{T.}}, \bauthor{\bsnm{Thielk},~\bfnm{Marvin}\binits{M.}} \AND \bauthor{\bsnm{Gentner},~\bfnm{Timothy~Q.}\binits{T.~Q.}}
(\byear{2020}).
\btitle{Finding, visualizing, and quantifying latent structure across diverse animal vocal repertoires}.
\bjournal{PLOS Computational Biology}
\bvolume{16(10)}.
\end{barticle}
\endbibitem

\bibitem[\protect\citeauthoryear{Shahin~Tavakoli and Coleman}{2019}]{tavakoli}
\begin{barticle}[author]
\bauthor{\bsnm{Shahin~Tavakoli},~\bfnm{John A. D.~Aston}\binits{J.~A. D.~A.} \bsuffix{Davide~Pigoli}} \AND \bauthor{\bsnm{Coleman},~\bfnm{John~S.}\binits{J.~S.}}
(\byear{2019}).
\btitle{A Spatial Modeling Approach for Linguistic Object Data: Analyzing Dialect Sound Variations Across Great Britain}.
\bjournal{Journal of the American Statistical Association}
\bvolume{114}
\bpages{1081--1096}.
\bdoi{10.1080/01621459.2019.1607357}
\end{barticle}
\endbibitem

\bibitem[\protect\citeauthoryear{Shirota and Gelfand}{2017}]{Shinichiro2017}
\begin{barticle}[author]
\bauthor{\bsnm{Shirota},~\bfnm{Shinichiro}\binits{S.}} \AND \bauthor{\bsnm{Gelfand},~\bfnm{Alan~E.}\binits{A.~E.}}
(\byear{2017}).
\btitle{{Space and circular time log Gaussian Cox processes with application to crime event data}}.
\bjournal{The Annals of Applied Statistics}
\bvolume{11}
\bpages{481 -- 503}.
\bdoi{10.1214/16-AOAS960}
\end{barticle}
\endbibitem

\bibitem[\protect\citeauthoryear{Sperber et~al.}{2017}]{sperber2017}
\begin{barticle}[author]
\bauthor{\bsnm{Sperber},~\bfnm{Anna~Lucia}\binits{A.~L.}}, \bauthor{\bsnm{Werner},~\bfnm{Lynne~M.}\binits{L.~M.}}, \bauthor{\bsnm{Kappeler},~\bfnm{Peter~M.}\binits{P.~M.}} \AND \bauthor{\bsnm{Fichtel},~\bfnm{Claudia}\binits{C.}}
(\byear{2017}).
\btitle{Grunt to go - Vocal coordinate of group movements in redfronted lemurs}.
\bjournal{Ethology}
\bvolume{123(12)}
\bpages{894-905}.
\end{barticle}
\endbibitem

\bibitem[\protect\citeauthoryear{Valente et~al.}{2019}]{valente2019}
\begin{barticle}[author]
\bauthor{\bsnm{Valente},~\bfnm{Daria}\binits{D.}}, \bauthor{\bsnm{De~Gregorio},~\bfnm{Chiara}\binits{C.}}, \bauthor{\bsnm{Torti},~\bfnm{Valeria}\binits{V.}}, \bauthor{\bsnm{Miaretsoa},~\bfnm{Longondraza}\binits{L.}}, \bauthor{\bsnm{Friard},~\bfnm{Olivier}\binits{O.}}, \bauthor{\bsnm{Randrianarison},~\bfnm{Rose~Marie}\binits{R.~M.}}, \bauthor{\bsnm{Giacoma},~\bfnm{Cristina}\binits{C.}} \AND \bauthor{\bsnm{Gamba},~\bfnm{Marco}\binits{M.}}
(\byear{2019}).
\btitle{Finding Meanings in Low Dimensional Structures: Stochastic Neighbor Embedding Applied to the Analysis of Indri indri Vocal Repertoire}.
\bjournal{Animals}
\bvolume{9}.
\bdoi{10.3390/ani9050243}
\end{barticle}
\endbibitem

\bibitem[\protect\citeauthoryear{Valente et~al.}{2022}]{valente22}
\begin{barticle}[author]
\bauthor{\bsnm{Valente},~\bfnm{Daria}\binits{D.}}, \bauthor{\bsnm{Miaretsoa},~\bfnm{Longondraza}\binits{L.}}, \bauthor{\bsnm{Anania},~\bfnm{Alessio}\binits{A.}}, \bauthor{\bsnm{Costa},~\bfnm{Francesco}\binits{F.}}, \bauthor{\bsnm{Mascaro},~\bfnm{Alessandra}\binits{A.}}, \bauthor{\bsnm{Raimondi},~\bfnm{Teresa}\binits{T.}}, \bauthor{\bsnm{De~Gregorio},~\bfnm{Chiara}\binits{C.}}, \bauthor{\bsnm{Torti},~\bfnm{Valeria}\binits{V.}}, \bauthor{\bsnm{Friard},~\bfnm{Olivier}\binits{O.}}, \bauthor{\bsnm{Ratsimbazafy},~\bfnm{Jonah}\binits{J.}}, \bauthor{\bsnm{Giacoma},~\bfnm{Cristina}\binits{C.}} \AND \bauthor{\bsnm{Gamba},~\bfnm{Marco}\binits{M.}}
(\byear{2022}).
\btitle{Comparative Analysis of the Vocal Repertoires of the Indri (Indri indri) and the Diademed Sifaka (Propithecus diadema)}.
\bjournal{International Journal of Primatology}
\bvolume{43}
\bpages{733--751}.
\bdoi{10.1007/s10764-022-00287-x}
\end{barticle}
\endbibitem

\bibitem[\protect\citeauthoryear{Wilkinson}{2019}]{WilkinsonW}
\begin{bphdthesis}[author]
\bauthor{\bsnm{Wilkinson},~\bfnm{W}\binits{W.}}
(\byear{2019}).
\btitle{Gaussian Process Modelling for Audio Signals},
\btype{PhD thesis},
\bpublisher{Queen Mary University of London},
\baddress{Example City, CA}.
\end{bphdthesis}
\endbibitem

\bibitem[\protect\citeauthoryear{Wilkinson and Davy}{2018}]{phylogram}
\begin{bmanual}[author]
\bauthor{\bsnm{Wilkinson},~\bfnm{Shaun~P.}\binits{S.~P.}} \AND \bauthor{\bsnm{Davy},~\bfnm{Simon~K.}\binits{S.~K.}}
(\byear{2018}).
\btitle{{phylogram}: an R package for phylogenetic analysis with nested lists}.
\bdoi{10.21105/joss.00790}
\end{bmanual}
\endbibitem

\bibitem[\protect\citeauthoryear{Zimmermann}{2017}]{Zimmermann2017}
\begin{binbook}[author]
\bauthor{\bsnm{Zimmermann},~\bfnm{Elke}\binits{E.}}
(\byear{2017}).
\btitle{Evolutionary Origins of Primate Vocal Communication: Diversity, Flexibility, and Complexity of Vocalizations in Basal Primates}
In \bbooktitle{Primate Hearing and Communication}
\bpages{109--140}.
\bpublisher{Springer International Publishing}, \baddress{Cham}.
\end{binbook}
\endbibitem

\end{thebibliography}


\begin{thebibliography}{}

\end{thebibliography}

\end{document}